\pdfoutput=1
\documentclass[aps,prd, amsmath,amssymb,superscriptaddress,showkeys,nofootinbib,reprint,preprintnumbers]{revtex4-1}
\usepackage{color}
\usepackage{aas_macros,graphicx}
\usepackage{hyperref,breakurl}
\usepackage[usenames,dvipsnames]{xcolor}
\usepackage{dcolumn}
\usepackage{bm}
\usepackage{hyperref}
\usepackage{array}
\usepackage{dcolumn}

\usepackage{amsmath,amssymb,latexsym,times}
\DeclareMathOperator\arctanh{arctanh}
 
\newcommand{\imshape}{\textsc{im3shape}}
\newcommand{\metacal}{\textsc{Metacalibration}}
\newcommand{\ngmix}{\textsc{ngmix}}

\newcommand{\healpix}{\textsc{HEALPix}}
\newcommand{\flask}{\textsc{FLASK}}
\newcommand{\mof}{\textsc{MOF}}
\newcommand{\bpz}{\textsc{BPZ}}
\newcommand{\cosmolike}{\textsc{CosmoLike}}
\newcommand{\cosmosis}{\textsc{CosmoSIS}}
\newcommand{\cosmos}{COSMOS}
\newcommand{\redmagic}{\textsc{redMaGiC}}
\newcommand{\photoz}{photo-$z$}

\def\Msun{\, h^{-1} \, {\rm M_{\odot}}}

\newcommand{\mpcoh}{\,h^{-1}\,{\rm Mpc}}

\newcommand{\bea}{\begin{eqnarray}}
\newcommand{\be}{\begin{equation}}
\newcommand{\ben}{\begin{enumerate}}
\newcommand{\bi}{\begin{itemize}}
\newcommand{\eea}{\end{eqnarray}}
\newcommand{\ee}{\end{equation}}
\newcommand{\ei}{\end{itemize}}
\newcommand{\een}{\end{enumerate}}

\newcommand{\om}{\Omega_\mr m}

\newcommand{\sig}{\sigma_8}

\newcommand{\mr}{\mathrm}

\newcommand{\mcalRg}{\mbox{\boldmath $R_\gamma$}}

\newcommand{\mcalRgmean}{\mbox{\boldmath $\langle R_\gamma \rangle$}}
\newcommand{\mcalRSmean}{\mbox{\boldmath $\langle R_S \rangle$}}

\def\bigstrutdown{\vrule width0pt height0truecm depth0.16truecm}

\begin{document}

\title{Dark Energy Survey Year 1 Results: Cosmological Constraints from Cosmic Shear}

\author{M.~A.~Troxel}
\email[Corresponding author: ]{troxel.18@osu.edu}
\affiliation{Center for Cosmology and Astro-Particle Physics, The Ohio State University, Columbus, OH 43210, USA}
\affiliation{Department of Physics, The Ohio State University, Columbus, OH 43210, USA}
\author{N.~MacCrann}
\affiliation{Center for Cosmology and Astro-Particle Physics, The Ohio State University, Columbus, OH 43210, USA}
\affiliation{Department of Physics, The Ohio State University, Columbus, OH 43210, USA}
\author{J.~Zuntz}
\affiliation{Institute for Astronomy, University of Edinburgh, Edinburgh EH9 3HJ, UK}
\author{T.~F.~Eifler}
\affiliation{Department of Physics, California Institute of Technology, Pasadena, CA 91125, USA}
\affiliation{Jet Propulsion Laboratory, California Institute of Technology, 4800 Oak Grove Dr., Pasadena, CA 91109, USA}
\author{E.~Krause}
\affiliation{Kavli Institute for Particle Astrophysics \& Cosmology, P. O. Box 2450, Stanford University, Stanford, CA 94305, USA}
\author{S.~Dodelson}
\affiliation{Fermi National Accelerator Laboratory, P. O. Box 500, Batavia, IL 60510, USA}
\affiliation{Kavli Institute for Cosmological Physics, University of Chicago, Chicago, IL 60637, USA}
\author{D.~Gruen}
\email[]{Einstein Fellow}
\affiliation{Kavli Institute for Particle Astrophysics \& Cosmology, P. O. Box 2450, Stanford University, Stanford, CA 94305, USA}
\affiliation{SLAC National Accelerator Laboratory, Menlo Park, CA 94025, USA}
\author{J.~Blazek}
\affiliation{Center for Cosmology and Astro-Particle Physics, The Ohio State University, Columbus, OH 43210, USA}
\affiliation{Institute of Physics, Laboratory of Astrophysics, \'Ecole Polytechnique F\'ed\'erale de Lausanne (EPFL), Observatoire de Sauverny, 1290 Versoix, Switzerland}
\author{O.~Friedrich}
\affiliation{Max Planck Institute for Extraterrestrial Physics, Giessenbachstrasse, 85748 Garching, Germany}
\affiliation{Universit\"ats-Sternwarte, Fakult\"at f\"ur Physik, Ludwig-Maximilians Universit\"at M\"unchen, Scheinerstr. 1, 81679 M\"unchen, Germany}
\author{S.~Samuroff}
\affiliation{Jodrell Bank Center for Astrophysics, School of Physics and Astronomy, University of Manchester, Oxford Road, Manchester, M13 9PL, UK}
% ...
\collaboration{DES Collaboration}
\noaffiliation

\date{\today}

\label{firstpage}

\begin{abstract}
We use 26 million galaxies from the Dark Energy Survey (DES) Year 1 shape catalogs over 1321 deg$^2$ of the sky to produce the most significant measurement of cosmic shear in a galaxy survey to date. We constrain cosmological parameters in both the flat $\Lambda$CDM and $w$CDM models, while also varying the neutrino mass density. These results are shown to be robust using two independent shape catalogs, two independent \photoz\ calibration methods, and two independent analysis pipelines in a blind analysis. We find a 3.5\% fractional uncertainty on $\sigma_8(\Omega_m/0.3)^{0.5} = 0.782^{+0.027}_{-0.027}$ at 68\% CL, which is a factor of 2.5 improvement over the fractional constraining power of our DES Science Verification results. In $w$CDM, we find a 4.8\% fractional uncertainty on $\sigma_8(\Omega_m/0.3)^{0.5} = 0.777^{+0.036}_{-0.038}$ and a dark energy equation-of-state $w=-0.95^{+0.33}_{-0.39}$. We find results that are consistent with previous cosmic shear constraints in $\sigma_8$ -- $\Omega_m$, and see no evidence for disagreement of our weak lensing data with data from the CMB. Finally, we find no evidence preferring a $w$CDM model allowing $w\ne -1$. We expect further significant improvements with subsequent years of DES data, which will more than triple the sky coverage of our shape catalogs and double the effective integrated exposure time per galaxy.
\end{abstract}

\pacs{Valid PACS appear here}

\preprint{DES-2016-0211}
\preprint{FERMILAB-PUB-17-279-PPD}

\maketitle


\section{Introduction}\label{sec:intro}

The study of cosmology over the previous few decades has been very successful at building a minimal model that is based on predictions from General Relativity at cosmological scales, and validated through a wide range of increasingly sophisticated experimental probes. In this model, $\Lambda$CDM, the gravitational dynamics of matter on large scales are dominated by a cold dark matter component that only interacts gravitationally (CDM) \cite{1984Natur.311..517B}, while the accelerated expansion of the Universe is driven by a cosmological constant $\Lambda$. These components make up about 25\% and 70\% of the Universe, respectively, while the remainder is composed of baryons, radiation, and neutrinos.
Despite the overall success of modern cosmological study, however, there remain several fundamental mysteries that enter the model as purely phenomenological parameters. These include our lack of understanding of the value of the cosmological constant or of any motivation for a different driver of cosmic acceleration. Further, although there are candidates within particle physics models for dark matter, there has been no detection of such a new particle. As our ability to constrain cosmological models at lower redshift continues to increase, we can also begin to explore subtle discrepancies between low- and high-redshift observations. This could indicate that $\Lambda$CDM, which has explained measurements like the power spectrum of the cosmic microwave background (CMB) so well, may not be sufficient to connect observations across cosmic times as the Universe undergoes significant evolution and becomes strongly inhomogeneous on smaller scales.

It is with these fundamental mysteries and potential for new physics in mind that several major observing programs have been undertaken to measure cosmic shear, a probe that is sensitive to both the expansion of the Universe and the growth and evolution of structure across vast volumes of space \cite{detf,esoesa,weinberg13}. Cosmic shear measures the correlated distortion of the envelopes of light bundles that are emitted from distant sources (i.e., galaxies) due to gravitational lensing by large-scale structure in the Universe. It is sensitive to both the growth rate and evolution of matter clustering as well as the relative distances between objects, and thus, the expansion history of the Universe. With the significant improvements in constraining power from the work described herein, cosmic shear has now become a leading probe of the nature of dark energy, dark matter, and astrophysical models of structure formation.

Cosmic shear directly measures inhomogeneities along the line-of-sight
to an observed galaxy, typically labelled a source (of the light bundle), with a weighted kernel that depends on the ratio of
distances between the lensing mass, the observed galaxy, and the observer.  
The distortion of light bundles due to cosmic shear
can be expressed in terms of the convergence,
\begin{equation}
\kappa = \frac{3}{2}\left(\frac{H_0}{c}\right)^2{\Omega_m }
\int_0^{r_{s}}dr\;  \frac{\delta(r)}{a(r)} \frac{r(r_{s}-r)}{ r_{s}},
\end{equation}
where $r$ is the comoving distance to an element of lensing mass, and
this is integrated from the observer to the source at $r_{s}$; the
matter density fluctuation is $\delta$, the fractional matter density parameter $\Omega_m$, $a$ is the scale factor, $c$ is the speed of light, and $H_0$ the Hubble parameter. 
Cosmic shear primarily probes the weak-lensing regime, $\kappa\ll1$, meaning
that distortions are small and linearly related to the
potential. In this regime the convergence ($\kappa$) and shear ($\gamma$) are simply related (see e.g., \cite{2001PhR...340..291B}). However, the associated density
fluctuations that are probed are not necessarily in the linear regime. 

The amplitude of the lensing signal is primarily sensitive to the
normalization of the matter fluctuations, $\sigma_8$, and to the matter density 
$\Omega_m$. The combination most tightly constrained is $S_8\equiv \sigma_8(\Omega_m/0.3)^\alpha$ \cite{jain97},
where empirically $\alpha\approx 0.5$ for cosmic shear. The degeneracy is not exact, so that contours in the $\sigma_8-\Omega_m$ plane take the
shape of a banana, but our most sensitive measurement is perpendicular
to the banana and is captured by the value of $S_8$.
This measurement via cosmic shear of inhomogeneities at $z < 1$ is
often compared to information on the amplitude of
primordial fluctuations at recombination, which is encoded in the
temperature fluctuations of the CMB. The CMB constraint on
fluctuations can be evolved forward in time, assuming a model such as
$\Lambda$CDM, to predict the lensing signal at $z < 1$ and allow a
direct comparison with the measurements of structure made using cosmic
shear.  A closely related and growing field of study is gravitational lensing
of the CMB (e.g., \cite{battye2014,leistedt2014,beutler2014}); this probes redshifts $z\le 2$, and has
demonstrated promising results with constraining power comparable to
previous cosmic shear results.

The statistical power of cosmic shear has increased rapidly over the last several years relative to the first detections in the 2000s \cite{Bacon:2000yp,kaiser:2000if,Wittman:2000tc,van_Waerbeke:2000rm,Hoekstra02,vanWaerbeke05,Jarvis06,Semboloni06,2007MNRAS.381..702B,Massey07,Hetterscheidt07,Schrabback10} as the current generation of surveys have produced their first results. More recent observations of cosmic shear include analyses of data from the Sloan Digital Sky Survey (SDSS), Canada-France-Hawaii-Telescope Legacy Survey (CFHTLS), the Deep Lens Survey (DLS), the Kilo-Degree Survey (KiDS)\footnote{http://kids.strw.leidenuniv.nl}, and the Dark Energy Survey (DES)\footnote{http://www.darkenergysurvey.org}. A portion of the SDSS Stripe 82 (S82) region was analyzed in \cite{Lin12,Huff14}. Cosmological constraints from cosmic shear using DLS data taken with the Mosaic Imager on the Blanco telescope between 2000 and 2003 were shown in \cite{jee2013,jee2015}. Several cosmological constraints have resulted from the CFHT Lensing Survey (CFHTLenS) \cite{heymans12,kilbinger13,heymans13,joachimicfht,kitching14}, which was recently re-analyzed in \cite{joudaki17}. Cosmic shear from 139 deg$^2$ of the DES Science Verification (SV) data was used to place the first constraint on cosmology with DES \cite{des2016} using both real- and harmonic-space measurements. Cosmic shear has also been recently measured using KiDS data \cite{kids2015,kids450}, which was used in \cite{kids450,kids450b,kids450c,kids450d,2017arXiv170706627J} to place tomographic constraints on cosmology.

There are currently three ongoing Stage III surveys designed for measuring cosmic shear: DES using the Blanco telescope, the Hyper-Suprime Cam (HSC)\footnote{http://hsc.mtk.nao.ac.jp/ssp/} survey using the Subaru telescope, and KiDS using the VLT Survey Telescope (VST). Preparations are also underway for four Stage IV weak lensing surveys to operate over the next decade:  the Large Synoptic Survey Telescope (LSST)\footnote{http://www.lsst.org}, Euclid\footnote{http://sci.esa.int/euclid}, the Square Kilometer Array (SKA)\footnote{http://skatelescope.org}, and the Wide Field InfraRed Survey Telescope (WFIRST).\footnote{http://wfirst.gsfc.nasa.gov} These surveys have significantly different observing strategies, and in most cases, complementary overlap in observing fields to allow for joint calibration and measurements. Two, Euclid and WFIRST, will use space-based telescopes to remove the obstacle of dealing with distortions in observed shapes due to the Earth's atmosphere.

While cosmic shear can now be said to be one of the most powerful probes of cosmology and the nature of dark energy and dark matter,
it is also a challenging measurement to make. The weak distortion in the shapes of objects that we measure is of the order of 1\%. In the presence of noise and the intrinsic scatter in galaxy shapes of individual objects, gravitational shear must be statistically measured typically over many millions of galaxies, each of which must have a robust shape measurement constrained to high accuracy, 
to precisely reconstruct the cosmic shear signal. Several technical advances in the robust measurement and
interpretation of galaxy shapes are discussed in \cite{shearcat}, and we exploit these in the current analysis. In order to interpret the measured cosmic shear signal, one also must have robust estimates of the distribution of galaxies in redshift, which is a challenging and evolving field of study. This is discussed further in \cite{photoz,redmagicpz,xcorrtechnique,xcorr}. Finally, one must also be able to interpret the measured signal in the presence of the correlated intrinsic shapes of galaxies (`intrinsic alignment') \cite{Troxel20151,Joachimi2015}, as well as other astrophysical effects that impact the measured cosmic shear signal, such as poorly understood baryonic physics (e.g., \cite{vandalen11,sembolini11, harnois14}). 

In this work, we present the first cosmological constraints from cosmic shear in the main DES wide-field survey, using data taken during its first year of observations. Preliminary constraints from DES were the result of an analysis of cosmic shear with data from the DES Science Verification (SV) observing period in \cite{becker2016,des2016}. This was followed by parameter constraints from weak lensing peak statistics \cite{Kacprzak2016} and the combination of galaxy-galaxy lensing and galaxy clustering \cite{kwan2016}. We combine galaxy-galaxy lensing and galaxy clustering with our cosmic shear results in a joint DES Y1 analysis \cite{keypaper}, which shares most components of the analysis pipeline used in this work. We present further information supporting the DES Y1 cosmological analyses in several concurrent papers: 
\begin{itemize}
\setlength\itemsep{-0.2em}
\item The construction and validation of the `Gold' catalog of objects in DES Y1 is described in \cite{y1gold}.
\item The DES Y1 \redmagic\ galaxy sample and clustering systematics that enter our clustering \photoz\ constraints are described in \cite{wthetapaper}. 
\item Shape measurement, calibration techniques, and validation of the two shape catalogs, \metacal\ and \imshape, are described in \cite{shearcat}. 
\item Further exploration of the \imshape\ image simulations is discussed in \cite{des_sim_2017}. 
\item Additional null tests of the reconstructed shear or convergence fields are discussed in \cite{des_mm_2017}. 
\item Construction and validation of the redshift distributions are discussed in \cite{photoz}. 
\item Constraints on the accuracy of the \redmagic\ redshifts that enter our clustering \photoz\ constraints are shown in \cite{redmagicpz}.
\item The accuracy of the clustering cross-correlation methods to constrain the source photometric redshifts is detailed in \cite{xcorrtechnique}.
\item The final construction of the clustering cross-correlation constraints on the source photometric redshifts using the DES Y1 \redmagic\ galaxy sample is described in \cite{xcorr}.
\item Galaxy-galaxy lensing measurements and further shape catalog tests and validation are shown in \cite{gglpaper}.
\item The general methodology, likelihood analysis, and covariance matrix used in the cosmological analyses shown in this work and \cite{keypaper} are described and validated in \cite{methodpaper}. 
\item Finally, this methodology is independently validated using complex simulations in 
\cite{simspaper}.
\end{itemize}
We encourage the reader to refer to these papers for further extensive information about the DES Y1 data production, testing, and analysis framework that is not repeated in detail in the current work.

The paper is organized as follows. We discuss in Sec. \ref{sec:data} the DES data, and our shape and photometric redshift (\photoz) catalogs.
Simulations and mock catalogs are discussed in Sec. \ref{sec:mocks}. We present our measurements of cosmic shear in Sec. \ref{sec:2pt} and covariance validation in Sec \ref{sec:covmatrix}. We discuss our blinding strategy in Sec. \ref{sec:blinding}, and describe our analysis choices in Sec. \ref{sec:model}. Cosmological parameter constraints are shown in Sec. \ref{sec:params} and further robustness tests in Sec. \ref{sec:modelchoice}. Finally, we conclude in Sec. \ref{sec:conclusion}.

\section{Dark Energy Survey Year 1 Data}
\label{sec:data}

The Dark Energy Survey (DES) is a five year observing program using the 570 megapixel DECam \cite{decam} on the Blanco telescope at the Cerro Tololo Inter-American Observatory (CTIO). The nominal DES wide-field survey images 5000 square degrees of the southern sky to 24th $i$-band limiting magnitude in the $grizY$ bands spanning 0.40-1.06 $\mu$m. The survey tiling strategy ultimately consists of ten overlapping 90 second exposures in each of $griz$ and 45s exposures in $Y$ over the full wide-field area.

The DES Year 1 (Y1) shape catalogs used for this analysis are based on observations taken between Aug. 31, 2013 and Feb. 9, 2014 during the first full season of DES operations. DES Y1 wide-field observations were targeted to a large region overlapping the South Pole Telescope (SPT) survey footprint extending between approximately $-60^{\circ} < \delta< -40^{\circ}$, and a much smaller area overlapping the `Stripe 82' region of the Sloan Digital Sky Survey (SDSS), which is not included in this analysis.
The observed area was limited in the DES Y1 period to reach a sufficient number of overlapping exposures across the observed footprint. In practice, this resulted in a total area of about 1514 deg$^2$ with a mean depth of three exposures, after masking potentially bad regions not used for weak lensing \cite{y1gold}.

The DES Y1 data incorporate a variety of improvements over the DES Science Verification (SV) data used in preliminary DES weak lensing analyses, including updates to the telescope and systems components and to data processing. These are discussed in detail in \cite{y1gold}, which describes the production and validation of a `Gold' catalog of 137 million objects prior to the `bad region' masking referred to above, and in \cite{shearcat}, where the shape catalog production and validation is described.

\subsection{Shape Catalogs}\label{sec:shapes}

We test the robustness of our results with two independent shape measurement pipelines, \metacal\ and \imshape, which are fully described and characterized in the accompanying catalog paper \citep{shearcat}. The pipelines use different subsets of the DES Y1 data, different measurement techniques, and different calibration strategies. Each was developed without direct comparison to the other at the two-point level -- blinded measurements of $\xi_{\pm}$ were compared only once the two catalogs were finalized. Unlike in the DES Science Verification (SV) analysis \cite{jarvis2016}, no effort was made to modify them to ensure they agreed prior to comparing cosmological constraints, beyond applying the same suite of null tests to both catalogs in \cite{shearcat}. This stems from the difficulty of comparing two shear measurement methods in a robust way, since any joint selection may bias both methods, even if separately they are each unbiased. We discuss this further in Sec. \ref{sec:sysshapecode}.

The median measured seeing (FWHM of stars selected for point-spread function (PSF) modeling) in the $riz$ bands is 0.96 arcsec for the DES Y1 shape catalog, which is an improvement over the DES SV seeing. This value is after routine nightly rejection of exposures \cite{y1gold} and the blacklisting of a small number of exposures during the PSF model building process \cite{jarvis2016} due to imaging and processing anomalies. At the catalog level, objects are removed based on a set of criteria unique to each shape catalog, but which generally satisfy a lower S/N cut and a rigorous size cut relative to the PSF size. More details on these selections are described in \cite{shearcat}, which discusses further the impact of data quality and various selections on the final catalog number density. 

For both catalogs, we use only objects that pass the default recommended selection $\textsc{flags\_select}$ (see selection criteria defined in \cite{shearcat}). We additionally limit objects to have \photoz\ point estimates within the redshift range 0.2 -- 1.3 (cf. \autoref{sec:photoz}) and to fall within the large, contiguous southern portion of the footprint ($\textrm{dec}<-35$) that overlaps with the SPT survey. 
Finally, we limit our study to objects that are contained within the \redmagic\ mask described in \cite{wthetapaper}, which additionally removes a few tens of deg$^2$ from the original shape catalog footprint, bringing its final effective area to 1321 deg$^2$. This final mask, while not strictly necessary for cosmic shear, is applied to make this work consistent with the joint cosmological constraints combining weak lensing and galaxy clustering in \cite{keypaper}, where the same footprint is assumed in our covariance calculation. This has the added benefit of reducing depth variation across the field, and thus spatial variations in our redshift distribution.

\subsubsection{\metacal~}\label{sec:mcal}

\metacal\ is a method to calibrate a shear statistic, such as a mean shear
estimate or shear two-point function, from available imaging data, without
requiring significant prior information about galaxy properties or calibration
from image simulations \cite{HuffMandelbaum2017,SheldonHuff2017}.  \metacal\
has been tested with complex image simulations and shown to be accurate at the
part per thousand level \citep{SheldonHuff2017}.  The implementation used in
DES is described in detail in \citep{shearcat}, where the ellipticity is measured 
using a single Gaussian model that is fit to the galaxy image in the $riz$ bands. The galaxy image is then artificially 
sheared and the ellipticity remeasured to construct the shear response matrix via numerical 
derivatives of the ellipticity. We do this by deconvolving the PSF, applying
a shear, and then reconvolving by a symmetrized version of the PSF. This results 
in one unsheared and four sheared versions of the shape catalog (one for each direction ($\pm$) and component of shear), all of which include flux measurements for photo-$z$ estimation.
Some limitations in the application of \metacal\ to DES Y1 data 
are discussed in \cite{shearcat}, which leads us to assign a non-zero mean for the
Gaussian prior assumed in this analysis on the shear calibration of $m = 0.012 \pm 0.013$. 
This error budget is dominated by our estimate of the unaccounted effects of contaminating light from neighboring objects on the shear estimation.

With \metacal, corrections are calculated for both the response of the shape
estimator to a shear and the response of object selections to a shear.  The
\metacal\ procedure produces a noisy estimate of the shear response \mcalRg\
for each galaxy, which is then averaged to produce \mcalRgmean. The induced
selection bias is calculated only in the mean \mcalRSmean. These quantities are in general 2x2 matrices of the ellipticity components. The explicit calculation of 
these corrections using the four sheared catalogs is described in Secs. 4.1 \& 7.4 of \cite{shearcat}.
The application of these corrections depends on the details of the shear
statistic that is being calibrated; some examples are derived in
\cite{SheldonHuff2017}. 

In this work we adopt a number of approximations that
simplify this process.  First, we assume that the shear response is independent
of environment, and thus not dependent on the separation of
galaxies.  
Under this assumption, the correction to the shear two-point
function is simply the square of the mean response (see section 3.2 in
\citep{SheldonHuff2017}).  
We further make the assumption that the correction is independent of the
relative orientation of galaxies, so that the mean response can be calculated
without the shape rotations that are applied when measuring the shear two-point
function.  We find that the mean response matrices are consistent
with being diagonal, which further simplifies the calibration procedure. 
While these assumptions appear to be valid for the current analysis, fully testing the propagation of the full rotated selection
response through the shear two-point function is left to a future
work. There should be no additive correction of the response necessary for \metacal, due to the symmetric reconvolution
function used during the metacalibration process \citep{shearcat}, though we discuss the impact of residual mean shears in Appendix \ref{sec:psf}.
The use of the response corrections is discussed further in Sec. \ref{sec:2pt}. 

\begin{figure*}
\begin{center}
\includegraphics[width=1.5\columnwidth]{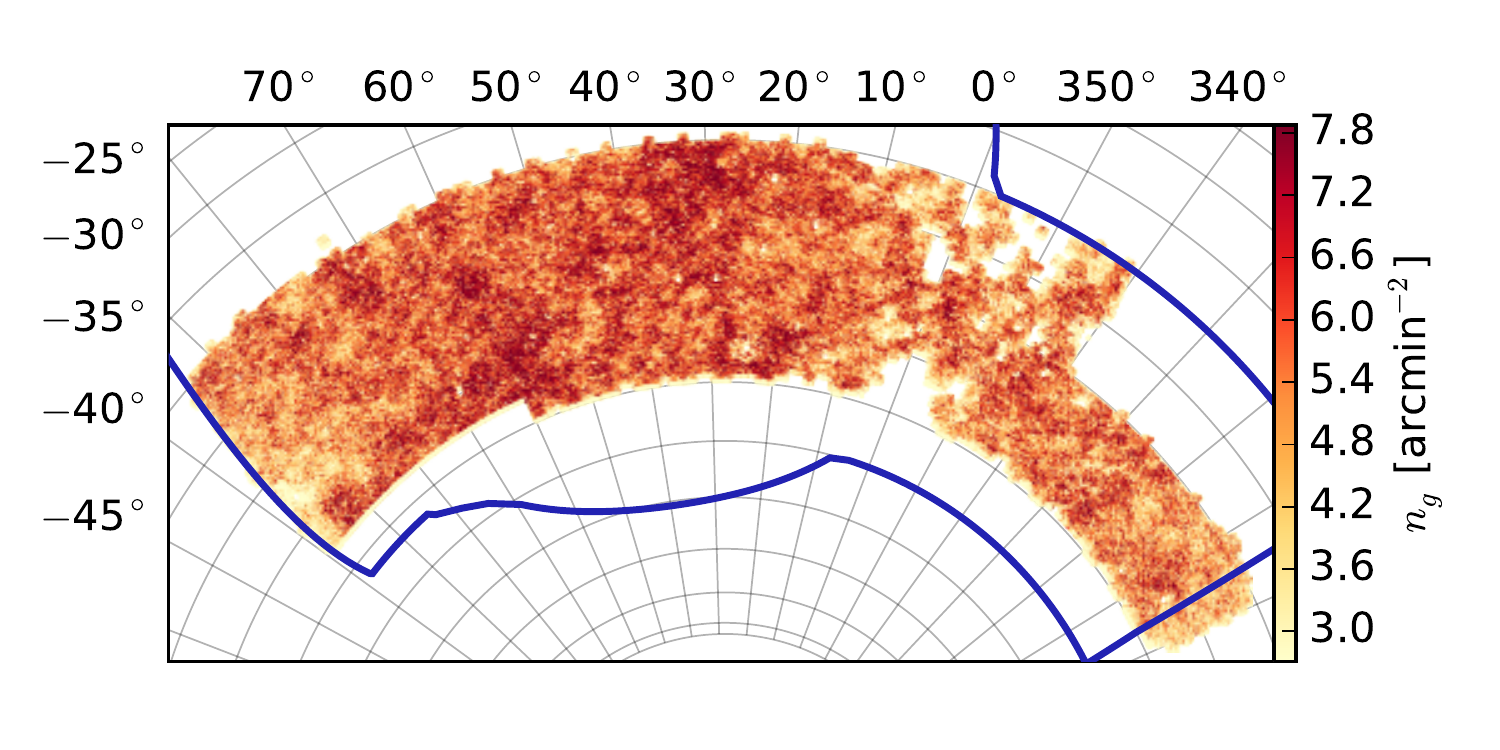}
\end{center}
\caption[]{The footprint of the DES Y1 \metacal\ catalog selection used in this work, covering 1321 deg$^2$. The joint \redmagic\ mask described in Sec. \ref{sec:shapes} is not included. The raw mean number density of objects drawn by \textsc{Skymapper} in \healpix\ cells of $N_{\mathrm{side}}$ 1024 is shown, which is uncorrected for the coverage fraction at subpixel scales. Overlaid are the bounds of the nominal five year DES survey footprint. The full shape catalog footprint, which includes the `Stripe 82' region, is shown in \cite{shearcat}. For the \metacal\ catalog, $n_g$ is equivalent to the H12 $n_{\mathrm{eff}}$ in Table \ref{table:neff}. 
\label{fig:footprint}}
\end{figure*}

The \metacal\ catalog yields a total of 35 million objects, 26 million of which are 
used in the selection for the current analysis. The final number density of the selection used in this analysis is 5.5 galaxies arcmin$^{-2}$.
The raw mean number density of objects are shown in Fig. \ref{fig:footprint} as a function of position on the sky, drawn by \textsc{Skymapper}\footnote{https://github.com/pmelchior/skymapper} in \healpix\footnote{http://healpix.sourceforge.net} \cite{gorski2005} cells of $N_{\mathrm{side}}$ 1024. Overlaid are the bounds of the nominal five year DES survey footprint. 

\subsubsection{\imshape~}\label{sec:i3}

\imshape\ measures galaxy shapes by fitting bulge and disc models to each object
with a Levenberg-Marquardt algorithm and then taking the best fitting of the two
models. This measurement process is unchanged from that used in the DES SV 
catalog \cite{jarvis2016}.  The code is described in detail in \cite{zuntz2013}. 
Noise, model, and selection biases on the galaxy shapes are
calibrated using a suite of simulations designed to closely reflect
real data, which are described in \cite{shearcat,des_sim_2017}.  

The calibration of \imshape\ produces a multiplicative $m$ and additive $c_i$ bias corrections per object, where $i$ is the ellipticity component, such that the observed ellipticity ($e^o$) is related to the true ellipticity ($e^t$) as $e^o_i=(1+m)e^t_i+c_i$. The multiplicative bias is assumed to be the same for both shear components. Based on the work described in \cite{shearcat}, we assign a Gaussian prior on the shear calibration of $m = 0 \pm 0.025$, which is wider than that obtained with \metacal\ due primarily to our estimates of uncertainties related to the accuracy of reproducing the real survey in our image simulations. The bias corrections $m$ and $c_i$ are applied in the same way as previous cosmic shear studies (e.g., \cite{des2016}), and thus not discussed in detail here. We propagate the impact of the shear calibration $m$ for \imshape\ through the two-point estimator to produce a two-point correction as a function of scale. This is described further in Sec. \ref{sec:2pt}.

\imshape\ was applied only to $r$ band images, yielding a smaller catalog of
22 million objects, which is reduced to 18 million with the selection for the current analysis. The final number density of the selection used in this analysis is 3.4 galaxies arcmin$^{-2}$. For this selection,  $\sigma_{\textrm{sh}}=0.27$.

\subsection{Photometric Redshift Estimates}\label{sec:photoz}

\begin{figure}
\begin{center}
\includegraphics[width=\columnwidth]{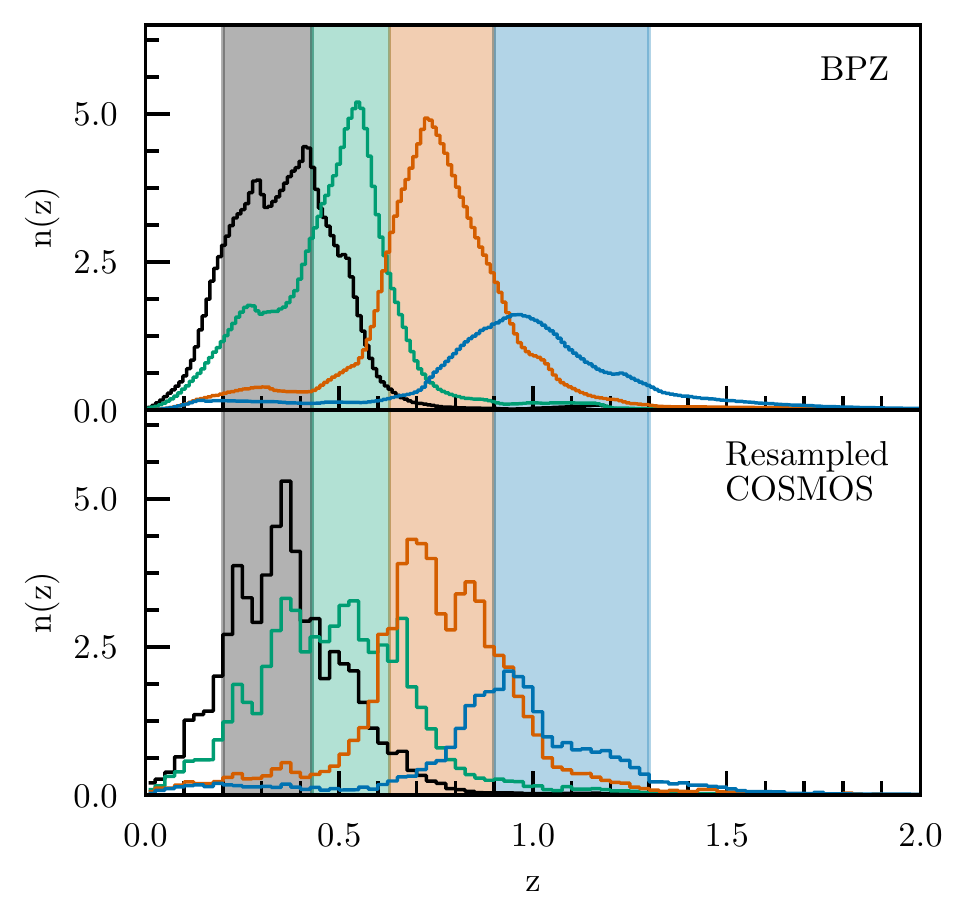}
\end{center}
\caption[]{The measured \bpz\  and resampled \cosmos\ redshift distributions for the \metacal\ shape catalog, binned by the means of the photo-$z$ posteriors into the four tomographic ranges in Table \ref{table:neff} and marked by the color shading. The normalization of each bin reflects their relative $n_{\textrm{eff}}$. The \bpz\ distributions are corrected by the mean of the redshift bias priors $\Delta z^i$. The contribution of each galaxy is weighted by $W_i S_i$, as defined in Sec. \ref{sec:2pt}. The \imshape\ redshift distributions are similar to those shown for \metacal. The second bin is clearly most different between the resampled COSMOS estimate and BPZ -- we explore this further in Sec. \ref{sec:sysphotoz} and show that it does not significantly impact the inferred cosmological parameters.
\label{fig:nofz}}
\end{figure}

A tomographic cosmic shear measurement requires an assignment of each
source galaxy to a redshift bin $i$, and its interpretation requires an accurate estimation
of the redshift distribution of galaxies
in each 
redshift bin, $n^i(z)$. The procedures for doing so, and for assigning uncertainties to
$n^i(z),$ are described fully in \cite{photoz} and the companion papers \cite{xcorrtechnique,xcorr}.
In this analysis, galaxies in the shape catalogs are
assigned to the four redshift bins listed in Table~\ref{table:neff}
by the mean of the \photoz\ posterior $p(z)$ estimated from DES $griz$
flux measurements. The redshift distribution of each 
bin is constructed by stacking a random sample from the $p(z)$ of each 
galaxy, weighted according to $W_i S_i$, which is defined in Sec. \ref{sec:2pt}. 
The \photoz\ posteriors used for
bin assignment and $n^i(z)$ estimation in the fiducial
analysis are derived using the Bayesian photometric redshift (\bpz)
methodology \cite{Benitez2000}. Details are given in Sec. 3.1 of \cite{photoz}. The estimated redshift distributions for
\metacal\ are shown in Fig. \ref{fig:nofz}.

One notable complication when compared to previous cosmic shear studies is the direct correction of photo-$z$ induced selection biases in \metacal, which requires calculating the impact that shearing a galaxy image has on the photometric redshift determination. We thus construct a total of six versions of our photo-$z$ estimates based on various photometric measurements: a) the original Multi-Epoch Multi-Object Fitting (\mof)\ $griz$-band photometry (see \cite{y1gold} for details on the MOF technique), b) the measurements of $griz$-band photometry from the unsheared \metacal\ galaxy fit, and c) four versions of the $griz$-band photometry from the four sheared \metacal\ galaxy fits. In all cases, the redshift distribution $n^i(z)$ of each bin is reconstructed using BPZ estimates from MOF, which gives a better estimate of the shape of the redshift distribution. This is because: 1) \mof\ fluxes are superior to those of \textsc{mag\_auto} because they properly account for PSF variations between images and impose a consistent galaxy model across bands, and 2) \mof\ fluxes are superior to those derived from metacalibration because the \metacal\ process adds extra noise to the image to correct for correlated noise, thus degrading \metacal\ flux measurements. 

To calculate the \metacal\ selection bias correction due to redshift selection, we then construct the galaxy selection in each tomographic bin from the photo-$z$ estimates using both the unsheared \metacal\ photometry and the four sheared photometries. We use these five selections, in addition to all other selection criteria such as signal-to-noise cuts, to construct the component of the selection bias correction \mcalRSmean. For more details on the mechanics of this calculation, see Secs. 4.1 \& 7.4 of \cite{shearcat}. For the \imshape\ catalog, BPZ redshifts estimated from \mof\ photometry are used both for binning and for reconstructing the redshift distribution, since \imshape\ does not actively calibrate for such biases via the data. It has been confirmed for \imshape\ that there is no apparent residual redshift-dependent bias $m$ in its image simulation, for the same redshift bins used in this work and using the associated COSMOS redshifts of each input object, as discussed in \cite{shearcat}. This test is performed by constructing the calibration from one half of the simulation and testing the residual in the other, split both randomly and by input COSMOS objects.

\begin{table}
\caption{Effective number density $n_{\textrm{eff}}$ (gal arcmin$^{-2}$) and ellipticity dispersion $\sigma_{\mathrm{e}}$ (per component) estimates for each tomographic redshift bin of the \metacal\ catalog. We include values for both the \cite{C13} (C13) and \cite{heymans12} (H12) definitions of these values. The ellipticity dispersion defined in H12 includes both shape noise ($\sigma_{\mathrm{sh}}$) and measurement noise ($\sigma_{\mathrm{m}}$), while that in C13 is purely $\sigma_{\mathrm{sh}}$. The shot noise $\sigma_{\mathrm{e}}^2/n_{\mathrm{eff}}$ for both definitions is the same. These quantities are discussed further in \cite{shearcat}.
\label{table:neff}}
\begin{center}
\begin{ruledtabular}
\begin{tabular}{ cccccc }
  Bin 		& Extent 	& \multicolumn{2}{c}{$n_\mathrm{eff}$} & $\sigma_{\mathrm{sh}}$ & $\sigma_{\mathrm{sh}}+\sigma_{\mathrm{m}}$  \\
      		&		& C13 & H12 & C13 & H12 \\
  \hline
  Full &	0.20 -- 1.30 & 5.14 & 5.50 & 0.27 & 0.27 \\
  1 	&      0.20 -- 0.43 & 1.47 & 1.52 & 0.25 & 0.26 \\
  2 	&      0.43 -- 0.63 & 1.46 & 1.55 & 0.28 & 0.29 \\
  3 	&      0.63 -- 0.90 & 1.50 & 1.63 & 0.26 & 0.27 \\
  4 	&      0.90 -- 1.30 & 0.73 & 0.83 & 0.27 & 0.29
\end{tabular}
\end{ruledtabular}
\end{center}
\end{table}

Our adopted model for the redshift distribution assumes
that the true redshift distribution in each bin is related to our measured distribution such that:
\begin{equation}
n^i(z) = n^{i}_{\textrm{PZ}}(z - \Delta z^i),
\label{eq:Dz}
\end{equation}
where $\Delta z$ is the difference in the mean redshift of the true and measured $n(z)$. This is a sufficient description of the \photoz\ uncertainty for the current cosmic shear analysis, as we demonstrate in Sec. \ref{sec:sysphotoz}. Deviations in the shape of the $n(z)$ are subdominant to the impact of the mean $z$, for reasonable variance in the shape at the level of precision necessary for the DES Y1 analysis.  
We derive constraints on $\Delta z^i$ for the estimated redshift distributions by comparison of the mean redshift in each bin to that from two independent methods:
\begin{enumerate}
\item The mean, high-quality photo-$z$ of a sample of galaxies from the COSMOS2015 catalog \cite{Laigle}, matched to resemble the source galaxies in $griz$ flux and pre-seeing size \cite{photoz}.
\item In the lowest three redshift bins, the clustering of source galaxies with \redmagic\ galaxies at $0.15<z<0.85$, for which accurate and precise photometric redshifts can be derived from DES photometry \cite{redmagicpz,xcorrtechnique,xcorr}.\footnote{The calculation of the difference of the mean redshift relative to the truncated redshift range of our reference \redmagic\ galaxies is described in detail in \cite{xcorr}, but briefly we calculate the mean in this cross-correlation comparison for both samples in a redshift window of $\pm 2\sigma$ from the mean of the $n(z)$ derived from the cross-correlation.}
\end{enumerate}
We will refer to these as the `COSMOS' and `WZ' redshift validation
methods, respectively.  Their constraints on $\Delta z^i$ are independent and consistent for the first three bins and of comparable uncertainty. We thus combine them to provide a prior on the systematic parameters $\Delta z^i$ at the level of $\pm0.02$ \cite{photoz}. For \metacal\, the $\Delta z^i$ are listed in Table \ref{table:params} and are consistent with the original BPZ estimate. The agreement between these
validation methods provides further justification of our reliance on the
accuracy of the COSMOS2015 30-band \photoz s.
The priors for alternate combinations of shear and photo-$z$ pipelines are given in \cite{photoz} and require statistically significant shifts to the redshift distributions in some cases.

\begin{table}
\caption{Summary of cosmological, systematic, and astrophysical parameters used in the fiducial analysis. In the case of flat priors, the prior is identical to the listed range. Gaussian priors are indicated by their mean and 1 $\sigma$ width listed in the prior column. In the case of $w$, it is fixed to $-1$ for $\Lambda$CDM and varies over the range given for $w$CDM. For $m^i$, the values listed are for \metacal, which are inflated from the original 1.3\% constraint to preserve the overall $m$ uncertainty when combining tomographic pairs in the likelihood analysis. }
\label{table:params}
\begin{center}
\begin{ruledtabular}
\begin{tabular}{ lccccccc }
  Parameter 		&  Range & Prior   \\
  \hline
\multicolumn{3}{l}{Cosmological} \\
  $A_s\times 10^{9}$             &  0.5 \ldots 5.0 & Flat   \\
  $\Omega_m$ 	&  0.1 \ldots 0.9  & Flat  \\
  $\Omega_b$ 	& 0.03 \ldots 0.07 & Flat  \\
  $\Omega_{\nu}h^2$ &  0.0006 \ldots 0.01 & Flat  \\
  $H_0$ (km s$^{-1}$ Mpc $^{-1}$) 		&  55 \ldots 90 & Flat  \\
  $n_s$ 		& 0.87 \ldots 1.07 & Flat  \\
  $w$ &   $-2.0$ \ldots $-0.333$  & Fixed/Flat  \\
  $\Omega_k$  & 0.0  & Fixed  \\
  $\tau$ &  0.08 & Fixed\\
  \hline
\multicolumn{3}{l}{Systematic} \\
  ($m^1$ -- $ m^4$)$\times10^2$ & $-10$ \ldots 10 & $1.2\pm 2.3$ \\
  $\Delta z^1\times10^2$ &  $-10$ \ldots 10 & $0.1\pm 1.6$ \\
  $\Delta z^2\times10^2$ &  $-10$ \ldots 10 & $-1.9\pm 1.3$ \\
  $\Delta z^3\times10^2$ &  $-10$ \ldots 10 & $0.9\pm 1.1$ \\
  $\Delta z^4\times10^2$ &  $-10$ \ldots 10 & $-1.8\pm 2.2$ \\
  \hline
\multicolumn{3}{l}{Astrophysical} \\
  $A$  &  $-5.0$ \ldots 5.0 & Flat \\
  $\eta$  &  $-5.0$ \ldots 5.0 & Flat \\
  $z_0$  &  0.62 & Fixed \\
\end{tabular}
\end{ruledtabular}
\end{center}
\end{table}

Estimation of the redshift distribution of the lensing source galaxies
is one of the most difficult components of a broadband
cosmic shear survey like DES.  Along with the use of two independent
methods to constrain $\Delta z^i$, we present in
Sec.~\ref{sec:sysphotoz} several tests of the robustness of our
cosmological results to the methods and assumptions of our $n^i(z)$ estimates.
One such test is to replace the shifted \bpz\
estimator in Eq. (\ref{eq:Dz}) with the resampled 
COSMOS $n^i(z)$ to confirm robustness to the shape of $n^i(z)$.  

\section{Simulations and Mock Catalogs}\label{sec:mocks}

In this analysis, we have employed both a limited number of mock shear catalogs produced from dark-matter-only $N$-body simulations, described in Secs.~\ref{sec:buzzard} \& \ref{sec:micegc}, and a large number of lognormal mock shear catalogs, described in Sec.~\ref{sec:lognormal}. The full $N$-body mock catalogs have been used to validate our analysis pipeline \citep{simspaper} and our covariance estimation \citep{methodpaper}. There is not a sufficient number of $N$-body mock catalogs to produce an independent covariance matrix for our data vector to compare to the halo model covariance described in Sec. \ref{sec:covmatrix}, 
so we also employ a large suite of lognormal mock catalogs to test certain pieces of the full halo model covariance calculation (also discussed in \cite{methodpaper}) and to construct covariance matrices for various null tests.

\subsection{Buzzard Mock Catalogs}\label{sec:buzzard}

The Buzzard simulations, built from dark-matter-only $N$-body simulations, are a suite of 18 mock realizations of the DES Y1 survey. The most important aspects of these simulations are summarized below, but we refer the reader to more detailed descriptions in \citep{DeRose2017, Wechsler2017, simspaper}. The 18 mock catalogs are composed of 3 sets of 6 catalogs each, where each set is built from a combination of three separate $N$-body simulations. These have box lengths of $1.05$, $2.6$, and $4.0\, \ensuremath{h^{-1}\mathrm{Gpc}}$, and $1400^3$, $2048^3$, and $2048^3$ particles, giving mass resolutions of $2.7\times10^{10}$, $1.3\times10^{11}$, and $4.8\times10^{11} h^{-1}M_{\odot}$, respectively. The simulations were run using the \textsc{L-Gadget2} code \citep{springel2005} using 2nd order Lagrangian perturbation theory initial conditions generated using \textsc{2LPTIC} \citep{crocce2006}. The light-cones are output on the fly -- the two highest resolution simulations are stitched together at redshift $z=0.34$ and the lowest resolution box is used for $z>0.9$. 

The \textsc{ADDGALS} \citep{Wechsler2017, DeRose2017} algorithm  is used to add galaxies to the simulations by assigning $r$-band absolute magnitudes to particles in the simulations based on large-scale density. The particles to which galaxies are assigned are not necessarily in resolved dark matter halos, but all resolved central dark matter halos have galaxies assigned to them. Galaxy spectral energy distributions (SEDs) are then assigned to the galaxies from a training set derived from SDSS DR7 \citep{Cooper2011}, and DES $griz$ magnitudes are generated by convolving the SEDs with the DES pass bands. Galaxy ellipticities and sizes are assigned by drawing from distributions fit to SuprimeCam $i^{'}$-band data. Before any cuts, we find $\sigma_e=0.31$, about $10\%$ larger than $\sigma_e$ for the \metacal\ catalog. Galaxy magnitudes, ellipticities, and sizes are then lensed using the multiple-plane ray-tracing algorithm \textsc{CALCLENS} \citep{Becker2013b}. To mimic DES depth fluctuations, we apply photometric errors using the DES Y1 \textsc{mag\_auto} depth maps according to angular position within the footprint and the true apparent magnitude of the galaxy.

Flux and size cuts are applied to the simulated galaxies to approximate the signal-to-noise distribution of a weak lensing sample roughly mimicking the selection of the \metacal\ shape catalog. To bring the shape noise in the simulated sample to that of the \metacal\ catalog, we then apply an additional redshift dependent flux cut to the mock catalogs. After final cuts and unblinding of the \metacal\ catalog, the effective number density of the mocks is about 7 per cent larger than in the data.

\subsection{MICE-GC Mock Catalogs}
\label{sec:micegc}

The MICE Grand Challenge (MICE-GC) simulation is a large $N$-body simulation which evolved $4096^3$ particles in a volume of $(3072\mpcoh)^3$ using the \textsc{gadget-2} code \citep{springel2005}. 
This results in a particle mass of $2.93\times10^{10}\Msun$. The initial conditions were generated at $z_i=100$ using the Zel'dovich approximation and a linear power spectrum generated with \textsc{camb}.\footnote{\texttt{http://camb.info}} On-the-fly light-cone outputs of dark-matter particles up to $z=1.4$ were produced without repetition in one octant.
A set of 256 all-sky maps with angular \healpix\ resolution $N_{\mathrm{side}}=8192$ of the projected mass density field in narrow redshift shells were measured. The process used to compute weak lensing maps from \healpix\ mass maps in $z$-slices was first discussed in \cite{2008MNRAS.391..435F}. These were used to derive the convergence field $\kappa$ in the Born approximation by integration along the line-of-sight. The convergence was transformed to harmonic space, converted to an E-mode shear map, and transformed back to angular space to obtain the Stokes $(\gamma_1,\gamma_2)$ shear fields, following \cite{MICEIII}. In this way 3D lensing maps of convergence and shear were produced.

Halos in the light-cone were identified using a Friends-of-Friends algorithm. A combination of Halo Occupation Distribution (HOD) and SubHalo Abundance Matching (SHAM) techniques were then implemented to populate halos with galaxies, assigning positions, velocities, luminosities and colors to reproduce the luminosity function, ($g-r$) color distribution, and clustering as a function of color and luminosity in SDSS \citep{2003ApJ...592..819B,2011ApJ...736...59Z}. Spectral energy distributions (SEDs) are then assigned to the galaxies resampling from the COSMOS catalog of \cite{ilbert2009}. Finally, DES $griz$ magnitudes are generated by convolving the SEDs with the DES pass bands. The catalogs are available at {\tt cosmohub.pic.es} and a detailed description is given in \cite{MICEI,MICEII,MICEIII,2015MNRAS.447..646C}. 

\subsection{Lognormal Mock Catalogs}\label{sec:lognormal}

In order to generate large numbers of realizations of mock shear data, we take advantage of the fact that a lognormal shear field can be produced quickly and with reasonable levels of non-Gaussianity. The potential use of lognormal random fields in cosmological analyses was first outlined in \cite{ColesJones1991}, and the lognormal distribution of shear fields has shown good agreement with $N$-body simulations and real data up to nonlinear scales \cite{Kayoetal2001,LahavSuto2004,Hilbertetal2011}. The production of such mock catalogs has a significantly smaller computational expense than a full $N$-body simulation and ray-tracing. Thus, lognormal mock simulations provide a compromise between accuracy and computational cost that allows us to quantify how the non-Gaussianity of cosmic fields and incomplete sky coverage propagate into the covariance of cosmic shear.

We use the publicly available code \flask\footnote{http://www.astro.iag.usp.br/$\sim$flask/} \cite{Xavieretal2016}, which generates consistent density and convergence fields, to produce 150 mock full-sky shear maps that reproduce a set of input power spectra that fit a fiducial cosmology\footnote{The cosmology used for the \flask\ simulations is described by a flat $\Lambda$CDM model with: $\Omega_m = 0.295$, $\Omega_b = 0.0468$, $A_s = 2.260574\times 10^{-9}$, $h = 0.6881$, $n_s = 0.9676$.} and the actual redshift distribution of sources in our data.  These maps are produced on a \healpix\ grid with resolution set by an $N_\mathrm{side}$ parameter of 4096. In this resolution, the typical pixel area is around 0.73 arcmin$^2$. The full sky mocks are then divided into eight non-overlapping DES Y1 footprints per full-sky simulation. To a good approximation, the footprints belonging to the same full-sky are uncorrelated for sufficiently high multipoles. This produces a total of 1,200 mock shear maps. For each mock realization, we simulate four shear fields corresponding to the redshift distributions of the four redshift bins shown in Fig. \ref{fig:nofz}.

To capture the expected noise properties of the shear fields, we then add appropriate shape-noise by sampling each pixel of the map to match the measured $n_\mathrm{eff}$ and $\sigma_e$ shown in Table \ref{table:neff} for each tomographic shear bin. Covariance validation was done using \metacal\ parameters, while the mock catalogs were remade for each null test in Sec. \ref{sec:syssurvey} to match the effective shape noise of either shape catalog after reweighting the objects to match redshift distributions of subsets of the catalogs. For further details of how this was implemented, see Appendix \ref{sec:syssurvey}.

\section{Cosmic Shear Measurement}\label{sec:2pt}

We present in this section the measurements of the real-space two-point correlation function $\xi_{\pm}$ from the \metacal\ and \imshape\ catalogs. These results are derived from measurements in a contiguous area 1321 deg$^2$ on the sky, which has been split into four tomographic bins as described in Sec. \ref{sec:photoz}.
These measurements are the highest signal-to-noise measurements of cosmic shear in a galaxy survey to date, with total detection significance ($S/N$) of $25.4$ $\sigma$ for the fiducial \metacal\ measurement using all angular scales and redshift bin pairs.\footnote{Signal-to-noise is derived here as in \cite{becker2016} as $S/N = \frac{\xi_{\textrm{data}}C^{-1}\xi_{\textrm{model}}}{\sqrt{\xi_{\textrm{model}}C^{-1}\xi_{\textrm{model}}}}$, where $C$ is the covariance described in Sec. \ref{sec:covmatrix} and $\xi_{\textrm{model}}$ is the best-fit model obtained from the analysis in Sec. \ref{sec:params}.}

Cosmic shear is a quantity with two components, based on two 2$^\mathrm{nd}$ order angular
derivatives of the lensing potential, $\psi$:
$\gamma_1=(\psi_{11}-\psi_{22})/2$; $\gamma_2=\psi_{12}$ (where the
angular deflection is $-\mbox{\boldmath$\nabla$}\psi$).
For points along the 1 axis,
these components give a simple definition of the
tangential- and cross-shear components:
$\gamma_t=-\gamma_1$; $\gamma_\times = -\gamma_2$. There are
thus three 2-point functions to consider, but in practice the
cross-correlation $\langle\gamma_t\gamma_\times\rangle$ vanishes,
leaving the two standard quantities that are the focus of
most weak-lensing studies \cite{schneider02}:
\begin{equation}
\xi_\pm = \langle\gamma_t\gamma_t\rangle \pm
\langle\gamma_\times\gamma_\times\rangle .
\end{equation}

We estimate $\xi_{\pm}$ for redshift bin pair $i,j$ as
\begin{equation}
\hat{\xi}^{ij}_{\pm}(\theta) = \frac{\sum_{ab} W_a W_b \left[ \hat{e}^i_{a,t}(\vec{\theta}) \hat{e}^j_{b,t}(\vec{\theta}) \pm \hat{e}^i_{a,\times}(\vec{\theta}) \hat{e}^j_{b,\times}(\vec{\theta}) \right]}{\sum_{ab}W_a W_b S_a S_b},
\label{eq:estimator}
\end{equation}
where $\hat{e}_{a,t}$ is the tangential component of the corrected ellipticity of galaxies $a$ along the direction towards galaxy $b$ and $\hat{e}_{a,\times}$ is the cross component, $W$ is a per-object weight, and $S$ is either a multiplicative bias correction (\imshape) or a shear response correction (\metacal). The sums are each computed for a subset of galaxy pairs $a,b$ within each angular separation $\Delta \theta$ for each $\theta=|\vec{\theta}_b-\vec{\theta}_a|$. These angular bins are chosen to be logarithmic with a total of 20 bins between 2.5 and 250 arcmin, though only a subset of these angular bins are used in parameter estimation, as discussed in Sec. \ref{sec:systheory}. All two-point calculations are done with the public code \textsc{TreeCorr}\footnote{https://github.com/rmjarvis/TreeCorr} \cite{treecorr}. The estimator for $\xi_{\pm}$ is in practice calculated quite differently for the two shape catalogs, because they each estimate the ellipticity of an object and any shear calibrations via fundamentally different processes. 

For \metacal, the $k$th component of the unrotated ellipticity is given by $\hat{e}_k=e_k- \langle e_k\rangle$, where $\langle e\rangle$ is the residual mean shear in a given tomographic bin. The \metacal\ catalog does not use a galaxy weight ($W=1$), and the shear response correction ($S$) is given by $S\equiv R=R_{\gamma} + R_{S}$. In general $R$ is a 2x2 matrix, where $R_{ii}=R_{\gamma,ii}+R_{S,ii}$ is the sum of the $ii$th element of the measured shear response and shear selection bias correction matrix for \metacal. We simply use the average of the components of $R$, where $R=(R_{11}+R_{22})/2$. For \imshape, $\hat{e}_k=e_k - c_k - \langle e_k - c_k\rangle$, where $c$ is the additive shear correction and $\langle e - c\rangle$ is the residual mean shear for a tomographic bin. The \imshape\ catalog uses an empirically derived weight ($W=w$), and a multiplicative shear correction $S=1+m$, where $m$ is defined irrespective of the ellipticity component. The residual mean shear is discussed in Appendix \ref{sec:psf}, with typical abs. values of $1$ to $9\times 10^{-4}$ per tomographic bin. For \imshape, the typical mean value of $c$ is 0.4 to $2.9\times10^{-4}$. For more details about the calculation of $c$, $w$, $m$, and $R$, see \cite{shearcat}.

The redshift distribution of each tomographic bin for the \metacal~measurements is shown in Fig. \ref{fig:nofz}. 
The redshift boundaries, effective number density,
and per component $\sigma_e$ of each tomographic bin for \metacal\ are given in Table \ref{table:neff}. Due to the inherent weighting of each object in the estimator in Eq. (\ref{eq:estimator}), the objects contributing to the $n(z)$ for a tomographic bin have been weighted by the factor $W_i S_i$.

\begin{figure}
\begin{center}
\includegraphics[width=\columnwidth]{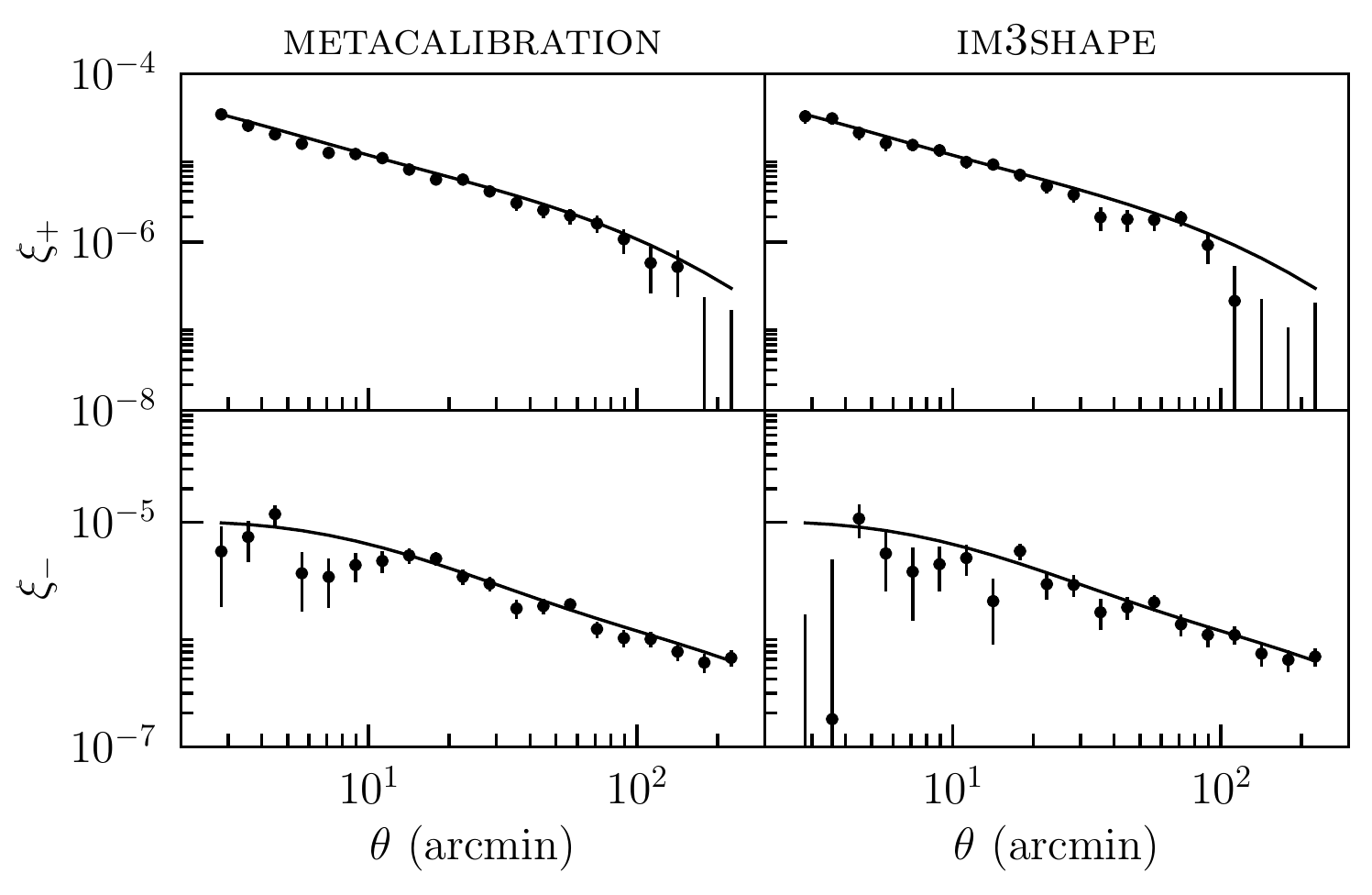}
\end{center}
\caption[]{The measured non-tomographic shear correlation function $\xi_{\pm}$ for the DES Y1 shape catalogs. The best-fit $\Lambda$CDM theory line from the fiducial tomographic analysis is shown as the same solid line compared to measurements from both catalogs.
\label{fig:xinotomo}}
\end{figure}

\begin{figure*}
\begin{center}
\includegraphics[width=\textwidth]{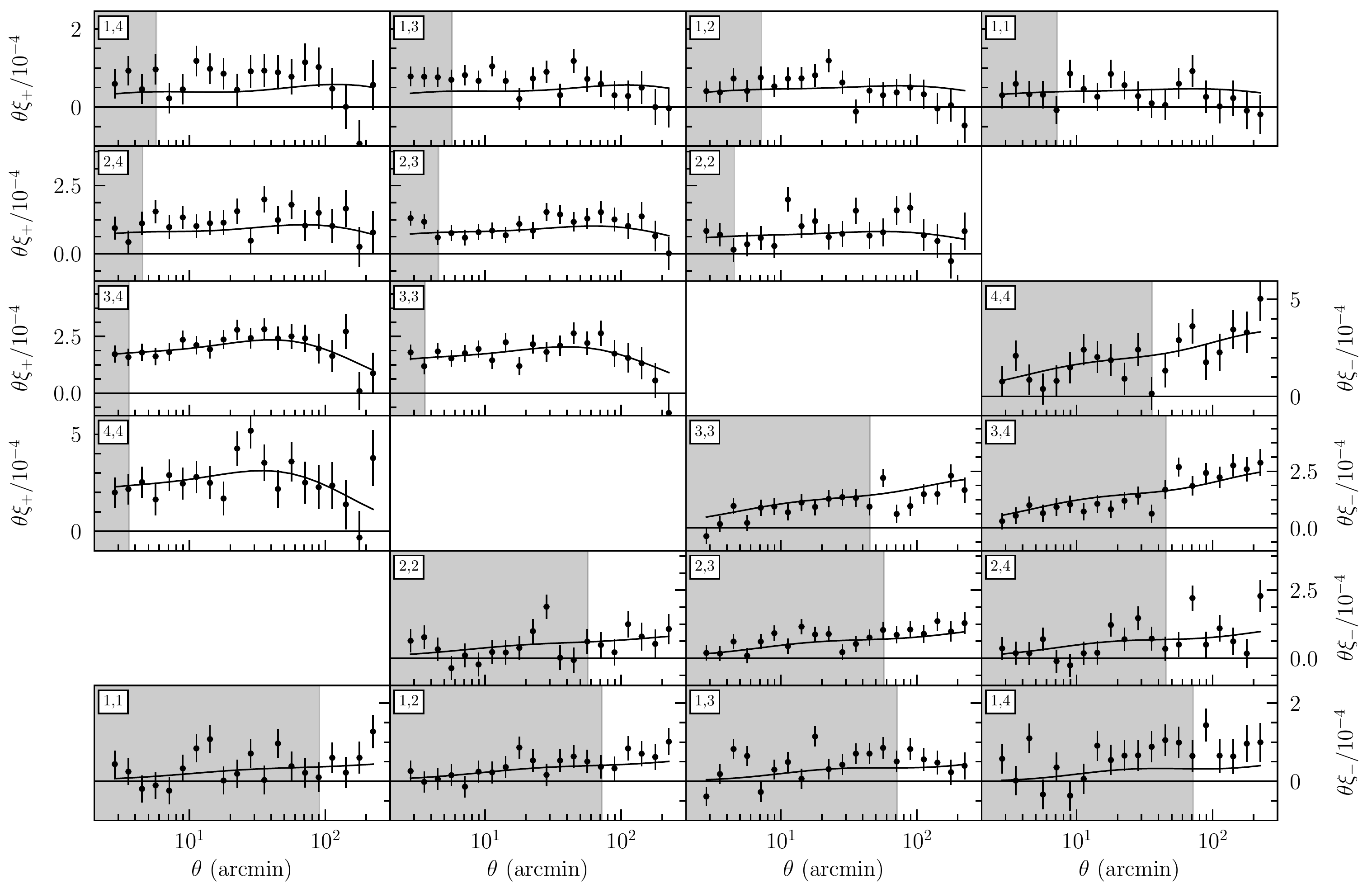}
\end{center}
\caption[]{The measured shear correlation function $\xi_{+}$ (top triangle) and $\xi_{-}$ (bottom triangle) for the DES Y1 \metacal\ catalog. Results are scaled by the angular separation ($\theta$) to emphasize features and differences relative to the best-fit model. 
The correlation functions are measured in four tomographic bins spanning the redshift ranges listed in Table \ref{table:neff}, with labels for each bin combination in the upper left corner of each panel.
The assignment of galaxies to tomographic bins is discussed in Sec. \ref{sec:photoz}. Scales which are not used in the fiducial analysis are shaded (see Sec. \ref{sec:systheory}).  The best-fit $\Lambda$CDM theory line from the full tomographic analysis is shown as the solid line. We find a $\chi^2$ of 227 for 211 degrees of freedom in the non-shaded regions. 
 \label{fig:ximcal}}
\end{figure*}

\begin{figure*}
\begin{center}
\includegraphics[width=\textwidth]{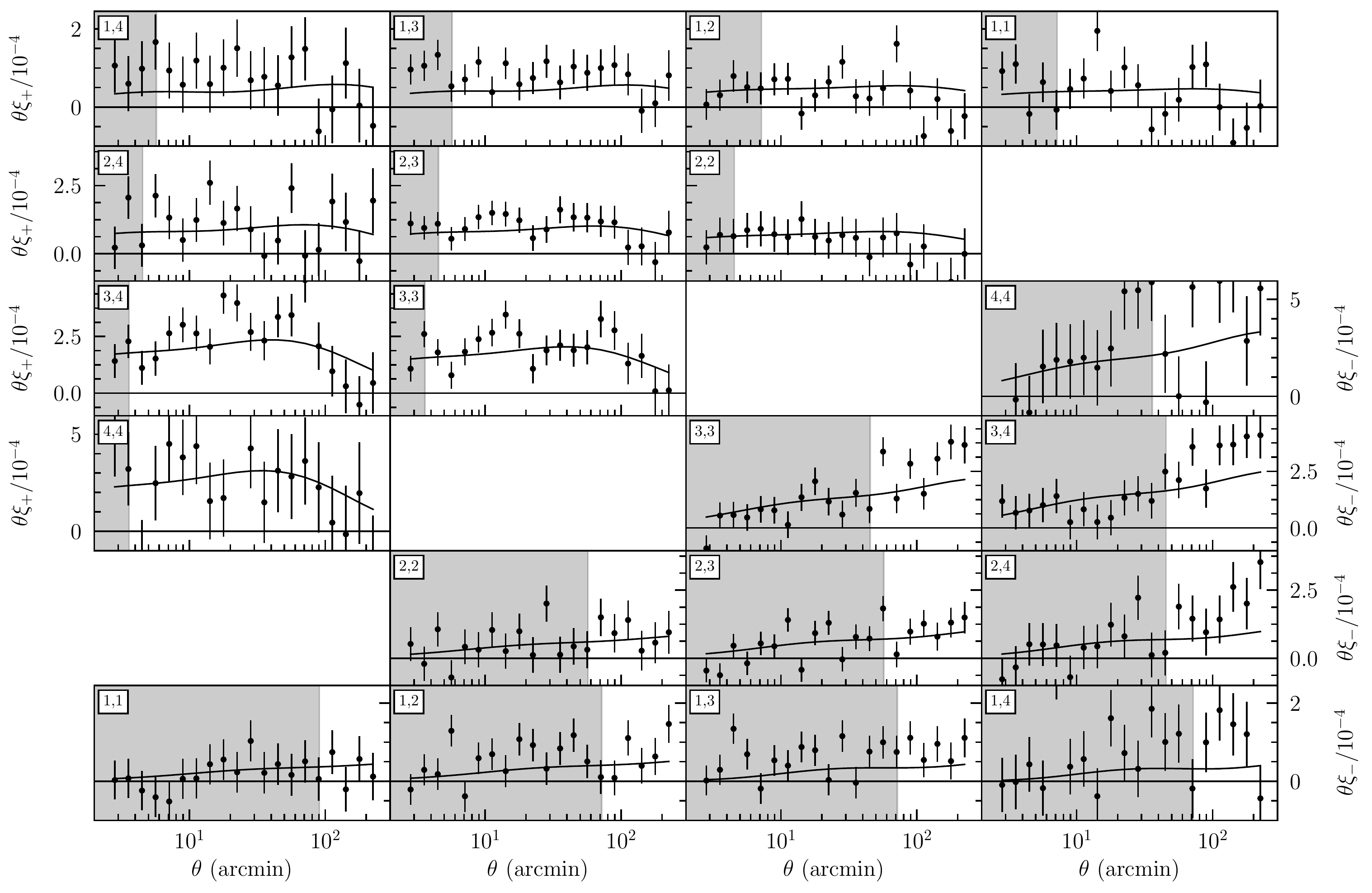}
\end{center}
\caption[]{The measured shear correlation function $\xi_{+}$ (top triangle) and $\xi_{-}$ (bottom triangle) for the DES Y1 \imshape\ catalog (see caption of Fig. \ref{fig:ximcal}). 
The uncertainty on $\xi_{\pm}$ is clearly larger for \imshape\ compared to \metacal\ in Fig. \ref{fig:ximcal} due to the lower number density of objects. 
We find a $\chi^2$ of 224 for 211 degrees of freedom in the non-shaded regions. 
\label{fig:xii3}}
\end{figure*}

We show the measured two-point correlation function $\xi_{\pm}$ for each shape catalog in Figs. \ref{fig:xinotomo} -- \ref{fig:xii3}. Scales not used to constrain cosmological parameters are shaded in Fig. \ref{fig:ximcal} \& \ref{fig:xii3}. This is the first measurement to correct, through the metacalibration process, the shear selection effects $R_S$, e.g., due to \photoz\ binning in the data. This effect can be only roughly approximated in traditional image simulation calibrations by assigning redshifts based on the original redshift measurement of the input objects, which is not the same as the redshift measurement used in the data and not even necessarily correlated with magnitude or color in a natural way in the simulation.\footnote{We preserve the original \cosmos\ magnitudes of objects in the simulations used to calibrate \imshape, so the assigned redshifts do correspond to the flux and morphology of the simulated image.}

The measured selection effects $R_S$ vary in each redshift bin from $0.007$ (lowest $z$-bin) to $0.014$ (highest $z$-bin), which can be compared to the shear response correction $R_\gamma$ in the four tomographic bins that ranges from 0.72 (lowest $z$-bin) to 0.56 (highest $z$-bin). The $R_S$ is comparable to the Gaussian prior width on the multiplicative bias of $0.013$ for the \metacal\ catalog. This effect can also be compared to the selection bias correction with no tomographic binning, which is 0.011. Thus, the inclusion of the selection bias correction calculated from the four versions of BPZ based on the sheared photometry from \metacal\ is likely a significant contribution to the corrected selection bias, and the additional computational resources and complexity introduced are warranted.

\section{Covariance Matrix}\label{sec:covmatrix}

The calculation of the covariance matrix of $\xi_\pm$ and tests to validate its quality can be found in \cite{methodpaper}. A large part of our covariance is caused by the shape-noise and Gaussian components of the covariance, i.e., covariance terms that involve at most two-point statistics of the cosmic shear fields. To guarantee that our covariance model captures these error contributions correctly, the Gaussian parts of the model are compared to a sample covariance from 1200 Gaussian random realizations of the shear fields in our tomographic bins. The uncertainties on cosmological parameters projected from each of these covariances agree very well \cite{methodpaper}. The non-Gaussian parts of our covariance, i.e., the parts involving higher order correlations of the shear field, are modeled in a halo model framework \cite{ke16}. To measure the influence of realistic survey geometry on the covariance matrices, covariance matrices determined in three different ways are compared: 1) the full halo model covariance, 2) a sample covariance from 1200 lognormal realizations (see Sec. \ref{sec:lognormal}) of the convergence field in our tomographic bins that assumes a circular survey footprint, and 3) a sample covariance from 1200 lognormal realizations using our actual DES Y1 footprint.

\begin{figure}
\begin{center}
\includegraphics[width=\columnwidth]{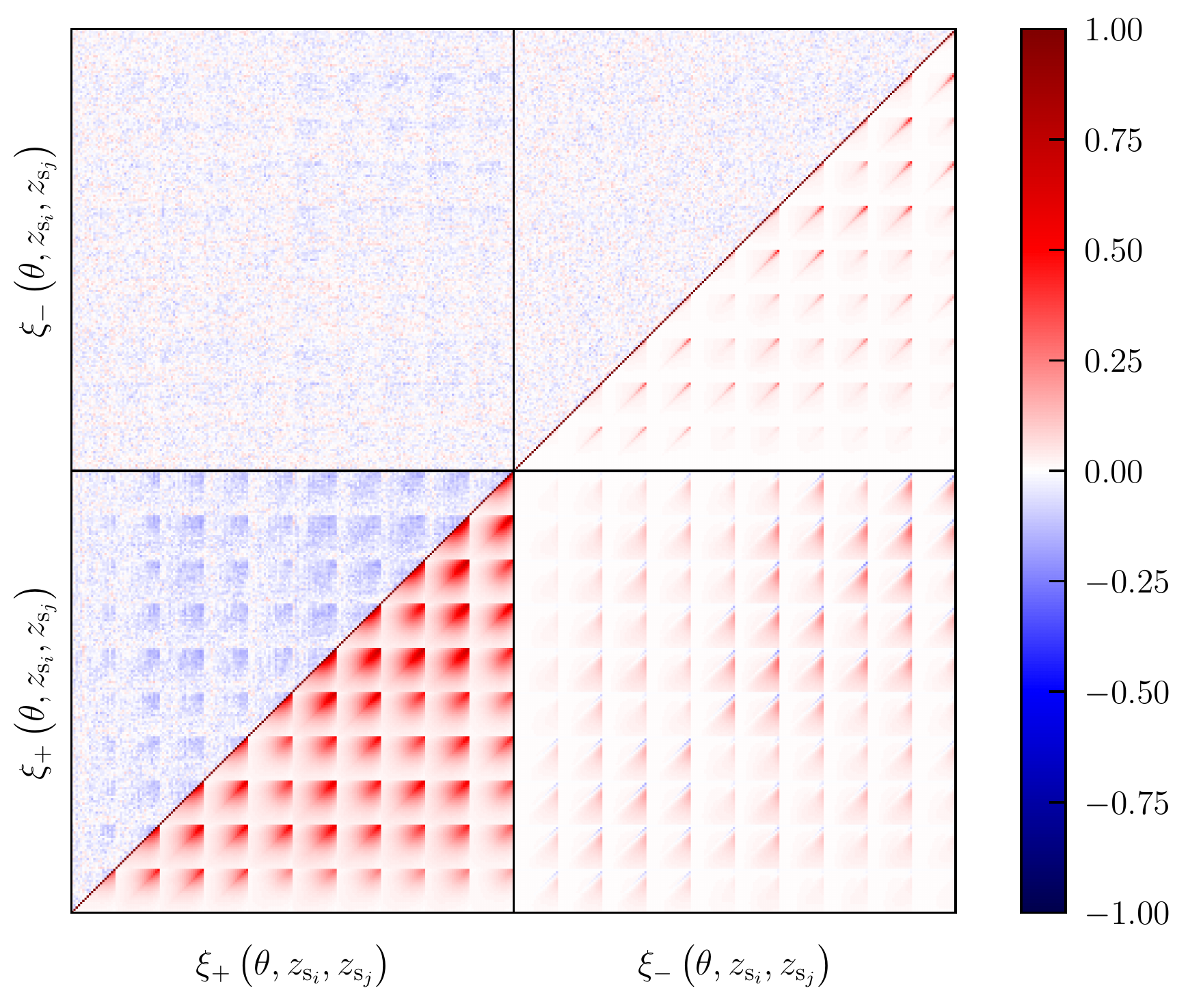}
\end{center}
\caption[]{The cosmic shear correlation matrix. The fiducial halo model correlation matrix is shown in the lower triangle, while the difference of this with the correlation matrix derived from 1200 \flask\ lognormal simulations with the DES Y1 mask applied is shown in the upper triangle. This shows primarily the noise in the \flask\ covariance, and any differences between the two derived covariance matrices were shown to be entirely negligible in \cite{methodpaper}. Elements are ordered to the right (upward) by increasing redshift bin pair index $ij$ with $i\le j$ (i.e., 11, 12, 13...). Within each $ij$ block, angular scales also increase to the right (upward). \label{fig:covmat}}
\end{figure}

We show the full halo model correlation matrix for $\xi_{\pm}$ as the lower triangle in Fig. \ref{fig:covmat}. The upper triangle is the difference of the full halo model correlation matrix and the correlation matrix resulting from the 1200 lognormal realizations masked by the DES Y1 footprint. Following the suggestion of an iterative approach to dealing with the cosmological dependence of covariance matrices proposed by \cite{2009A&A...502..721E}, an initial covariance matrix was calculated using an arbitrary cosmology, but the final covariance matrix used in this work was recalculated with the best-fit cosmology of the initial fiducial result from \cite{keypaper}. We found no significant change in our inferred cosmology due to this covariance change. 

We also test the amplitude of the diagonal of the covariance matrix by comparing the halo model prediction for the variance of $\xi_{-}$ on small scales ($2.5 < \theta < 10$ arcmin) to the variance of  $\xi_{-}$ directly estimated from DES Y1 data. To obtain the latter, we divide the shape catalogs into 200 patches of similar area using the \texttt{kmeans} algorithm\footnote{https://github.com/esheldon/kmeans\_radec} and take the variance of the  $\xi_{-}$ measurement in each of them. We find good agreement between these two approaches within the uncertainty of the estimate of the variance of the $\xi_{-}$ measurement.

\section{Blinding}\label{sec:blinding}

For the DES Y1 analysis, we have maintained a catalog-level blinding scheme similar to the DES SV analyses, but rescaling $|\eta| = 2 \arctanh{|e|}$ by a factor between 0.9 and 1.1 (see \cite{2005ARNPS..55..141K} for a review of blinding in general). This catalog blinding\footnote{During the internal review process for \cite{shearcat}, it was discovered that separate, but equivalent, oversights in the shear calibration of the two catalogs led to a substantial fraction (e.g., the linear part in $e$) of the blinding factor being calibrated. This was undiscovered until the catalogs were finalized, and thus had no impact on catalog-level choices. It is valid to question whether this invalidated our blinding strategy at the parameter estimation level. It did not, for two reasons: 1) only a few people in the collaboration were aware of the potential issue until after we unblinded the cosmological parameters, minimizing any impact, and 2) The secondary blinding enforced at the two-point and parameter level ensured that even had we become aware of this oversight much sooner, it could not have led to experimenter bias in our analyses.} 
was preserved until the catalogs and primary DES Y1 cosmological analyses and papers (this work and \cite{keypaper}) completed a first round of the DES internal review process. 
All calculations were then repeated with the unblinded catalogs for the final version of this paper.

In addition to this catalog-level blinding, no comparison to theory at the two-point level ($\xi_{\pm}$) or of cosmological contours was made, nor were central values of any cosmological inferences revealed, until after the shape catalogs and priors were finalized. A qualitative comparison of results from the two shape catalogs with axes and values suppressed was performed to confirm that they produced consistent results after their development was complete and before finalizing the shear priors. The results of this test were acceptable, and no modification to the shape catalogs or priors was necessary. All measurement, processing, and plotting routines were tested either on measurements of the mock catalogs or on synthetic data vectors before use on the DES data. Only after analysis plans were finalized was any comparison to theory allowed or inferred cosmological parameter values revealed. Several negligible updates to the precise values of the shear and photo-z priors, and some bugs in the \imshape\ catalog selection related to blacklisted images, were approved after unblinding the catalogs but before unblinding the parameter values. These changes did not have an impact on the final results -- changes due to updates to the \imshape\ selection, for example, were entirely negligible at the two-point function level. 

Following initial submission of this paper, further investigations were performed to identify the source of an initially large reported $\chi^2$ for the DES Y1 analyses. During the course of this investigation we identified three further modifications that we have implemented in the final published analysis. The first is a change to use the actual number of pairs of galaxies that enter the two-point function to evaluate the shape-noise component of the covariance matrix. The second change was to properly account for the measured variation in $\sigma_e$ between redshift bins, which was not propagated to the covariance originally. This impacts the reported cosmic shear results by reweighting information between tomographic bins.  These changes to the covariance significantly improve the initially large $\chi^2$. The impact of survey geometry on the shape noise part of the covariance was intended to be tested in our covariance validation scheme, but we did not examine the resulting $\chi^2$ in these tests -- doing so with our \flask\ realizations indicates that correcting how we calculate the shape noise resolves a clear offset in the $\chi^2$ distribution from the mocks, while inflating our constraints only slightly. Finally, we corrected a minor bug in the \metacal\ selection processes, which increases the number density of objects at the 2\% level. The last two corrections were identified while trying to isolate the cause of the initially large $\chi^2$. 

\section{Modeling Choices}\label{sec:model}

The measured $\xi_{\pm}$ for tomographic bins $i$ and $j$ can be related to the angular convergence power spectrum in a flat universe through the integral
\begin{equation}
\hat{\xi}^{ij}_{\pm}(\theta) = \frac{1}{2\pi}\int d\ell \ell J_{0/4}(\theta \ell) P^{ij}_{\kappa}(\ell),
\label{eq:Pkappa}
\end{equation}
where $J_{n}$ is the $n$th order Bessel function of the first kind. $P_{\kappa}$ is then related to the matter power spectrum $P_\delta$ with the harmonic-space version \cite{limberkaiser1,limberkaiser2} of the Limber approximation \cite{limber,PhysRevD.78.123506}
\begin{equation}
P^{ij}_{\kappa}(\ell) = \int_0^{\chi_H}d\chi \frac{q^i(\chi)q^j(\chi)}{\chi^2} P_{\textrm{NL}}\left(\frac{\ell+1/2}{\chi},\chi\right),
\label{eq:Pdelta}
\end{equation}
where $\chi$ is radial comoving distance, $\chi_H$ is the distance to the horizon, and $q(\chi)$ is the lensing efficiency function
\begin{equation}
q^i(\chi) = \frac{3}{2}\Omega_m \left(\frac{H_0}{c}\right)^2 \frac{\chi}{a(\chi)} \int_{\chi}^{\chi_H} d\chi' n^i(\chi') \frac{\chi'-\chi}{\chi'},
\end{equation}
where $\Omega_m$ is the matter density parameter, $H_0$ is the Hubble constant, $c$ is the speed of light, $a$ is the scale factor, and $n^i(\chi)$ is the effective number density of galaxies as a function of comoving radial distance normalized such that $\int d\chi n^i(\chi)=1$. The appropriateness of the Limber and flat-sky approximations in these relationships is tested in \cite{methodpaper} for DES Y1 statistical precision.

Our data vector $D$ (i.e., Figs. \ref{fig:ximcal} \& \ref{fig:xii3}) contains 227 data points after the cuts described in Sec. \ref{sec:systheory}. We sample the likelihood, which is assumed to be Gaussian in the multi-dimensional parameter space:
\begin{equation}
\ln\mathcal{L}(\bm{p}) = -\frac{1}{2}\sum_{ij} \left(D_i-T_i(\bm{ p})\right) C^{-1}{}_{ij} \left(D_j-T_j(\bm{ p})\right)
\end{equation}
where $\bm{p}$ is the full set of parameters and $T_i(\bm{ p})$ are the theoretical predictions for $\xi_{\pm}$ as given above. The likelihood also depends on the covariance matrix $C$ from Sec. \ref{sec:covmatrix}, which describes how the measurement in each angular and redshift bin is correlated with every other measurement. The covariance matrix should also depend on the model parameters $\bm{p}$, but we assume a fiducial set of parameters and use a fixed covariance. This has been shown to not impact the inferred cosmology (see Sec. \ref{sec:covmatrix}). The posterior is then the product of the likelihood with the priors, $\mathcal{P}(\bm{ p})$, as given in Table~\ref{table:params}.   

Results are derived via two analysis pipelines: \cosmolike\ \cite{cosmolike} and \cosmosis\footnote{https://bitbucket.org/joezuntz/cosmosis/wiki/Home} \cite{cosmosis}. These pipelines were validated against each other in \cite{methodpaper} and through an analysis by \cite{simspaper} on simulations. To calculate the matter power spectrum $P_{\mathrm{NL}}(k,z)$, \cosmolike\ uses \textsc{CLASS} \cite{class}, while \cosmosis\ uses \textsc{CAMB} \cite{Lewis2000,howlett2012}. We sample the parameter space using both $\textsc{MultiNest}$ \cite{mn1,mn2,mn3} and $\textsc{emcee}$ \cite{emcee} -- $\textsc{MultiNest}$ in particular produces estimates for the Bayesian evidence that we use for model and data comparison. All constraints shown are produced with $\textsc{MultiNest}$ results, primarily due to the speed of convergence and availability of the Bayesian evidence estimate.\footnote{We find a minor difference in results in poorly constrained directions of parameter space, however, with $\textsc{emcee}$ giving slightly broader results for the longest chain we compared. In general, these do not matter for the interpretation of our results (e.g., the edges of the $\Omega_b$ prior range), and in the primary constraint direction $S_8$, this amounts to a change of 0.5\% or less in the fractional constraint.}

We perform likelihood evaluation using the cosmic shear measurements described in
Sec. \ref{sec:2pt}, the redshift distributions described in
Sec. \ref{sec:photoz}, and the covariance matrix described in
Sec. \ref{sec:covmatrix}. Cosmological, astrophysical, and systematic
parameters are constrained for both the $\Lambda$CDM model and the $w$CDM
model, where the equation of state of dark energy is described by a single
parameter $w$. We leave exploration of models with non-zero curvature to future work. 
A varying neutrino mass density is included in both models, which we believe is strongly motivated, but is one reason we must recompute the likelihood of some external data (see Sec. \ref{sec:fidother}) to compare directly to our own. The parameters varied in the fiducial analysis are listed in Table \ref{table:params}, along with their range and any priors applied. For the $\Lambda$CDM model, $w$ is fixed to $-1$, while in $w$CDM it is allowed to vary in the given range. For those cosmological parameters we expect to constrain well with cosmic shear, we choose flat priors that are wide enough to be uninformative using forecasts of DES Y1 constraints. 
For $w$, we exclude regions with $w>-1/3$ that do not produce acceleration and impose a limit of $w>-2$, which is a broad enough range to allow our 1 $\sigma$ contour to close assuming we had found a posterior centered at $w=-1$. Those parameters that are not constrained well by cosmic shear have informative priors that widely bracket allowed values from external experiments. In particular, for $\Omega_{\nu}h^2$ we take a lower limit obtained from oscillation experiments \cite{pdg}, and an upper limit of $0.01$ that is roughly five times the 95\% confidence limit (CL) of the typical limiting value found by external data, $\Omega_{\nu}h^2 \approx 0.002$ (see e.g., \citet{planck2015cosmo}).\footnote{The Planck limit is derived assuming the validity of $\Lambda$CDM. The conservative inflation of this limit in our prior leaves us confident in using the prior for tests of $\Lambda$CDM.}

Though we sample over the normalization of the matter power spectrum $A_s$, we present results in terms of the commonly used parameter $S_8\equiv\sigma_8 (\Omega_m/0.3)^\alpha$. Choosing $\alpha=0.5$ largely decorrelates $S_8$ and $\Omega_m$ in constraints from cosmic shear experiments. The amplitude of the shear correlation function is roughly $\propto S_8^2$. We will refer to both the 68\% confidence limit, which is the area around the peak of the posterior within which 68\% of the probability lies, as well as the figure of merit (FoM), which is useful for comparing the relative constraining power of results in 2D. In two parameter dimensions, the FoM is defined for parameters $p_1$ and $p_2$ as \cite{HutTur01,wang08}:
\begin{equation}
\mathrm{FoM}_{p_1-p_2} = \frac{1}{\sqrt{\det{\mathrm{Cov}(p_1,p_2)}}},\label{eq:fom}
\end{equation}
which is a generalization of the Dark Energy Task Force (DETF) recommendation for the dark energy FoM \cite{detf}. This kind of statistic is naturally motivated by the form taken by a change in relative entropy driven by a gain in information. 

\subsection{Matter Power Spectrum Modeling and Baryonic Effects}\label{sec:systheory}

Approximations in the nonlinear clustering of matter on small scales, including the impact of baryonic effects, is a key modeling choice for the cosmic shear signal. The discussion in \cite{des2016}, to which we refer the reader, remains applicable to the current analysis, though some updates to scale selection are necessary and are discussed further in Sec. \ref{sec:scales}. We also explore the impact of these modeling choices in \cite{methodpaper}. 

To model the nonlinear matter power spectrum, we use \textsc{HALOFIT} \cite{SPJ+03} with updates from \cite{takahashi2012}. The impact of neutrino mass on the matter power spectrum is implemented in \textsc{HALOFIT} from \cite{nu}, which introduces some additional uncertainty of potentially up to 20\% (e.g., \cite{2010ApJ...715..104H,2016JCAP...04..047S,2017ApJ...847...50L}). This is not a significant concern for this analysis,  however, given our scale cuts and the fact that cosmic shear alone is insensitive to the effects of neutrino mass (see Appendix \ref{sec:fullresults}). 

The fiducial analysis removes scales that could be significantly biased by baryonic effects. For scale selection, these effects are modeled as a rescaling of the nonlinear matter power spectrum 
\begin{equation}
P_{\textrm{NL}}(k,z) \rightarrow \frac{P_{\textrm{DM}+\textrm{Baryon}}}{P_{\textrm{DM}}}P_{\textrm{NL}}(k,z),
\end{equation}
where `DM' refers to the power spectrum from the OWLS (OverWhelmingly Large Simulations project) dark-matter-only simulation, while `DM+Baryon' refers to the power spectrum from the  OWLS AGN simulation \cite{schaye10,vandalen11}. OWLS is a suite of hydrodynamic simulations with different sub-grid prescriptions for baryonic effects. We use this particular OWLS simulation for two reasons. First, it is the one which deviates most from the dark matter-only case in the relevant scales of the matter power spectrum; given we are cutting scales based on the size of this deviation, this is a conservative choice. Secondly, \citet{mccarthy11} find that of the OWLS, the AGN simulation best matches observations of galaxy groups in the X-ray and optical, so arguably it is the most realistic. 

We remove any scales from the $\xi_{\pm}$ data vector that would have a fractional contribution from baryonic effects exceeding 2\% at any physical scale. This removes a significant number of data points, particularly in $\xi_{-}$, on small scales. In general we find that our cuts in scale to remove parts of the cosmic shear data vector contaminated by potential baryonic effects are sufficient to alleviate any potential bias due to uncertainties in modeling nonlinear matter clustering. This can be seen in \cite{mead}, which compares inaccuracies in \textsc{HALOFIT} relative to \textsc{COSMIC EMU} \cite{emu,emu2}, a power spectrum emulator, with the impact of baryonic effects from OWLS AGN, which is comparable or larger at all $k$. 

\subsection{Intrinsic Alignment Modeling}\label{sec:ia}

In addition to coherent shape distortions induced by lensing, galaxies can exhibit physical shape correlations due to their formation and evolution in the same large-scale gravitational environment. Along with baryonic effects, `intrinsic alignment' (IA) constitutes the most significant astrophysical systematic to cosmic shear. IA includes both an `Intrinsic-Intrinsic' (II) term due to physically nearby galaxy pairs \citep{HRH2000,croft00,CKB01,CNP+01} and a `Gravitational-Intrinsic' (GI) term from the correlation of galaxies that are aligned with those that are lensed by the same structure \citep{HS04}. The total measured cosmic shear signal is the sum of the pure lensing contribution and the two IA terms:
\begin{equation}
P^{ij}_{\rm obs}(\ell) = P^{ij}_{\rm GG}(\ell) + P^{ij}_{\rm GI}(\ell) + P^{ij}_{\rm IG}(\ell) + P^{ij}_{\rm II}(\ell),
\end{equation}
where `GG' refers to the cosmic shear signal. Neglecting IA can lead to significantly biased cosmological constraints \citep{HRH2000,JMA+11,KRH+12,KEB16}. See the reviews \cite{Troxel20151,Joachimi2015} for further information on these effects.

For our fiducial analysis, we treat IA in the `tidal alignment' paradigm, which assumes that intrinsic galaxy shapes are linearly related to the tidal field \citep{CKB01}. While a complete understanding of alignment mechanism(s) remains a topic of active study, tidal alignment has been shown to accurately describe red/elliptical galaxy alignment and is expected to dominate the IA signal on linear scales \citep{JMA+11,BMS11,BVS15}. We also perform an analysis that includes the potential impact of alignments from angular momentum correlations (e.g., tidal torque theory \cite{LP00}) which are expected to contribute for spiral galaxies, although the amplitude is likely smaller than tidal alignment of ellipticals \cite{MBB+11}.
This work is thus the first to include both tidal alignment and tidal torque-type alignments, as well as their cross-correlation \cite{bmt17}. This more general, `mixed' model is completely analogous to a perturbative expansion of galaxy bias beyond linear order and is thus expected to capture the relevant alignment effects at next-to-leading order, even if they are not due to classical `tidal torquing.' 

The amplitude ($A$) of the `nonlinear alignment' (NLA) model \cite{bk07} and its redshift evolution ($\eta$) are allowed to vary in our fiducial analysis, such that the amplitude is described by $A\equiv A[(1+z)/(1+z_0)]^{\eta}$, where the pivot redshift is chosen to be approximately the mean redshift of sources $z_0=0.62$. This is an improvement over the fiducial analyses of previous cosmic shear studies, which fixed this power-law dependence (or neglected it entirely), and this fiducial model is the one employed in our combined probes analysis \cite{keypaper}. The amplitude of the terms are then scaled as $P_{\rm GI}(\ell)\propto -A$ ($\rm GI\leftrightarrow \rm IG$) and $P_{\rm II}(\ell)\propto A^2$, following the standard tidal alignment sign convention. In Sec.~\ref{sec:sysia} we vary the IA model to demonstrate robustness to the specific modeling choice.

\subsection{Modeling Shear Systematics}

The shear multiplicative bias
is modeled as \cite{Heymans06,HTBJ06}
\begin{equation}
\xi^{ij} = (1+m^i)(1+m^j) \xi^{ij}_{\mathrm{true}},
\end{equation}
where $m^{i}$ are free to independently vary in each tomographic bin. We do not explicitly marginalize over the potential impact of additive systematics. We use a Gaussian prior on $m^i$ of $0.012\pm 0.023$ for \metacal, given in Table \ref{table:params}, which is rescaled from the non-tomographic prior $m=0.012\pm 0.013$ due to potential correlations between tomographic bins as discussed in Appendix D of \cite{shearcat}. The equivalent rescaled \imshape\ prior on $m^i$ is $0.0\pm0.035$. Both are allowed to vary independently in each tomographic bin.

The only potential source of additive systematics we have identified in \cite{shearcat} is related to incorrect modeling of the PSF. We can model the impact of the PSF model errors in cosmic shear and this is described in detail in Appendix \ref{sec:psf} along with a discussion of the residual mean shear in each tomographic bin, which is not fully described by PSF model errors. We find that after correcting the signal for the mean shear, the effect of PSF modeling errors is negligible. 

\subsection{Modeling Photo-$z$ Systematics}

The \photoz\ bias is modeled as an additive shift of the $n(z)$ as shown in Eq. \ref{eq:Dz}, where $\Delta z^i$ are free to independently vary in each tomographic bin. As discussed in Sec. \ref{sec:photoz}, this is a sufficient approximation for the DES Y1 cosmic shear analysis, and this is further validated in Sec. \ref{sec:sysphotoz}. The Gaussian priors on $\Delta z^i$ for the \metacal\ measurements are listed in Table \ref{table:params}. We separately calibrate priors for the \imshape\ measurements, which have Gaussian priors of $\Delta z^i = (0.004\pm 0.015; -0.024\pm 0.013; -0.003\pm 0.011; -0.057\pm 0.022)$ \cite{photoz,xcorr}. When using the resampled \cosmos\ $n^i(z)$, the same width for the prior on $\Delta z^i$ is used, but it is centered at zero. All $\Delta z^i$ are allowed to vary independently in each tomographic bin. As in the case of shear calibration, the width of these priors accounts for correlations between tomographic bins as described in Appendix A of \cite{photoz}.

\begin{figure}
\begin{center}
\includegraphics[width=0.95\columnwidth]{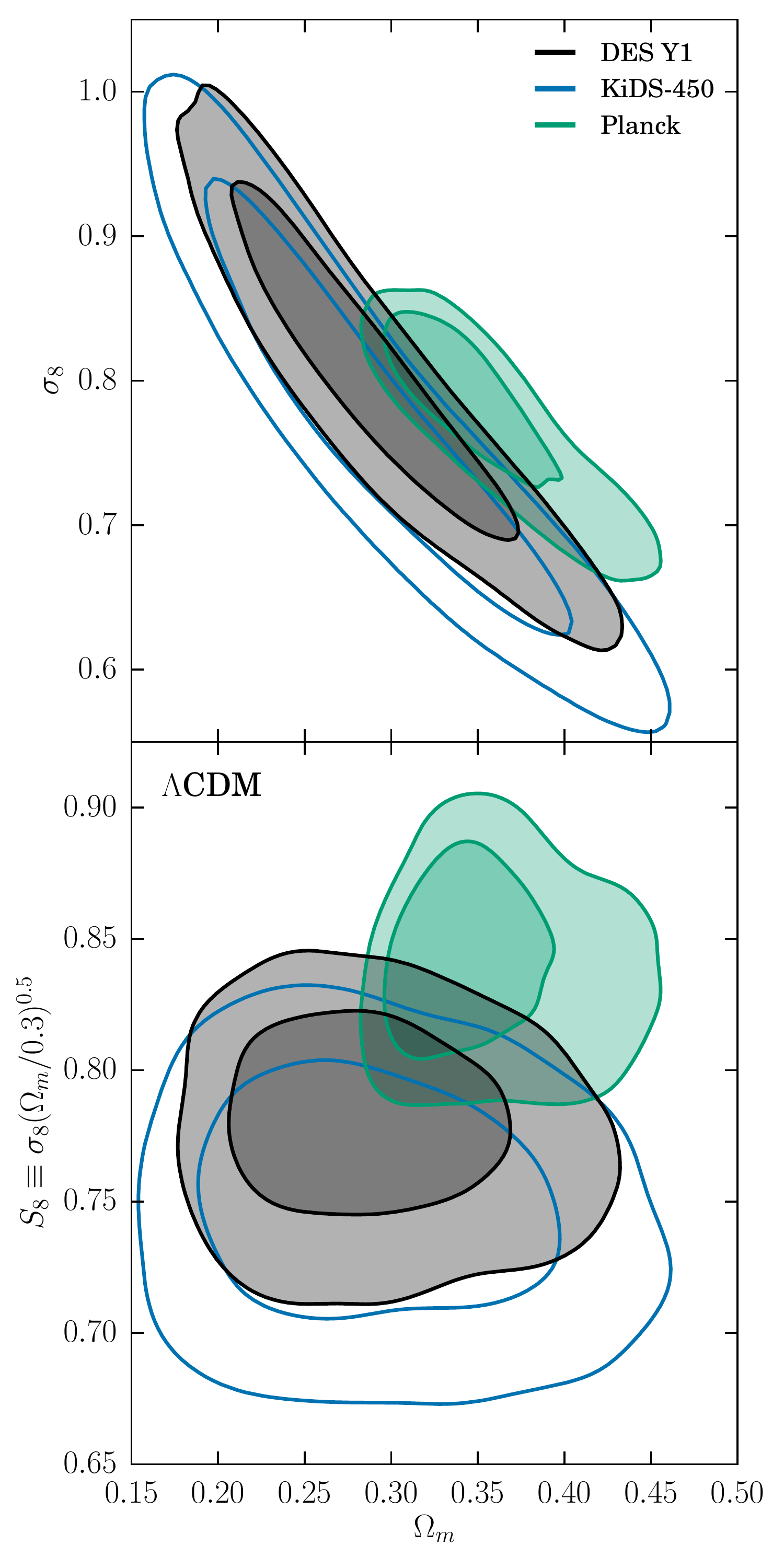}
\end{center}
\caption[]{Fiducial constraints on the clustering amplitude $\sigma_8$ and $S_8$ with the matter density $\Omega_m$ in $\Lambda$CDM. The fiducial DES Y1 cosmic shear constraints are shown by the gray filled contours, with Planck CMB constraints given by the filled green contours, and cosmic shear constraints from KiDS-450 by unfilled blue contours. In all cases, 68\% and 95\% confidence levels are shown. External data have been reanalyzed in our model space, as described in Sec. \ref{sec:fidother}. 
\label{fig:fidsig8lcdm}
}
\end{figure}

\begin{figure}
\begin{center}
\includegraphics[width=\columnwidth]{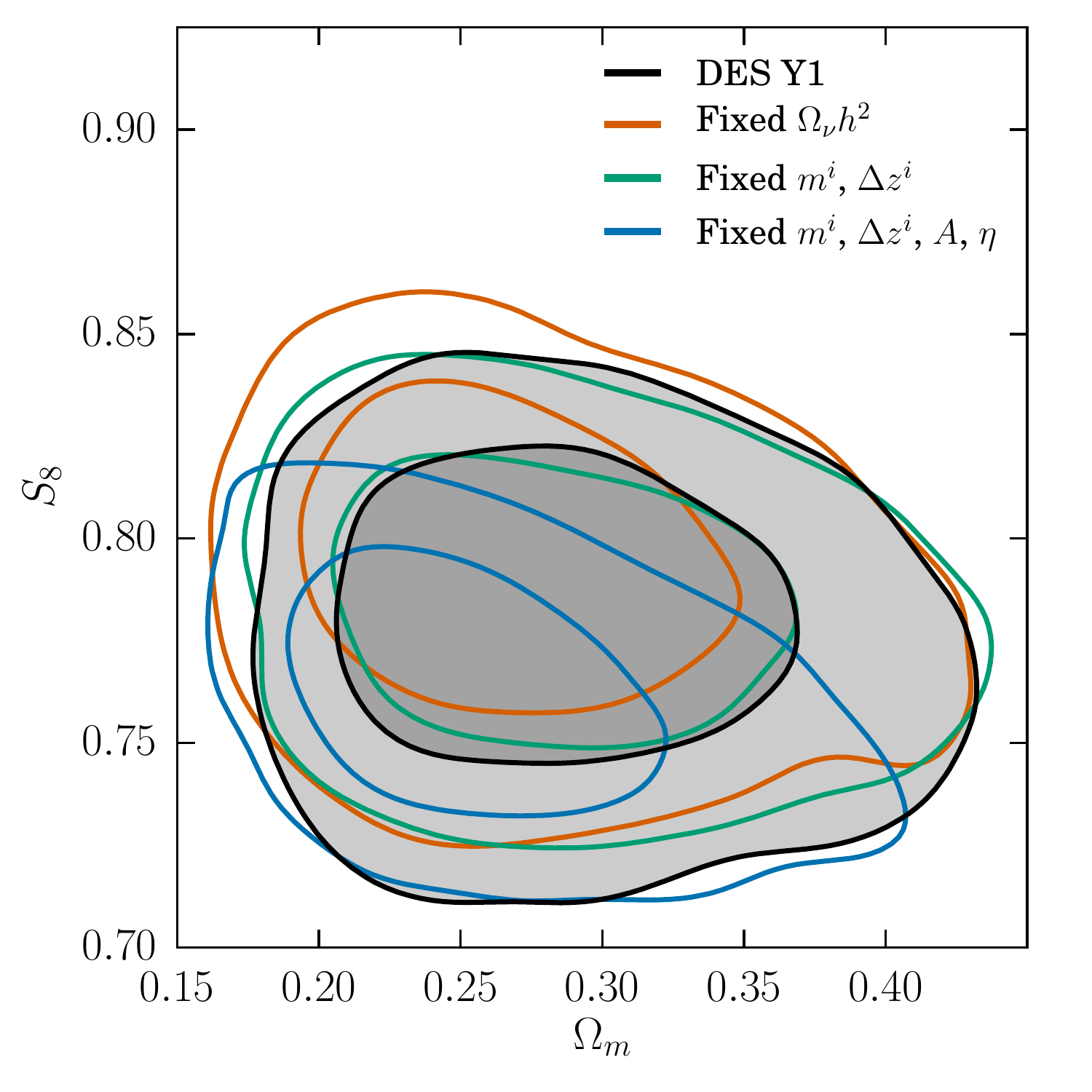}
\end{center}
\caption[]{A comparison of the fiducial constraints in $\Lambda$CDM (filled
gray contours) to constraints where we: 1) fix $\Omega_{\nu}h^2$ (orange contours), 2) fix all photo-z and shear systematic parameters (green), and 3) fix all systematic parameters and intrinsic alignment (IA) parameters (green). We find no visually significant bias correction or decrease in constraining power including systematics parameters, but varying $\Omega_{\nu}h^2$ and IA parameters both shift and enlarge the resulting contours.  Both 68\% and 95\% confidence levels are shown. 
 \label{fig:nosys}}
\end{figure}

\section{Cosmological Parameter Constraints}\label{sec:params}

Given the size and quality of the DES Y1 shape catalogs, we are able to make a highly significant statement about the robustness of the standard $\Lambda$CDM cosmological model. Our measurements of cosmic shear probe the evolution of nonlinear fluctuations in the underlying matter field and expansion of space across a very large volume around $z\approx 0.6$. By comparison, equally constraining measurements of the CMB at $z=1100$ use information from linear perturbations in the radiation field to constrain the same model eight billion years before light left the galaxies we now observe in DES. Comparing the prediction of these very different probes at the same redshift via the parameter $S_8$ allows us to test whether these results are consistent within the $\Lambda$CDM model to high precision. 

Using the fiducial modeling choices described in the previous section, we use cosmic shear from the first year of the Dark Energy Survey to constrain both the $\Lambda$CDM and $w$CDM models with varying neutrino mass to produce the tightest cosmological constraints from cosmic shear to date. In \cite{keypaper}, our cosmic shear results are further combined with galaxy-galaxy lensing and galaxy clustering to significantly improve these constraints. When comparing with external data, it is important to note that we vary $\Omega_{\nu}h^2$ in our fiducial analysis, and thus all results we compare to, and so the central values and uncertainties of parameters may differ from those previously published for these data.

\begin{table*}
\caption{
Summary of constraints on the 1D peak value of $S_8$ and $\Omega_m$ in $\Lambda$CDM. The FoM in Eq. (\ref{eq:fom}) for the $S_8-\Omega_m$ plane is also shown. In the $w$CDM model, we show the 1D peak value of $S_8$, $\Omega_m$, and $w$, along with the FoM for the $S_8-w$ plane. 68\% CL are included for each parameter, which are not symmetric about the peak in most cases. We distinguish variations on the fiducial model that are not required to give consistent results (e.g., by neglecting astrophysical systematics) by an asterisk. The constraints on $S_8$ are also visually summarized in Fig. \ref{fig:fids8}.}
\label{table:s8}
\begin{ruledtabular}
\begin{tabular}{lccccccc}
& \multicolumn{3}{c}{$\Lambda$CDM} & \multicolumn{4}{c}{$w$CDM} \\
\bigstrutdown Model & $S_8$ & $\Omega_m$ & FoM$_{S_8 - \Omega_m}$ & $S_8$ & $\Omega_m$ & $w_0$ & FoM$_{S_8 - w_0}$\\
\hline 
        \bigstrutdown \quad \\
	\bigstrutdown Fiducial & 							$0.782^{+0.027}_{-0.027}$  & $0.260^{+0.065}_{-0.037}$  & $626$  		& $0.777^{+0.036}_{-0.038}$  & $0.274^{+0.073}_{-0.042}$  & $-0.95^{+0.33}_{-0.39}$  & $106$  \\
        \bigstrutdown *Fixed neutrino mass density & 			$0.789^{+0.031}_{-0.019}$  & $0.248^{+0.065}_{-0.036}$  & $675$  		& $0.791^{+0.031}_{-0.044}$  & $0.264^{+0.067}_{-0.049}$  & $-0.94^{+0.28}_{-0.44}$  & $105$  \\
        \bigstrutdown *No photo-$z$ or shear systematics & 	$0.786^{+0.020}_{-0.028}$  & $0.248^{+0.080}_{-0.025}$  & $667$  		& $0.771^{+0.040}_{-0.040}$  & $0.276^{+0.068}_{-0.048}$  & $-1.07^{+0.41}_{-0.39}$  & $108$  \\
        \bigstrutdown *Only cosmological parameters & 	$0.760^{+0.023}_{-0.021}$  & $0.250^{+0.039}_{-0.046}$  & $970$  		& $0.733^{+0.036}_{-0.034}$  & $0.229^{+0.062}_{-0.038}$  & $-1.37^{+0.43}_{-0.35}$  & $138$  \\
        \hline \vspace{0.1cm}
{\it Shape Measurement} &&&& \vspace{0.15cm} \\
\bigstrutdown *No shear systematics & 				$0.783^{+0.025}_{-0.025}$  & $0.269^{+0.068}_{-0.038}$  & $700$  		& $0.766^{+0.049}_{-0.029}$  & $0.285^{+0.069}_{-0.055}$  & $-0.98^{+0.31}_{-0.44}$  & $105$  \\
        \bigstrutdown {\bf \imshape } 				& 	$0.799^{+0.048}_{-0.045}$  & $0.302^{+0.072}_{-0.057}$  & $279$  		& $0.778^{+0.053}_{-0.050}$  & $0.314^{+0.061}_{-0.069}$  & $-1.51^{+0.47}_{-0.32}$  & $52$  \\
        \hline \vspace{0.1cm}
{\it Photometric Redshifts} &&&& \vspace{0.15cm} \\
\bigstrutdown *No photo-$z$ systematics & 			$0.778^{+0.026}_{-0.026}$  & $0.272^{+0.053}_{-0.047}$  & $676$  		& $0.762^{+0.044}_{-0.034}$  & $0.270^{+0.067}_{-0.045}$  & $-0.99^{+0.30}_{-0.46}$  & $105$  \\
        \bigstrutdown Cosmos matched photo-$z$s & 		$0.776^{+0.022}_{-0.029}$  & $0.260^{+0.053}_{-0.034}$  & $739$  		& $0.772^{+0.038}_{-0.043}$  & $0.279^{+0.071}_{-0.037}$  & $-1.06^{+0.45}_{-0.29}$  & $96$  \\
        \bigstrutdown Removing highest $z$-bin & 		$0.776^{+0.032}_{-0.043}$  & $0.256^{+0.071}_{-0.042}$  & $381$  		& $0.784^{+0.038}_{-0.063}$  & $0.290^{+0.073}_{-0.052}$  & $-0.72^{+0.22}_{-0.52}$  & $58$  \\
        \hline \vspace{0.1cm}
{\it Data Vector Choices} &&&& \vspace{0.15cm} \\
\bigstrutdown *Extended angular scales & 			$0.758^{+0.019}_{-0.024}$  & $0.270^{+0.067}_{-0.033}$  & $915$  		& $0.761^{+0.027}_{-0.042}$  & $0.273^{+0.064}_{-0.043}$  & $-0.97^{+0.36}_{-0.34}$  & $128$  \\
        \bigstrutdown Large angular scales & 			$0.799^{+0.049}_{-0.046}$  & $0.307^{+0.087}_{-0.049}$  & $292$  		& $0.767^{+0.069}_{-0.051}$  & $0.324^{+0.082}_{-0.065}$  & $-1.41^{+0.60}_{-0.28}$  & $52$  \\
        \bigstrutdown Small angular scales & 			$0.775^{+0.031}_{-0.040}$  & $0.242^{+0.075}_{-0.042}$  & $419$  		& $0.794^{+0.029}_{-0.066}$  & $0.314^{+0.063}_{-0.074}$  & $-0.70^{+0.24}_{-0.48}$  & $72$  \\
        \hline \vspace{0.1cm}
{\it Intrinsic Alignment Modelling} &&&& \vspace{0.15cm} \\
\bigstrutdown *No IA modeling & 					$0.759^{+0.021}_{-0.023}$  & $0.256^{+0.044}_{-0.040}$  & $1006$  	& $0.752^{+0.032}_{-0.039}$  & $0.249^{+0.063}_{-0.037}$  & $-1.02^{+0.29}_{-0.42}$  & $126$  \\
        \bigstrutdown *NLA & 	$0.784^{+0.020}_{-0.029}$  & $0.281^{+0.062}_{-0.054}$  & $655$  		& $0.784^{+0.035}_{-0.048}$  & $0.303^{+0.060}_{-0.063}$  & $-0.99^{+0.41}_{-0.34}$  & $106$  \\
        \bigstrutdown NLA w/ free amp. per $z$-bin & 	$0.779^{+0.032}_{-0.042}$  & $0.278^{+0.046}_{-0.053}$  & $497$  		& $0.770^{+0.039}_{-0.054}$  & $0.266^{+0.071}_{-0.043}$  & $-1.25^{+0.52}_{-0.31}$  & $69$  \\
        \bigstrutdown Mixed Alignment Model & 			$0.764^{+0.027}_{-0.037}$  & $0.283^{+0.041}_{-0.044}$  & $552$  		& $0.724^{+0.047}_{-0.040}$  & $0.261^{+0.049}_{-0.051}$  & $-1.59^{+0.55}_{-0.21}$  & $79$  \\
        \hline \vspace{0.1cm}
{\it Baryonic Effects} &&&& \vspace{0.15cm} \\
\bigstrutdown Baryonic $P(k)$ model & $0.798^{+0.026}_{-0.028}$  & $0.268^{+0.053}_{-0.040}$  & $786$  & $0.794^{+0.06}_{-0.032}$  & $0.319^{+0.048}_{-0.074}$  & $-0.77^{+0.30}_{-0.37}$  & $100$  \\
        \hline \vspace{0.1cm}
{\it Other Lensing Data} &&&& \vspace{0.15cm} \\
\bigstrutdown {\bf DES SV } & 					$0.769^{+0.062}_{-0.072}$  & $0.268^{+0.057}_{-0.049}$  & $256$  		& $0.758^{+0.068}_{-0.109}$  & $0.264^{+0.068}_{-0.040}$  & $-1.19^{+0.37}_{-0.60}$  & $27$  \\
        \bigstrutdown KiDS-450 & 					$0.754^{+0.029}_{-0.037}$  & $0.261^{+0.087}_{-0.050}$  & $424$           & $0.759^{+0.044}_{-0.042}$  & $0.326^{+0.061}_{-0.078}$  & $-0.64^{+0.24}_{-0.38}$  & $78$  \\
        \bigstrutdown Planck TT + lowP & 				$0.841^{+0.027}_{-0.025}$  & $0.334^{+0.037}_{-0.026}$  & $1092$  	& $0.810^{+0.029}_{-0.036}$  & $0.222^{+0.069}_{-0.024}$  & $-1.47^{+0.31}_{-0.22}$  & $160$  \\
        \bigstrutdown Planck (+lensing) + BAO + JLA & 	$0.815^{+0.015}_{-0.013}$  & $0.306^{+0.007}_{-0.007}$  & $10607$  	& $0.816^{+0.014}_{-0.013}$  & $0.303^{+0.010}_{-0.008}$  & $-1.020^{+0.049}_{-0.046}$  & $1506$  \\
\end{tabular}
\end{ruledtabular}
\end{table*}

\begin{figure*}
\begin{center}
\includegraphics[width=\textwidth]{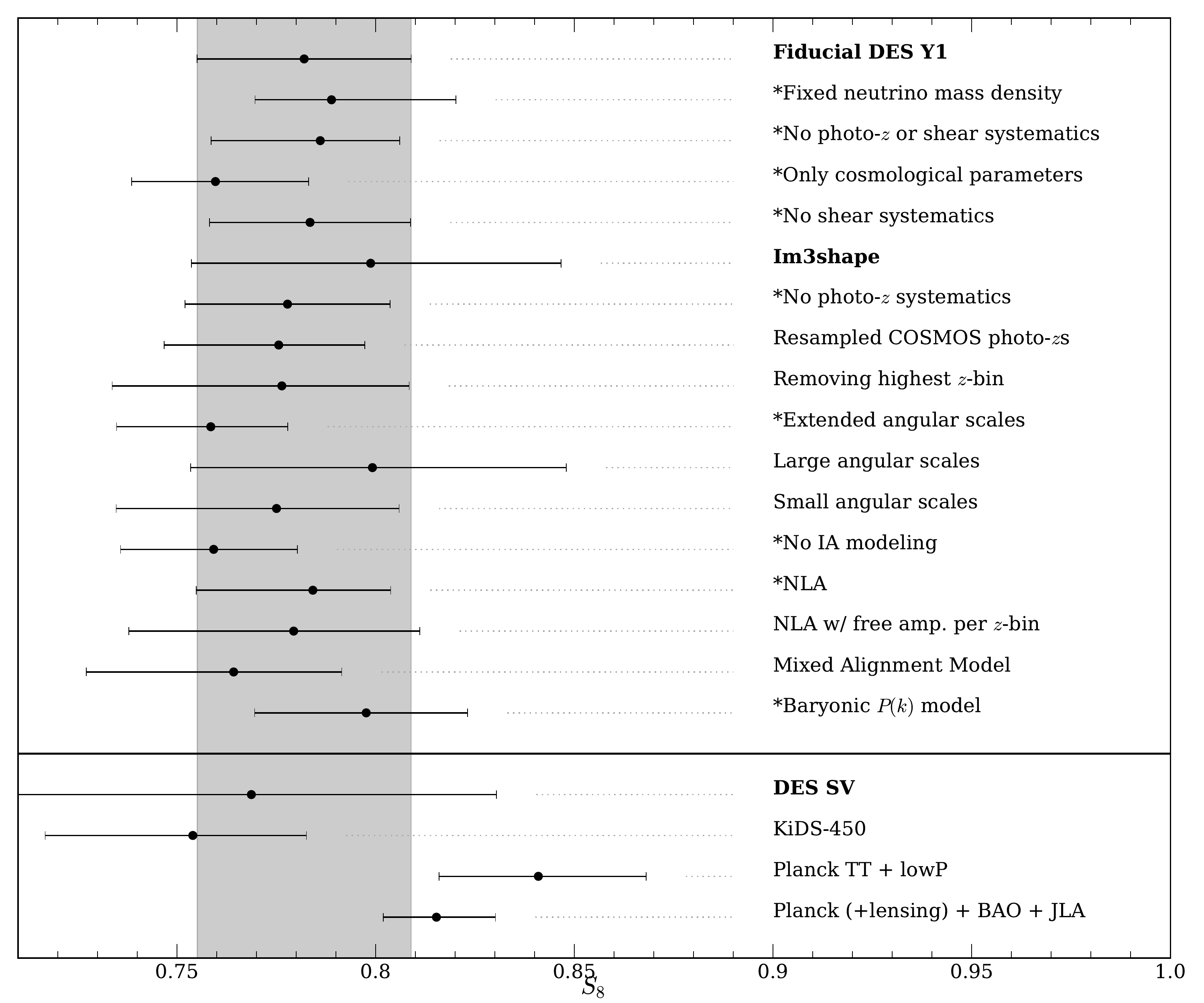}
\end{center}
\caption[]{Summary of constraints on the 1D peak value of $S_8\equiv \sig(\om/0.3)^{0.5}$ in $\Lambda$CDM. The 68\% CL are shown as horizontal bars. We distinguish variations on the fiducial setup that are not necessarily required to give consistent results (e.g., by neglecting astrophysical systematics) by an asterisk. The numerical parameter values are listed in Table \ref{table:s8}. Especially for external data, it is important to remember that we vary $\Omega_{\nu}h^2$ in our fiducial analysis, and thus all results we compare to, and so the central values and uncertainties of parameters may not follow intuition from previous results.
\label{fig:fids8}}
\end{figure*}

\begin{figure}
\begin{center}
\includegraphics[width=0.95\columnwidth]{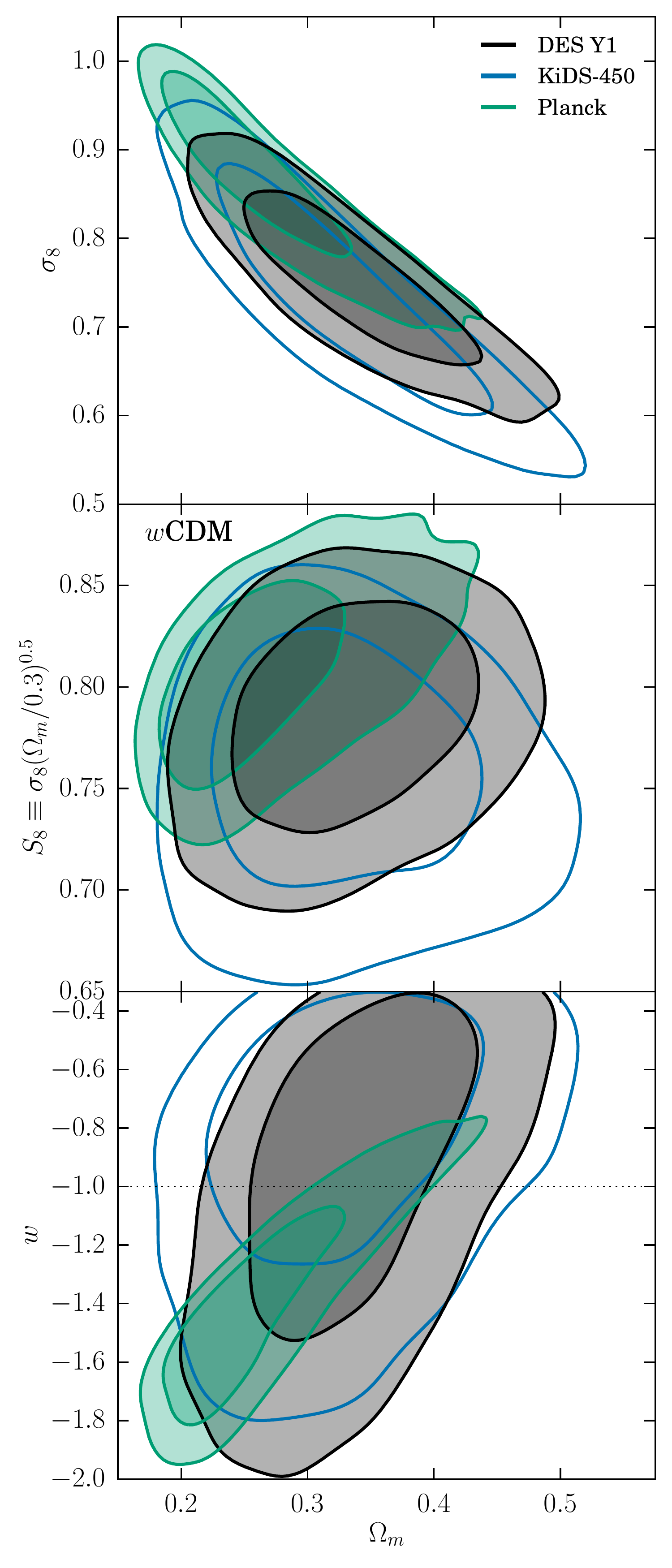}
\end{center}
\caption[]{Fiducial constraints on the clustering amplitude $\sigma_8$, $S_8$, and $w$ with the matter density $\Omega_m$ in $w$CDM. The fiducial DES Y1 cosmic shear constraints are shown by the gray filled contours, with Planck CMB constraints given by the filled green contours, and cosmic shear constraints from KiDS-450 by unfilled blue contours. In all cases, 68\% and 95\% confidence levels are shown. External data have been reanalyzed in our model space, as described in Sec. \ref{sec:fidother}. A dotted line at $w=-1$ indicates the $\Lambda$CDM value. \label{fig:fidsig8wcdm}}
\end{figure}

\subsection{Fiducial $\Lambda$CDM Results}\label{sec:lcdm}

We marginalize over a total of 6 cosmological parameters in the fiducial $\Lambda$CDM model, including a free neutrino mass density, and 10 systematic or astrophysical parameters. These are listed in Table \ref{table:params}.  Our fiducial $\Lambda$CDM constraints in the  $\sigma_8$ -- $\Omega_m$ and $S_8$ -- $\Omega_m$ planes are shown in Fig. \ref{fig:fidsig8lcdm}. The DES Y1 cosmic shear constraints are shown by the gray filled contours, while the previous best real-space cosmic shear constraints from the KiDS survey are shown in blue, and Planck constraints from the CMB in filled green, for comparison. Both 68\% and 95\% confidence levels are shown. For consistency, previous constraints have been reanalyzed in the parameter space used in this work (see Table \ref{table:params}), including varying $\Omega_{\nu}h^2$, which is discussed further in Sec. \ref{sec:fidother}. We show the impact of fixing $\Omega_{\nu}h^2$ in our fiducial $\Lambda$CDM analysis in Fig. \ref{fig:nosys}.

The 1D peak value of $S_8$ and $\Omega_m$ are listed in Table \ref{table:s8} along with 68\% CL about the peak for both our fiducial $\Lambda$CDM results and a large variety of consistency checks and constraints from external data. The FoM in Eq. (\ref{eq:fom}) for the $S_8$ -- $\Omega_m$ plane is also listed. These constraints on $S_8$ are visually summarized in Fig. \ref{fig:fids8}. In both, we distinguish variations on the fiducial setup that are not necessarily expected to give consistent results (e.g., by neglecting astrophysical systematics) by an asterisk. We find a 3.5\% fractional uncertainty on $S_8 = 0.782^{+0.027}_{-0.027}$ at 68\% CL, which is a factor of 2.5 improvement over the constraining power of our Science Verification results. We see similar improvements in the constraint on $\Omega_m$, which is more representative of the gain in the direction of degeneracy. We expect further significant improvements with subsequent years of DES data, which will more than triple the sky coverage of our shape catalogs and double the effective integrated exposure time per galaxy.

We have plotted the best-fit $\Lambda$CDM prediction in Figs. \ref{fig:ximcal} \& \ref{fig:xii3} for both shape catalogs. We find a total $\chi^2$ for the fiducial \metacal\ measurement of 227 with 211 degrees of freedom (227 data points and 16 free parameters\footnote{It is worth noting that half of the fitted parameters are tightly constrained by priors, such that fixing them does not significantly alter the final constraint, and thus the number of degrees of freedom may be underestimated.}) for the $\Lambda$CDM best-fit model.
The probability $p$ of getting a higher $\chi^2$ value can be derived assuming our data vector is drawn from a multi-variate Gaussian likelihood around the best-fit theory vector and that our covariance matrix is precisely and fully characterized. We find for our fiducial result $p=0.21$. 

\subsection{Fiducial $w$CDM Results}

We marginalize over a total of 7 cosmological parameters in the fiducial $w$CDM model, including a free neutrino mass density, and 10 systematic or astrophysical parameters. These are listed in Table \ref{table:params}.  Our fiducial $w$CDM constraints are shown in Fig. \ref{fig:fidsig8wcdm}. We find a 4.8\% fractional uncertainty on $S_8 = 0.777^{+0.036}_{-0.038}$ at 68\% CL, which is more than a factor of 2 improvement over the constraining power of our Science Verification results. We find a dark energy equation-of-state $w=-0.95^{+0.33}_{-0.39}$ using DES cosmic shear alone.

We find an equally good fit to the $w$CDM model as for $\Lambda$CDM, with best-fit $\chi^2$ of 228 for the 227 data points in the non-shaded region for the \metacal\ measurement. We can further compare the relative Bayesian evidence for each model via the Bayes factor. The Bayesian evidence, or probability of observing a dataset $\bm{D}$ given a model $M$ with parameters $\bm{p}$, is
\begin{equation}
P(\bm{D} | M) = \int d^N p P(\bm{D} | \bm{p}, M) P(\bm{ p }| M).
\end{equation}
The Bayes factor comparing the evidence for the $w$CDM and $\Lambda$CDM models is then
\begin{equation}
K=\frac{P(\bm{D}|w\textrm{CDM})}{P(\bm{D}|\Lambda\textrm{CDM})}.
\end{equation}
The interpretation of the Bayes factor can be characterized in multiple ways \cite{jeffreys61,doi:10.1080/01621459.1995.10476572}.
We find $\log(K) = -1.4$, which indicates no preference for a model which allows $w\ne -1$.

\subsection{Comparison to External Measurements}\label{sec:fidother}

In order to place our results in the context of both other cosmic shear constraints and those from complementary probes of the Universe, we recompute the posterior of external results in our fiducial parameter spaces for both $\Lambda$CDM and $w$CDM with a varying neutrino mass density. For recent real-space cosmic shear data, including DES SV \cite{des2016} and KiDS-450 \cite{kids450}, we use the original $\xi_{\pm}$, $n^i(z)$, covariance matrix, and priors on shear or photo-$z$ systematics from these works that inform limitations to the data sets. We have not corrected a bug in the published angular values $\theta$ in the original KiDS-450 measurements that was reported in Footnote 1 of \cite{2017arXiv170706627J}, instead using the published $\xi_{\pm}$. We explore this change and other updates to the KiDS-450 cosmic shear analysis and how they impact comparison with the DES Y1 results shown here in \cite{troxel2018}. We enforce our fiducial model choices for intrinsic alignment and baryon scale cuts, in addition to the recommended scale cuts in the original analysis. 

These modifications to the original analyses change the precise appearance of contours and quoted parameter values relative to the original works. In particular, the choice of priors (e.g., $\Omega_m$ vs. $\Omega_m h^2$) and their ranges can strongly impact the behavior of cosmic shear constraints along the degeneracy direction between $\Omega_m$ and $\sigma_8$. For the KiDS-450 analysis we show here, the contour is slightly better constrained asymmetrically relative to the original results in \cite{kids450}, which we attribute primarily to changes in parameterization and prior choices. We also sample over an effective $\Delta z^i$ parameter as used in our fiducial analysis and assume no correlation between the systematic photo-$z$ errors in each tomographic bin, which may also result in slightly better constrained parameters, though this effect should be minor compared to that of the prior changes.

We also recompute the posterior of other external data sets of complementary probes in our parameter spaces, including varying neutrino mass. These include:
\begin{itemize}
\item Constraints on the angular diameter distance from baryon acoustic oscillation (BAO) measurements -- the 6dF Galaxy Survey~\citep{Beutler:2011hx}, the SDSS Data Release 7 Main Galaxy Sample~\citep{Ross:2014qpa}, and the BOSS Data Release 12~\citep{Alam:2016hwk}. The BAO distances are measured relative to a true physical BAO scale $r_{\mathrm{d}}$, which leads to a factor that depends on the cosmological model, which must be calculated at each likelihood step (see \cite{Alam:2016hwk}).
\item Luminosity distance measurements -- the Joint Lightcurve Analysis of Type Ia Supernovae (SNe) from \citet{Betoule:2014frx}.
\item Cosmic microwave background (CMB) temperature and polarization measurements -- Planck~\cite{Ade:2015xua} (`Planck TT + lowP'), using TT ($\ell = 30$ -- $2508$) and TT+TE+EE+BB ($\ell = 2$ -- $29$), and additionally including lensing when combined with other external data.
\end{itemize}

\begin{figure}
\begin{center}
\includegraphics[width=\columnwidth]{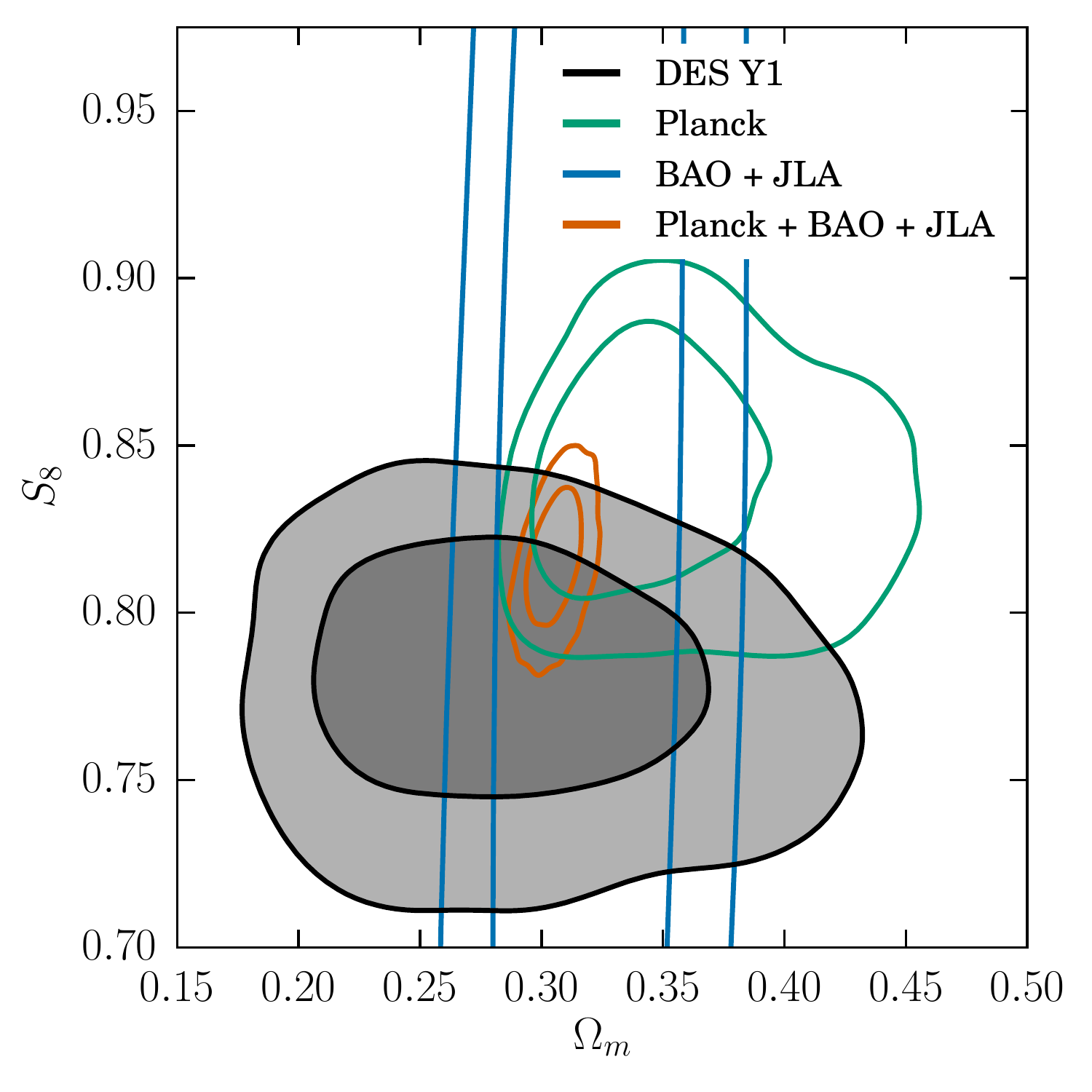}
\end{center}
\caption[]{A comparison of the DES Y1 constraints on $S_8$ and the matter density $\Omega_m$ in $\Lambda$CDM with external data sets excluding cosmic shear. The fiducial DES Y1 cosmic shear constraints are shown by the gray filled contours, with Planck CMB constraints given by the green contours, the combination of BAO and JLA (SNe) constraints in blue, and the combination Planck + BAO + JLA in orange. In all cases, 68\% and 95\% confidence levels are shown. 
\label{fig:ext}
}
\end{figure}

We show a comparison of our fiducial $\Lambda$CDM results to KiDS-450 and Planck in Fig. \ref{fig:fidsig8lcdm}, excluding CMB lensing to attempt to emphasize any differences between our low-$z$ measurements and Planck CMB predictions from high-$z$. We find our cosmic shear constraint is consistent with that of KiDS-450. There has been significant discussion in the literature regarding consistency of cosmic shear constraints with the CMB. 
We find no evidence of inconsistency between the DES Y1 cosmic shear results and constraints from Planck CMB data, with the cosmic shear contours overlapping constraints from Planck at the 1$\sigma$ level.
It is worth noting that the inclusion of CMB lensing, not included in Fig. \ref{fig:fidsig8lcdm}, lowers the Planck estimate of $S_8$ and $\Omega_m$ (Table 4 of \cite{Ade:2015xua}), thus further reducing any minor differences with our results in this plane.
We leave a detailed discussion and interpretation of consistency of our data with external probes to Dark Energy Survey et al. \cite{keypaper}, where significantly tighter cosmological constraints are presented when combining galaxy clustering and galaxy-galaxy lensing with our cosmic shear results. We also compare our fiducial results in $w$CDM to results from KiDS-450 and Planck in Fig. \ref{fig:fidsig8wcdm}, where our conclusion about consistency is unchanged. Finally, we show in Fig. \ref{fig:fids8} and Table \ref{table:s8} a summary of comparisons in $\Lambda$CDM and $w$CDM of our fiducial constraints with DES SV, KiDS-450, Planck, and Planck + BAO + JLA. We find our 1D constraints in $S_8$ and $\Omega_m$ agree well with the combination Planck + BAO + JLA. We directly compare these external data sets to our fiducial $\Lambda$CDM constraints in Fig. \ref{fig:ext}.

\section{Robustness to Modeling and Data Choices}\label{sec:modelchoice}

Validating the robustness of the cosmic shear signal to various potential residual systematic effects is technically challenging. The two primary measurement uncertainties in the cosmic shear signal, multiplicative shear and photometric redshift biases, are each difficult to constrain directly. The DES Y1 shape catalogs have undergone extensive null testing, both during catalog development and at the level of specific probes or measurements, including those demonstrated below for cosmic shear. Primary catalog-level tests are discussed in detail in \cite{shearcat}, which includes tomographic constraints on the shear B-mode signal, while other tests of the shape and \photoz\ catalogs have been carried out in \cite{wthetapaper,photoz,redmagicpz,xcorrtechnique,xcorr,gglpaper}. \cite{shearcat} do identify both a significant PSF model residual that we are able to model and an unidentified source of mean shear, which we correct. These effects and their impact on the cosmic shear signal are discussed in Appendix \ref{sec:psf}. Further successful null tests of the cosmic shear signal are discussed in Appendix \ref{sec:syssurvey}.

In designing a fiducial analysis, many choices are made in data selection and model design that can have a significant impact on any conclusions drawn from the analysis. 
Our choices are informed by an extensive battery of null tests, which were performed while blind to the consequences on cosmological parameters. We also performed many robustness tests blinded, where relative deviations in constraints are examined without knowledge of the absolute value of any parameter. Some of these variations and tests are summarized in Fig. \ref{fig:fids8} and Table \ref{table:s8}, and we reproduce many of them here unblinded.

\subsection{Impact of Nuisance Parameter Marginalization}

While we marginalize over a large number of non-cosmological parameters in the fiducial analysis, nearly twice as many as cosmological parameters, we find that our constraining power is not strongly impacted by marginalizing over shear and \photoz\ systematic parameters in the $S_8$ -- $\Omega_m$ plane. In terms of $S_8$ alone, marginalizing over shear and \photoz\ systematic parameters degrades our constraint by about 10\%. Our constraints are more significantly degraded by marginalizing over a model for intrinsic alignment (IA), which is one reason that combining cosmic shear with the other large-scale structure probes in \cite{keypaper} is so powerful. We also see a significant bias when ignoring intrinsic alignment. This is demonstrated in Fig. \ref{fig:nosys}, where we compare the fiducial analysis (DES Y1 - filled gray contours) in $\Lambda$CDM to the cases where we sample only over cosmological parameters (blue contours) and both cosmological and astrophysical parameters (green contours). Further discussion of the impact of intrinsic alignment modeling can be found in Sec. \ref{sec:sysia}. We compare the priors and posteriors for the non-cosmological parameters in Table \ref{table:syspost} and Appendix \ref{sec:fullresults}, and find that we do not gain a significant amount of information in constraining shear and \photoz\ bias parameters in the fiducial analysis, while providing significant constraints on the intrinsic alignment model.

\begin{table}
\caption{A comparison of the priors and posteriors of non-cosmological parameters in the fiducial analysis.}
\label{table:syspost}
\begin{center}
\begin{ruledtabular}
\begin{tabular}{ lccccccc }
  Parameter 		& Prior & Posterior  \\
  \hline
 \multicolumn{3}{l}{ Systematic} \\
  $m^1\times10^2$ 		& $1.2^{+2.3}_{-2.3}$ 	& $1.3^{+1.8}_{-1.8}$ \\
  $m^2\times10^2$ 		& $1.2^{+2.3}_{-2.3}$ 	& $1.1^{+2.1}_{-2.0}$ \\
  $m^3\times10^2$ 		& $1.2^{+2.3}_{-2.3}$ 	& $0.4^{+1.9}_{-1.8}$ \\
  $m^4\times10^2$ 		& $1.2^{+2.3}_{-2.3}$ 	& $1.4^{+2.1}_{-1.5}$ \\
  $\Delta z^1\times10^2$ 	& $0.1^{+1.6}_{-1.6} $ 	& $0.1^{+1.3}_{-1.3}$ \\
  $\Delta z^2\times10^2$ 	&$-1.9^{+1.3}_{-1.3} $ 	&$-2.0^{+1.1}_{-0.9}$ \\
  $\Delta z^3\times10^2$ 	& $0.9^{+1.1}_{-1.1} $ 	& $0.9^{+0.8}_{-0.9}$ \\
  $\Delta z^4\times10^2$ 	&$-1.8^{+2.2}_{-2.2} $ 	&$-1.6^{+1.6}_{-2.0}$ \\
  \hline
\multicolumn{3}{l}{ Astrophysical} \\
  $A$  	&  $0.0^{+5.0}_{-5.0}$ & $1.0^{+0.4}_{-0.7}$ \\
  $\eta$  	&  $0.0^{+5.0}_{-5.0}$ & $2.8^{+1.7}_{-2.0}$ \\
\end{tabular}
\end{ruledtabular}
\end{center}
\end{table}

\subsection{Shear Pipeline Comparison}\label{sec:sysshapecode}

In the DES SV shape catalog paper \cite{jarvis2016}, we made explicit comparisons between the two shape measurement methods \ngmix\ \& \imshape\ in simulations and at the two-point estimator level. This was informed by simultaneous measurements on simulated data by both shape measurement methods, which gave us an estimate of the relative selection bias, and resulted in choices that made the catalogs more similar. Residual differences ultimately provided the basis for the final prior for $m$. This is even more complicated to do with \metacal\ and \imshape\ due to the very different ways each are calibrated. Instead, \cite{shearcat} perform detailed independent, \textit{ab initio} estimations of uncertainty in $m$ for each pipeline. 

The paper \cite{shearcat} also demonstrates that there is no significant B-mode signal in our shear data. The (null) $B$-mode measurement is performed in harmonic space, where the E-mode and B-mode signals can be naturally separated. We note that this null result does not formally guarantee that the real-space correlation functions used in this work are B-mode-free. For example, B-mode power above $\ell=1000$ (the maximum $\ell$ used in \cite{shearcat}) could in principle contribute to the real-space statistics used here. However, \cite{kitching17} (Fig. 2) demonstrate that although $\xi_+$ has sensitivity up to $\ell\sim 10^4$ for a minimum angular scale of $1$ arcmin (the minimum scale used for $\xi_+$ in this work is $3.6$ arcmin), the contribution from $\ell>1000$ is small, so B-mode power at $\ell>1000$ would have to be extreme to significantly affect our measurements.

\begin{figure}
\begin{center}
\includegraphics[width=\columnwidth]{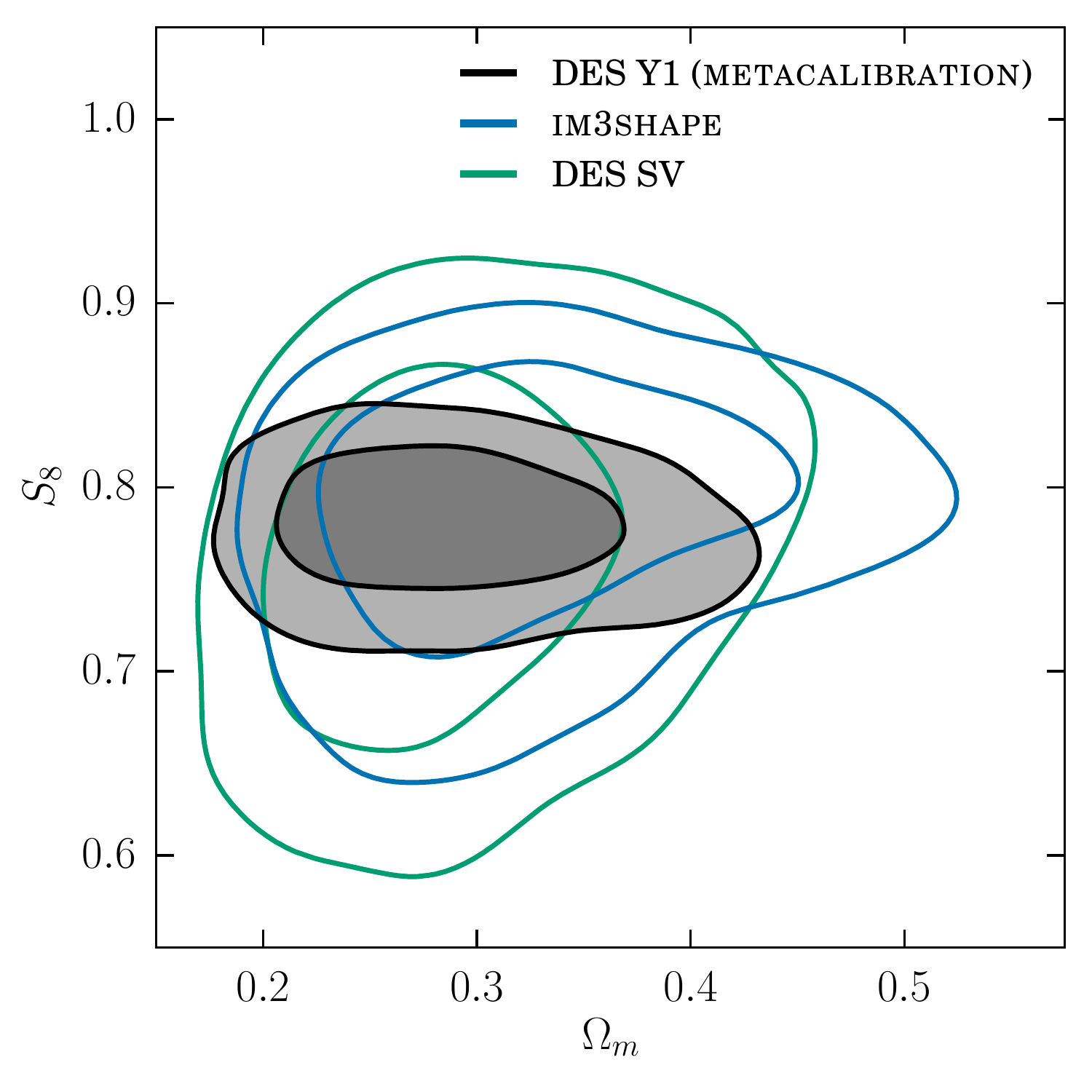}
\end{center}
\caption[]{A comparison of $\Lambda$CDM constraints in the $S_8$ -- $\Omega_m$ plane from the two shape measurement pipelines, \metacal\ (gray filled contours) and \imshape\ (blue contours). This is a strong test of robustness to assumptions and differences in measurement and calibration methodology.  Each pipeline utilizes very different and independent methods of shape measurement and shear calibration. We also compare the DES SV results (from \ngmix) in green. Both 68\% and 95\% confidence levels are shown.
\label{fig:shapecomp}}
\end{figure}

To confirm that the two shear measurements from \metacal\ and \imshape\ agree, we have relied on a quantitative comparison of their agreement only at the level of cosmological parameter constraints, where the differing selection of objects in each catalog is naturally accounted for. This comparison was performed only once the two shape catalogs were finalized based on results of tests in \cite{shearcat}, and is shown in Fig. \ref{fig:shapecomp} for $\Lambda$CDM. The resulting contours in the $S_8$ -- $\Omega_m$ plane are entirely consistent, though the mean of the \imshape\ constraint in $S_8$ is shifted to slightly higher values. The weaker constraint for \imshape\ is due primarily to using only the $r$ band for shape measurement, relative to $riz$ bands for \metacal, and additional necessary catalog selections to remove objects that cannot be calibrated accurately due to limitations of galaxy morphology in the input COSMOS catalog. These contribute to a significantly smaller $n_{\mathrm{eff}}$ for \imshape. This agreement, reached through two very different and independent shape measurement and calibration strategies, is thus a very strong test of robustness to shape measurement errors in the final cosmological constraints.

\subsection{Photometric Redshift Comparison}\label{sec:sysphotoz}

\begin{figure}
\begin{center}
\includegraphics[width=\columnwidth]{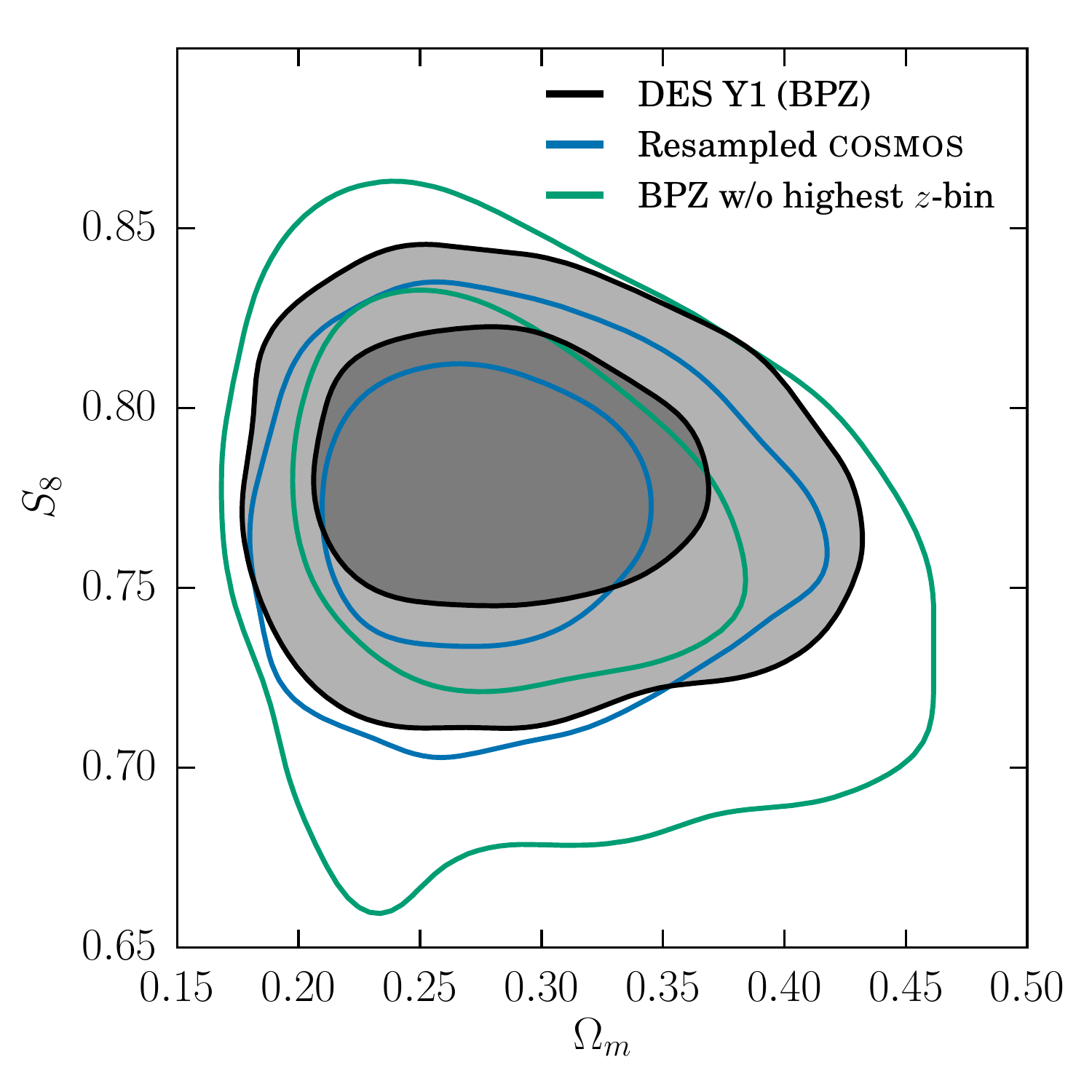}
\end{center}
\caption[]{A comparison of $\Lambda$CDM constraints in the $S_8$ -- $\Omega_m$ plane for different \photoz\ choices. The gray filled contours show the fiducial analysis using the BPZ redshift distribution shown in Fig. \ref{fig:nofz} with the offset priors listed in Table \ref{table:params}. The green contours are the result of the same analysis, but removing the fourth tomographic bin. The blue contours use the COSMOS $n(z)$ from which part of the BPZ prior is derived. By design, the mean redshift of each tomographic bin agrees between the resampled COSMOS $n(z)$ and the BPZ redshift distribution used in the fiducial analysis, but the shapes of the $n(z)$ are significantly different in some tomographic bins, providing a test of whether parameterizing the \photoz\ bias as only a shift in the mean redshift is sufficient. Both 68\% and 95\% confidence levels are shown. \label{fig:pzbias}}
\end{figure}

As discussed in Sec. \ref{sec:photoz}, we rely on a combination of 1) comparisons to redshift distributions of resampled COSMOS objects and 2) constraints due to clustering cross-correlations between source galaxies and \redmagic\ galaxies with very good photometric redshifts. We parameterize corrections to the $n(z)$ as a shift in the mean redshift of the distribution of galaxies. As an independent test of whether shifting the mean of the redshift
distribution captures the full effect of photo-$z$ bias uncertainty,
we show cosmological constraints directly using the resampled COSMOS $n^i(z)$ redshift distributions measured from COSMOS (see Fig. \ref{fig:nofz}). The resulting constraints are shown in Fig. \ref{fig:pzbias}, illustrating that differences in the shape of the redshift distribution are sub-dominant for cosmic shear when matching the mean at the current statistical precision (see also \cite{bonnett2016}). The independent constraint from clustering cross-correlations is unavailable in the fourth redshift bin, because the \redmagic\ sample ends at redshift $z=0.9$. Thus, we also tested removing the fourth bin from our analysis and confirmed in Fig. \ref{fig:pzbias} that it does not produce a significant shift in the inferred cosmology. Both tests were performed before unblinding. We also performed a non-tomographic analysis, and find consistent results, though the cosmological constraining power is severely degraded without tomographic information, which is necessary to constrain the IA model with cosmic shear alone. 

\subsection{Scale Selection}\label{sec:scales}

\begin{figure}
\begin{center}
\includegraphics[width=\columnwidth]{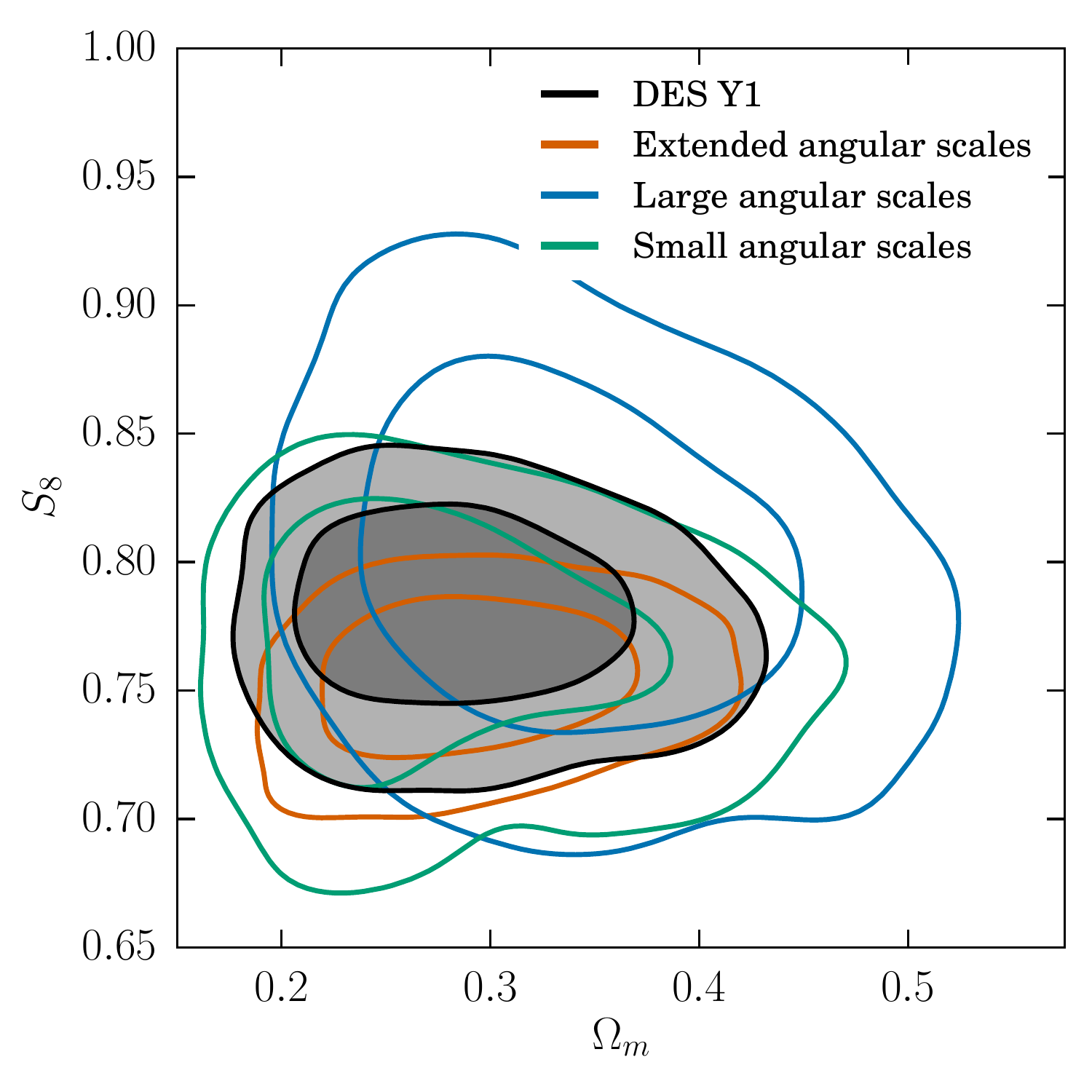}
\end{center}
\caption[]{A comparison of $\Lambda$CDM constraints in the $S_8$ -- $\Omega_m$ plane for different data vector choices. The filled gray contours show the fiducial analysis, while the orange contours small scales outside the fiducial selection, the blue contours use fiducial angular scales for $\xi_{+}$ ($\xi_{-}$) with $\theta>20'$ ($\theta>150'$), and the green contours use fiducial scales for $\xi_{+}$ ($\xi_{-}$) with $\theta<20'$ ($\theta<150'$). The orange contours should not necessarily agree with the fiducial case due to the impact of baryonic effects, while we find consistent results using subsets of our fiducial angular scale range. Both 68\% and 95\% confidence levels are shown.\label{fig:datavectorchoice}}
\end{figure}

We remove any scales from the $\xi_{\pm}$ data vector that would have a fractional contribution from baryonic interactions exceeding 2\%. We use results from the OWLS `AGN' simulation, discussed in Sec. \ref{sec:systheory}, to determine this limit. This removes a significant number of data points on small scales, particularly in $\xi_{-}$ where the impact of baryonic interactions is larger. Similarly, we correct a significant bias due to residual mean shear (discussed further in Appendix \ref{sec:psf}) in the signal that is partly due to PSF modeling errors. This impacts the signal primarily on the largest scales. To verify that our scale cuts are robust, we repeat the fiducial analysis by splitting the angular range in two, separately constraining cosmology with the smaller and larger scales. This split is at $\theta=20'$ for $\xi_{+}$ and $\theta=150'$ for $\xi_{-}$. The results of this test are shown in Fig. \ref{fig:datavectorchoice}, with the fiducial analysis shown as the filled gray contours, along with results from the smaller (green contours) and larger (blue contours) scale selection. We find consistent results in all three cases. To demonstrate the potential degradation in constraining power due to our baryon cuts on small scales, we also use the full data vector from $2.5'<\theta<250'$. This is shown as the orange contours in Fig. \ref{fig:datavectorchoice}, which also demonstrates the likely bias due to ignoring baryonic effects on these scales. Finally, we verify (see Fig. \ref{fig:fids8} and Table \ref{table:s8}) that replacing our power spectrum calculation with that from \cite{mead} and marginalizing additionally over the included two-parameter baryon feedback model in $\Lambda$CDM does not significantly change our inferred cosmology. Since the \cosmosis\ interface to the model in \cite{mead} does not yet incorporate more recent changes that account more accurately for massive neutrinos, the neutrino mass density cannot be properly marginalized over and this result should be compared to the fixed neutrino mass density constraint.

\subsection{Intrinsic Alignment Modeling}\label{sec:sysia}

\begin{figure}
\begin{center}
\includegraphics[width=\columnwidth]{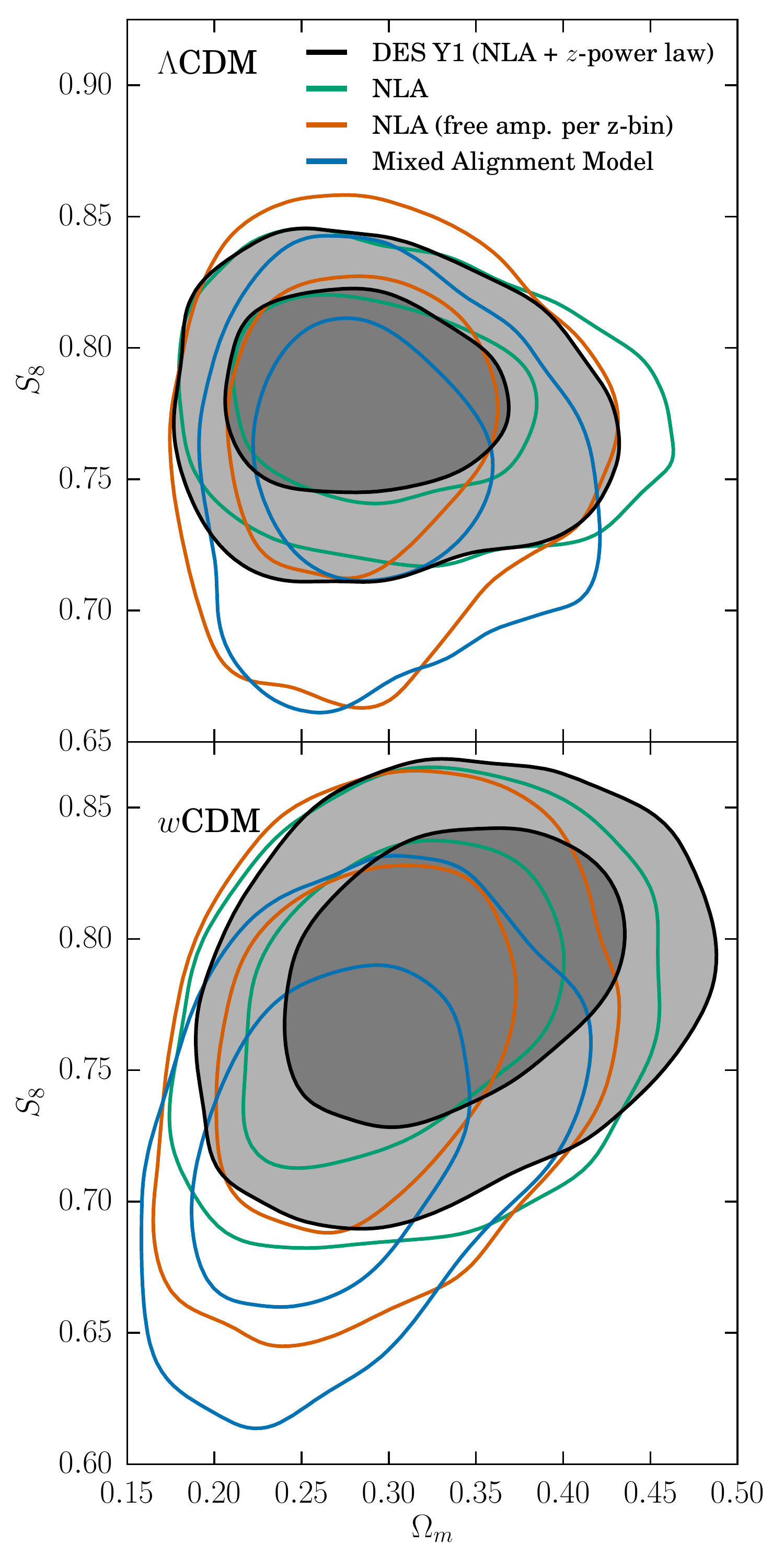}
\end{center}
\caption[]{A comparison of the impact of different intrinsic alignment (IA) models on $\Lambda$CDM and $w$CDM constraints in the $S_8$ -- $\Omega_m$ plane. The fiducial model (NLA + $z$-power law -- gray contours), is compared to the single-parameter NLA model (green contours), the NLA model with a free amplitude in each tomographic bin (orange contours), and the mixed alignment model (blue contours). There is no significant difference in inferred cosmology between the first three models, which are well-tested and have been implemented in the literature before. The mixed alignment model, which includes alignment due to tidal torquing (or other nonlinear contributions), does cause a non-negligible shift in inferred parameters in $w$CDM, which is discussed further in Sec. \ref{sec:sysia}. Both 68\% and 95\% confidence levels are shown.
\label{fig:IAimpact}}
\end{figure}

Unlike for astrophysical contaminants like the impact of baryonic effects, intrinsic alignment (IA) impacts the measured signal at all scales. In addition to the fiducial intrinsic alignment model, we also consider several variants to test the robustness of our results with respect to the choice of intrinsic alignment model over which we marginalize. These include: 1) fixing the power-law redshift scaling of the fiducial model to have $\eta=0$, leaving a single-parameter ($A$) model; 2) removing the power-law dependence of redshift evolution to marginalize over four free amplitudes in each redshift bin ($A^i$); 3) allowing for both tidal alignment and tidal torquing alignment amplitudes (`mixed' model, \cite{bmt17}). Note that the mixed model includes IA $B$-mode contributions, which are incorporated through $P_{\kappa} \rightarrow P_E \pm P_B$ in Eq.~\ref{eq:Pkappa}. This model also has mild dependence on the source galaxy bias, which we assume to be 1. Fig. \ref{fig:IAimpact} shows constraints in $\Lambda$CDM and $w$CDM for the fiducial model (NLA + $z$-power law -- gray contours), compared to the single-parameter NLA model (green contours), the NLA model with a free amplitude in each tomographic bin (orange contours), and the mixed alignment model (blue contours).  
There is no significant difference in inferred cosmology between these models in $\Lambda$CDM. In $w$CDM, the mixed alignment model, which includes alignment due to nonlinear effects in the tidal field, including tidal torquing, does cause a clearly non-negligible shift in inferred parameters. 

We caution against concluding that the fiducial results presented here are biased due to the shift in cosmology observed in Fig. \ref{fig:IAimpact} when using the mixed alignment model, however, because we have seen similar trends to lower $S_8$ and $\Omega_m$ in less constraining data sets when marginalizing over too flexible an intrinsic alignment model. For example, the DES SV (and to a lesser degree \imshape) result in Fig. \ref{fig:shapecomp} (see also IA discussion in \cite{des2016}), shows a similar trend toward this area of parameter space with even the fiducial IA model in this work, which disappears with our more constraining DES Y1 data. We further see much less significant an impact on cosmology in the full combined clustering and weak lensing analysis when injecting a tidal torque signal of greater amplitude than we find here into a pure lensing signal \cite{methodpaper}. It is also worth noting that there is no significant difference in $\chi^2$ or Bayesian evidence whether we include or not the tidal torque contribution of the mixed alignment model. We thus conclude that while this is an interesting result, it requires further exploration that we defer to a future work. Nevertheless, this result highlights the importance of considering the impact of IA models beyond the tidal alignment (linear) paradigm in future cosmic shear studies, and it may indicate a real bias in cosmic shear at our statistical precision when using the fiducial tidal alignment model. A more conclusive answer for this question will require more constraining data, which we are analyzing with DES Y3+ results, or better external priors on the amplitude of the tidal torquing component (and orientation).

\begin{figure}
\begin{center}
\includegraphics[width=\columnwidth]{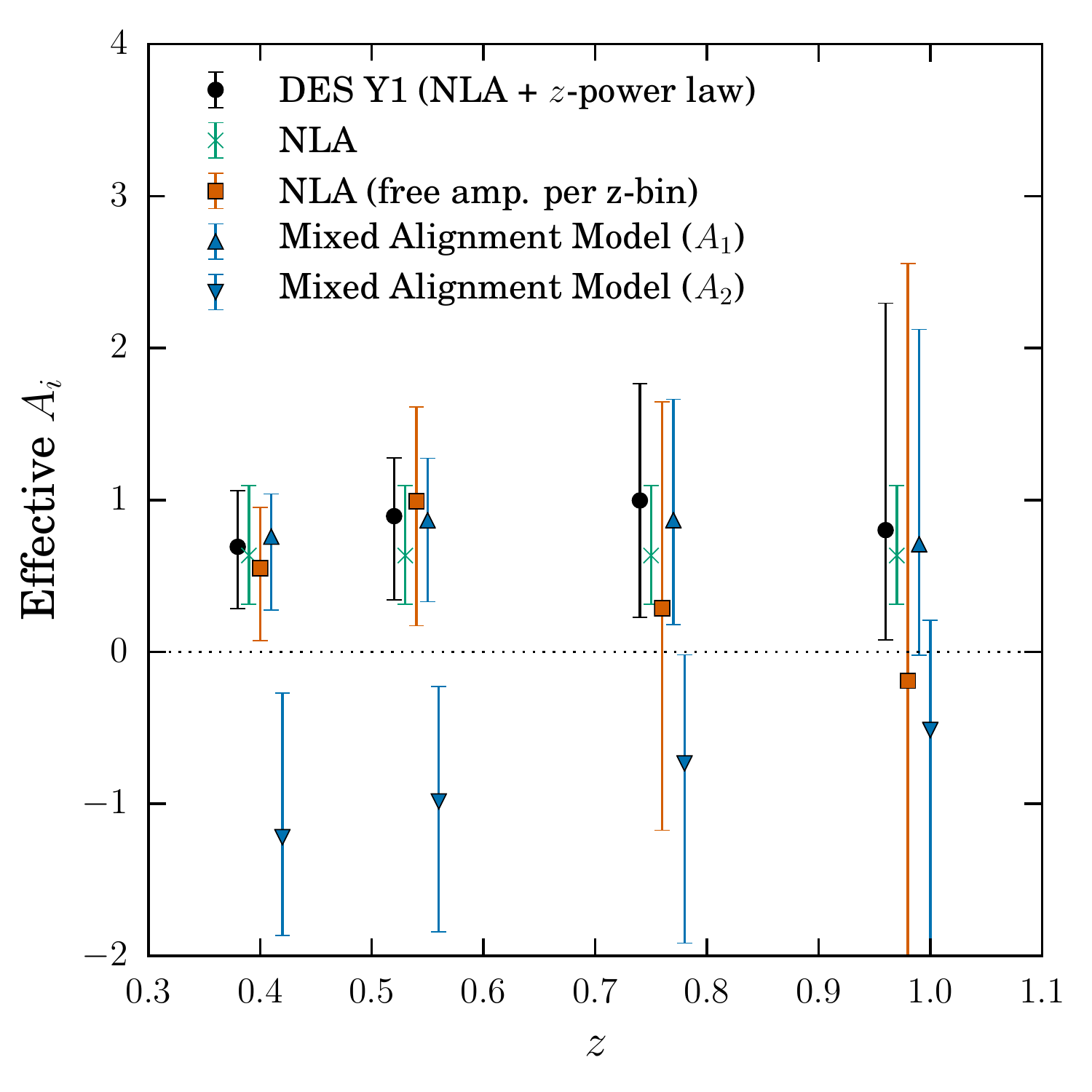}
\end{center}
\caption[]{The constraint on the intrinsic alignment amplitude $A$ as a function of redshift in the four models considered. For all models, we show $A_1$, the amplitude of the tidal alignment (TA) model, while for the mixed alignment model, we also show the constraint on $A_2$, the amplitude of the tidal torquing (TT) component of the model. We find good agreement between the redshift evolution of the tidal alignment amplitude in the four models. 
\label{fig:IAimpact2}
}
\end{figure}

Given the constraining power of the DES Y1 analysis, it is clear that we can learn not just about cosmology, but also interesting astrophysical effects like IA. In Fig. \ref{fig:IAimpact2} we compare the recovered value of $A$, the amplitude of the tidal alignment (TA) model as a function of redshift in the four models considered in this analysis. For the mixed alignment model, we also show the constraint on $A_2$, the amplitude of the tidal torquing (TT) component of the model. Note that subscripts are used with the amplitudes in the mixed alignment case and that $A_1$ corresponds to the fiducial $A$ parameter. We find good agreement in the TA amplitude between all four models, including the mixed alignment case, where the contributions from TT terms appear largely independent from the TA amplitude. For the fiducial IA model and the mixed alignment model, which have a smooth functional form with redshift, we derive the amplitude at the mean redshifts of each redshift bin and report this value and its uncertainty. This analysis provides a significant improvement in IA constraining power compared to previous analyses, with detection of nonzero $A=1.0$ at the 89\% CL when allowing a power-law redshift scaling, which is comparable to that when assuming a fixed $\eta$. The fiducial power-law $\eta=2.8$ is constrained to be non-zero at the 83\% CL. In the mixed model, $A_1=0.9$ is still constrained to be non-zero at 83\% CL with $\eta_1=2.3$ constrained to be positive at the 79\% CL. The tidal torque amplitude $A_2=-0.9$ is nonzero at the 84\% CL, with a negative amplitude, and power-law $\eta_2=0.4$, which is consistent with zero at 1$\sigma$. As discussed in \cite{bmt17}, the sign convention for $A_1$ and $A_2$ is such that positive values correspond to galaxy alignment towards overdense regions and thus a negative GI term.

The measured fiducial IA amplitude is in agreement with our prediction of $A \approx 0.5$ at $z_{\rm piv} = 0.62$, obtained from extrapolating IA amplitude scalings calibrated on galaxies that are significantly more luminous than our lensing sample \cite{methodpaper}. This prediction assumes that only red/elliptical galaxies contribute to the fiducial IA signal and accounts for the approximate red fraction of the source sample.
Our analysis thus provides significant improvement in constraining the IA signal in weak lensing measurements. Moreover, it is the first indication of nonlinear alignment mechanisms, such as tidal torquing, in a general weak lensing sample. Previous weak lensing studies (e.g., \cite{des2016,kids450}) did not account for the potential presence of these higher-order effects, while spectroscopic alignment studies on blue/spiral galaxies have placed comparatively weak constraints on these contributions (e.g., \cite{MBB+11}). Recent hydrodynamic simulations have also examined the expected alignment of both disk and elliptical galaxies (e.g., \cite{clc16,tmd16,hxs17}). These simulations consistently find an overall alignment towards overdense regions, dominated by elliptical galaxies, in agreement with the sign of our measured $A$. However, they disagree on the IA behavior of spiral galaxies (as well as other kinematic properties), with \cite{clc16} finding tangential alignment of the major axis with overdensities, consistent with our tentative measurement of $A_2$, while others find radial alignment (see \cite{tmd16} for a comparison). Improved observational data and hydrodynamic simulations, along with advances in analytic modeling, will clarify this issue. Finally, we note that our inferred redshift evolution of IA, characterized by $\eta$ or the per-bin amplitudes $A_i$, captures both the true underlying redshift evolution as well as the luminosity and galaxy-type dependence of IA, since these properties of the source sample evolve with redshift. Moreover, the IA redshift evolution is partially degenerate with the assumed source redshift distribution, and thus $\eta$ could absorb contributions both from IA and systematics in the source $n(z)$.

\section{Conclusions}\label{sec:conclusion}

We have used 26 million galaxies from Dark Energy Survey (DES Y1) shape catalogs over 1321 deg$^2$ of the southern sky to produce the most significant measurement of cosmic shear in a galaxy survey to date. We constrain cosmological parameters in both the $\Lambda$CDM and $w$CDM models, while also varying the neutrino mass density. We find a 3.5\% fractional uncertainty on $S_8 = 0.782^{+0.027}_{-0.027}$ at 68\% CL, which is a factor of 3 improvement over the constraining power of our SV results \cite{jarvis2016}. In $w$CDM, we find a 4.8\% fractional uncertainty on $S_8 = 0.777^{+0.036}_{-0.038}$ and $w=-0.95^{+0.33}_{-0.39}$. We find no evidence preferring a model allowing $w\ne -1$ using cosmic shear alone, and no constraint beyond our prior on the neutrino mass density. 

Our constraints from cosmic shear agree incredibly well with previous cosmic shear results from KiDS-450 (and DES SV). Despite significant discussion in previous literature, we find no evidence that any of the cosmic shear results from DES or KiDS analyzed here are in disagreement with CMB data from Planck. Significantly tighter cosmological constraints when galaxy clustering and galaxy-galaxy lensing are added to our fiducial cosmic shear measurements are discussed in Dark Energy Survey et al. \cite{keypaper}, and we expect further significant improvements with subsequent years of DES data, which will more than triple the sky coverage of our shape catalogs and double the effective integrated exposure time per galaxy.

We have detailed a suite of rigorous null tests at the data and catalog level \cite{photoz,redmagicpz,xcorrtechnique,xcorr,y1gold,shearcat,wthetapaper,gglpaper}, along with robustness and validation checks of our measurements in this work and \cite{keypaper}, that provide us with confidence in the accuracy of our results. We employ two independent and very different shape measurement and calibration methods to measure cosmic shear, \metacal\ and \imshape, which give consistent results. We further employ two independent methods of calibrating our photometric redshift distributions, including clustering cross-correlations with a photometric sample of galaxies with precise and accurate \photoz s. We account for the intrinsic alignment of galaxies, finding evidence for tidal alignment in our fiducial analysis and for tidal torquing (quadratic) alignment in an extended analysis. We employ two independently developed parameter inference pipelines, \cosmolike\ and \cosmosis, that have been validated against one another, and we validate components of our covariance matrix using a limited number of $N$-body mock catalogs with ray-traced shear, a large number of lognormal simulations, and jackknife measurements in our data to validate shape-noise contributions to the covariance. Finally, we present a series of robustness checks to variations on the fiducial analysis to demonstrate that our analysis choices are secure and well motivated. The statistical power of our DES weak lensing data set, particularly when combined with galaxy clustering and galaxy-galaxy lensing measurements in \cite{keypaper}, now constrains low-redshift clustering as strongly as it has been predicted by previous CMB measurements.

\section*{Acknowledgements}

We thank C.~M.~Hirata, S.~Joudaki, and H.~Hildebrandt for useful conversations in the course of preparing this work. TE is supported by NASA ROSES ATP 16-ATP16-0084 grant and by NASA ROSES 16-ADAP16-0116; TE's research was carried out at the Jet Propulsion Laboratory, California Institute of Technology, under a contract with the National Aeronautics and Space Administration. EK was supported by a Kavli Fellowship at Stanford University.
OF was supported by SFB-Transregio 33 `The Dark Universe' of the Deutsche Forschungsgemeinschaft (DFG) and by the DFG Cluster of Excellence `Origin and Structure of the Universe'.
Support for DG was provided by NASA through Einstein Postdoctoral Fellowship grant
number PF5-160138 awarded by the Chandra X-ray Center, which is
operated by the Smithsonian Astrophysical Observatory for NASA under
contract NAS8-03060. MJ, BJ, and GB are partially supported by the US Department of Energy
grant DE-SC0007901 and funds from the University of Pennsylvania.
SS acknowledges receipt of a Doctoral Training Grant from the UK Science and Technology Facilities Council. 

Funding for the DES Projects has been provided by the U.S. Department of Energy, the U.S. National Science Foundation, the Ministry of Science and Education of Spain, 
the Science and Technology Facilities Council of the United Kingdom, the Higher Education Funding Council for England, the National Center for Supercomputing 
Applications at the University of Illinois at Urbana-Champaign, the Kavli Institute of Cosmological Physics at the University of Chicago, 
the Center for Cosmology and Astro-Particle Physics at the Ohio State University,
the Mitchell Institute for Fundamental Physics and Astronomy at Texas A\&M University, Financiadora de Estudos e Projetos, 
Funda{\c c}{\~a}o Carlos Chagas Filho de Amparo {\`a} Pesquisa do Estado do Rio de Janeiro, Conselho Nacional de Desenvolvimento Cient{\'i}fico e Tecnol{\'o}gico and 
the Minist{\'e}rio da Ci{\^e}ncia, Tecnologia e Inova{\c c}{\~a}o, the Deutsche Forschungsgemeinschaft and the Collaborating Institutions in the Dark Energy Survey. 

The Collaborating Institutions are Argonne National Laboratory, the University of California at Santa Cruz, the University of Cambridge, Centro de Investigaciones Energ{\'e}ticas, 
Medioambientales y Tecnol{\'o}gicas-Madrid, the University of Chicago, University College London, the DES-Brazil Consortium, the University of Edinburgh, 
the Eidgen{\"o}ssische Technische Hochschule (ETH) Z{\"u}rich, 
Fermi National Accelerator Laboratory, the University of Illinois at Urbana-Champaign, the Institut de Ci{\`e}ncies de l'Espai (IEEC/CSIC), 
the Institut de F{\'i}sica d'Altes Energies, Lawrence Berkeley National Laboratory, the Ludwig-Maximilians Universit{\"a}t M{\"u}nchen and the associated Excellence Cluster Universe, 
the University of Michigan, the National Optical Astronomy Observatory, the University of Nottingham, The Ohio State University, the University of Pennsylvania, the University of Portsmouth, 
SLAC National Accelerator Laboratory, Stanford University, the University of Sussex, Texas A\&M University, and the OzDES Membership Consortium.

Based in part on observations at Cerro Tololo Inter-American Observatory, National Optical Astronomy Observatory, which is operated by the Association of 
Universities for Research in Astronomy (AURA) under a cooperative agreement with the National Science Foundation.

The DES data management system is supported by the National Science Foundation under Grant Numbers AST-1138766 and AST-1536171.
The DES participants from Spanish institutions are partially supported by MINECO under grants AYA2015-71825, ESP2015-88861, FPA2015-68048, SEV-2012-0234, SEV-2016-0597, and MDM-2015-0509, 
some of which include ERDF funds from the European Union. IFAE is partially funded by the CERCA program of the Generalitat de Catalunya.
Research leading to these results has received funding from the European Research
Council under the European Union's Seventh Framework Program (FP7/2007-2013) including ERC grant agreements 240672, 291329, and 306478.
We  acknowledge support from the Australian Research Council Centre of Excellence for All-sky Astrophysics (CAASTRO), through project number CE110001020.

This manuscript has been authored by Fermi Research Alliance, LLC under Contract No. DE-AC02-07CH11359 with the U.S. Department of Energy, Office of Science, Office of High Energy Physics. The United States Government retains and the publisher, by accepting the article for publication, acknowledges that the United States Government retains a non-exclusive, paid-up, irrevocable, world-wide license to publish or reproduce the published form of this manuscript, or allow others to do so, for United States Government purposes.

We use cosmic shear measurements from the Kilo-Degree Survey \cite{kids2015,kids450,2017MNRAS.467.1627F}, hereafter referred to as KiDS. The KiDS data are processed by THELI \cite{2013MNRAS.433.2545E} and Astro-WISE \cite{2013ExA....35....1B,2015A&A...582A..62D}. Shears are measured using lensfit \cite{2013MNRAS.429.2858M}, and photometric redshifts are obtained from PSF-matched photometry and calibrated using external overlapping spectroscopic surveys (see \cite{kids450}). Based on data products from observations made with ESO Telescopes at the La Silla Paranal Observatory under programme IDs 177.A-3016, 177.A-3017 and 177.A-3018.

Plots in this manuscript were produced partly with \textsc{Matplotlib} \cite{Hunter:2007}, and it has been prepared using NASA's Astrophysics Data System Bibliographic Services.  This research used resources of the National Energy Research Scientific Computing Center, a DOE Office of Science User Facility supported by the Office of Science of the U.S. Department of Energy under Contract No. DE-AC02-05CH11231. This work also used resources at the Ohio Supercomputing Center \cite{OhioSupercomputerCenter1987} and computing resources at SLAC National Accelerator Laboratory.

\bibliographystyle{apsrev4-1}
\bibliography{y1a1_cosmic_shear,../../des_bibtex/des_y1kp_short}

\begin{thebibliography}{158}%
\makeatletter
\providecommand \@ifxundefined [1]{%
 \@ifx{#1\undefined}
}%
\providecommand \@ifnum [1]{%
 \ifnum #1\expandafter \@firstoftwo
 \else \expandafter \@secondoftwo
 \fi
}%
\providecommand \@ifx [1]{%
 \ifx #1\expandafter \@firstoftwo
 \else \expandafter \@secondoftwo
 \fi
}%
\providecommand \natexlab [1]{#1}%
\providecommand \enquote  [1]{``#1''}%
\providecommand \bibnamefont  [1]{#1}%
\providecommand \bibfnamefont [1]{#1}%
\providecommand \citenamefont [1]{#1}%
\providecommand \href@noop [0]{\@secondoftwo}%
\providecommand \href [0]{\begingroup \@sanitize@url \@href}%
\providecommand \@href[1]{\@@startlink{#1}\@@href}%
\providecommand \@@href[1]{\endgroup#1\@@endlink}%
\providecommand \@sanitize@url [0]{\catcode `\\12\catcode `\$12\catcode
  `\&12\catcode `\#12\catcode `\^12\catcode `\_12\catcode `\%12\relax}%
\providecommand \@@startlink[1]{}%
\providecommand \@@endlink[0]{}%
\providecommand \url  [0]{\begingroup\@sanitize@url \@url }%
\providecommand \@url [1]{\endgroup\@href {#1}{\urlprefix }}%
\providecommand \urlprefix  [0]{URL }%
\providecommand \Eprint [0]{\href }%
\providecommand \doibase [0]{http://dx.doi.org/}%
\providecommand \selectlanguage [0]{\@gobble}%
\providecommand \bibinfo  [0]{\@secondoftwo}%
\providecommand \bibfield  [0]{\@secondoftwo}%
\providecommand \translation [1]{[#1]}%
\providecommand \BibitemOpen [0]{}%
\providecommand \bibitemStop [0]{}%
\providecommand \bibitemNoStop [0]{.\EOS\space}%
\providecommand \EOS [0]{\spacefactor3000\relax}%
\providecommand \BibitemShut  [1]{\csname bibitem#1\endcsname}%
\let\auto@bib@innerbib\@empty
\bibitem [{\citenamefont {{Blumenthal}}\ \emph {et~al.}(1984)\citenamefont
  {{Blumenthal}}, \citenamefont {{Faber}}, \citenamefont {{Primack}},\ and\
  \citenamefont {{Rees}}}]{1984Natur.311..517B}%
  \BibitemOpen
  \bibfield  {author} {\bibinfo {author} {\bibfnamefont {G.~R.}\ \bibnamefont
  {{Blumenthal}}}, \bibinfo {author} {\bibfnamefont {S.~M.}\ \bibnamefont
  {{Faber}}}, \bibinfo {author} {\bibfnamefont {J.~R.}\ \bibnamefont
  {{Primack}}}, \ and\ \bibinfo {author} {\bibfnamefont {M.~J.}\ \bibnamefont
  {{Rees}}},\ }\href {\doibase 10.1038/311517a0} {\bibfield  {journal}
  {\bibinfo  {journal} {Nature}\ }\textbf {\bibinfo {volume} {311}},\ \bibinfo
  {pages} {517} (\bibinfo {year} {1984})}\BibitemShut {NoStop}%
\bibitem [{\citenamefont {{Albrecht}}\ \emph {et~al.}(2006)\citenamefont
  {{Albrecht}}, \citenamefont {{Bernstein}}, \citenamefont {{Cahn}},
  \citenamefont {{Freedman}}, \citenamefont {{Hewitt}}, \citenamefont {{Hu}},
  \citenamefont {{Huth}}, \citenamefont {{Kamionkowski}}, \citenamefont
  {{Kolb}}, \citenamefont {{Knox}}, \citenamefont {{Mather}}, \citenamefont
  {{Staggs}},\ and\ \citenamefont {{Suntzeff}}}]{detf}%
  \BibitemOpen
  \bibfield  {author} {\bibinfo {author} {\bibfnamefont {A.}~\bibnamefont
  {{Albrecht}}}, \bibinfo {author} {\bibfnamefont {G.}~\bibnamefont
  {{Bernstein}}}, \bibinfo {author} {\bibfnamefont {R.}~\bibnamefont {{Cahn}}},
  \bibinfo {author} {\bibfnamefont {W.~L.}\ \bibnamefont {{Freedman}}},
  \bibinfo {author} {\bibfnamefont {J.}~\bibnamefont {{Hewitt}}}, \bibinfo
  {author} {\bibfnamefont {W.}~\bibnamefont {{Hu}}}, \bibinfo {author}
  {\bibfnamefont {J.}~\bibnamefont {{Huth}}}, \bibinfo {author} {\bibfnamefont
  {M.}~\bibnamefont {{Kamionkowski}}}, \bibinfo {author} {\bibfnamefont
  {E.~W.}\ \bibnamefont {{Kolb}}}, \bibinfo {author} {\bibfnamefont
  {L.}~\bibnamefont {{Knox}}}, \bibinfo {author} {\bibfnamefont {J.~C.}\
  \bibnamefont {{Mather}}}, \bibinfo {author} {\bibfnamefont {S.}~\bibnamefont
  {{Staggs}}}, \ and\ \bibinfo {author} {\bibfnamefont {N.~B.}\ \bibnamefont
  {{Suntzeff}}},\ }\href@noop {} {\bibfield  {journal} {\bibinfo  {journal}
  {ArXiv e-prints}\ } (\bibinfo {year} {2006})},\ \Eprint
  {http://arxiv.org/abs/0609591} {arXiv:0609591 [astro-ph.IM]} \BibitemShut
  {NoStop}%
\bibitem [{\citenamefont {{Peacock}}\ and\ \citenamefont
  {{Schneider}}(2006)}]{esoesa}%
  \BibitemOpen
  \bibfield  {author} {\bibinfo {author} {\bibfnamefont {J.}~\bibnamefont
  {{Peacock}}}\ and\ \bibinfo {author} {\bibfnamefont {P.}~\bibnamefont
  {{Schneider}}},\ }\href@noop {} {\bibfield  {journal} {\bibinfo  {journal}
  {The Messenger}\ }\textbf {\bibinfo {volume} {125}},\ \bibinfo {pages} {48}
  (\bibinfo {year} {2006})}\BibitemShut {NoStop}%
\bibitem [{\citenamefont {{Weinberg}}\ \emph {et~al.}(2013)\citenamefont
  {{Weinberg}}, \citenamefont {{Mortonson}}, \citenamefont {{Eisenstein}},
  \citenamefont {{Hirata}}, \citenamefont {{Riess}},\ and\ \citenamefont
  {{Rozo}}}]{weinberg13}%
  \BibitemOpen
  \bibfield  {author} {\bibinfo {author} {\bibfnamefont {D.~H.}\ \bibnamefont
  {{Weinberg}}}, \bibinfo {author} {\bibfnamefont {M.~J.}\ \bibnamefont
  {{Mortonson}}}, \bibinfo {author} {\bibfnamefont {D.~J.}\ \bibnamefont
  {{Eisenstein}}}, \bibinfo {author} {\bibfnamefont {C.}~\bibnamefont
  {{Hirata}}}, \bibinfo {author} {\bibfnamefont {A.~G.}\ \bibnamefont
  {{Riess}}}, \ and\ \bibinfo {author} {\bibfnamefont {E.}~\bibnamefont
  {{Rozo}}},\ }\href {\doibase 10.1016/j.physrep.2013.05.001} {\bibfield
  {journal} {\bibinfo  {journal} {\physrep}\ }\textbf {\bibinfo {volume}
  {530}},\ \bibinfo {pages} {87} (\bibinfo {year} {2013})}\BibitemShut
  {NoStop}%
\bibitem [{\citenamefont {{Bartelmann}}\ and\ \citenamefont
  {{Schneider}}(2001)}]{2001PhR...340..291B}%
  \BibitemOpen
  \bibfield  {author} {\bibinfo {author} {\bibfnamefont {M.}~\bibnamefont
  {{Bartelmann}}}\ and\ \bibinfo {author} {\bibfnamefont {P.}~\bibnamefont
  {{Schneider}}},\ }\href {\doibase 10.1016/S0370-1573(00)00082-X} {\bibfield
  {journal} {\bibinfo  {journal} {\physrep}\ }\textbf {\bibinfo {volume}
  {340}},\ \bibinfo {pages} {291} (\bibinfo {year} {2001})}\BibitemShut
  {NoStop}%
\bibitem [{\citenamefont {{Jain}}\ and\ \citenamefont
  {{Seljak}}(1997)}]{jain97}%
  \BibitemOpen
  \bibfield  {author} {\bibinfo {author} {\bibfnamefont {B.}~\bibnamefont
  {{Jain}}}\ and\ \bibinfo {author} {\bibfnamefont {U.}~\bibnamefont
  {{Seljak}}},\ }\href@noop {} {\bibfield  {journal} {\bibinfo  {journal}
  {Astrophys. J.}\ }\textbf {\bibinfo {volume} {484}},\ \bibinfo {pages} {560}
  (\bibinfo {year} {1997})}\BibitemShut {NoStop}%
\bibitem [{\citenamefont {{Battye}}\ and\ \citenamefont
  {{Moss}}(2014)}]{battye2014}%
  \BibitemOpen
  \bibfield  {author} {\bibinfo {author} {\bibfnamefont {R.~A.}\ \bibnamefont
  {{Battye}}}\ and\ \bibinfo {author} {\bibfnamefont {A.}~\bibnamefont
  {{Moss}}},\ }\href {\doibase 10.1103/PhysRevLett.112.051303} {\bibfield
  {journal} {\bibinfo  {journal} {Phys. Rev. Lett.}\ }\textbf {\bibinfo
  {volume} {112}},\ \bibinfo {eid} {051303} (\bibinfo {year}
  {2014})}\BibitemShut {NoStop}%
\bibitem [{\citenamefont {{Leistedt}}\ \emph {et~al.}(2014)\citenamefont
  {{Leistedt}}, \citenamefont {{Peiris}},\ and\ \citenamefont
  {{Verde}}}]{leistedt2014}%
  \BibitemOpen
  \bibfield  {author} {\bibinfo {author} {\bibfnamefont {B.}~\bibnamefont
  {{Leistedt}}}, \bibinfo {author} {\bibfnamefont {H.~V.}\ \bibnamefont
  {{Peiris}}}, \ and\ \bibinfo {author} {\bibfnamefont {L.}~\bibnamefont
  {{Verde}}},\ }\href {\doibase 10.1103/PhysRevLett.113.041301} {\bibfield
  {journal} {\bibinfo  {journal} {Phys. Rev. Lett.}\ }\textbf {\bibinfo
  {volume} {113}},\ \bibinfo {eid} {041301} (\bibinfo {year}
  {2014})}\BibitemShut {NoStop}%
\bibitem [{\citenamefont {{Beutler}}\ \emph {et~al.}(2014)\citenamefont
  {{Beutler}} \emph {et~al.}}]{beutler2014}%
  \BibitemOpen
  \bibfield  {author} {\bibinfo {author} {\bibfnamefont {F.}~\bibnamefont
  {{Beutler}}} \emph {et~al.},\ }\href {\doibase 10.1093/mnras/stu1702}
  {\bibfield  {journal} {\bibinfo  {journal} {Mon. Not. R. Astron. Soc.}\
  }\textbf {\bibinfo {volume} {444}},\ \bibinfo {pages} {3501} (\bibinfo {year}
  {2014})}\BibitemShut {NoStop}%
\bibitem [{\citenamefont {{Bacon}}\ \emph {et~al.}(2000)\citenamefont
  {{Bacon}}, \citenamefont {{Refregier}},\ and\ \citenamefont
  {{Ellis}}}]{Bacon:2000yp}%
  \BibitemOpen
  \bibfield  {author} {\bibinfo {author} {\bibfnamefont {D.~J.}\ \bibnamefont
  {{Bacon}}}, \bibinfo {author} {\bibfnamefont {A.~R.}\ \bibnamefont
  {{Refregier}}}, \ and\ \bibinfo {author} {\bibfnamefont {R.~S.}\ \bibnamefont
  {{Ellis}}},\ }\href@noop {} {\bibfield  {journal} {\bibinfo  {journal} {Mon.
  Not. R. Astron. Soc.}\ }\textbf {\bibinfo {volume} {318}},\ \bibinfo {pages}
  {625} (\bibinfo {year} {2000})}\BibitemShut {NoStop}%
\bibitem [{\citenamefont {{Kaiser}}\ \emph {et~al.}(2000)\citenamefont
  {{Kaiser}}, \citenamefont {{Wilson}},\ and\ \citenamefont
  {{Luppino}}}]{kaiser:2000if}%
  \BibitemOpen
  \bibfield  {author} {\bibinfo {author} {\bibfnamefont {N.}~\bibnamefont
  {{Kaiser}}}, \bibinfo {author} {\bibfnamefont {G.}~\bibnamefont {{Wilson}}},
  \ and\ \bibinfo {author} {\bibfnamefont {G.~A.}\ \bibnamefont {{Luppino}}},\
  }\href@noop {} {\bibfield  {journal} {\bibinfo  {journal}
  {arXiv:astro-ph/0003338}\ } (\bibinfo {year} {2000})}\BibitemShut {NoStop}%
\bibitem [{\citenamefont {Wittman}\ \emph {et~al.}(2000)\citenamefont
  {Wittman}, \citenamefont {Tyson}, \citenamefont {Kirkman}, \citenamefont
  {Dell'Antonio},\ and\ \citenamefont {Bernstein}}]{Wittman:2000tc}%
  \BibitemOpen
  \bibfield  {author} {\bibinfo {author} {\bibfnamefont {D.~M.}\ \bibnamefont
  {Wittman}}, \bibinfo {author} {\bibfnamefont {J.~A.}\ \bibnamefont {Tyson}},
  \bibinfo {author} {\bibfnamefont {D.}~\bibnamefont {Kirkman}}, \bibinfo
  {author} {\bibfnamefont {I.}~\bibnamefont {Dell'Antonio}}, \ and\ \bibinfo
  {author} {\bibfnamefont {G.}~\bibnamefont {Bernstein}},\ }\href@noop {}
  {\bibfield  {journal} {\bibinfo  {journal} {Nature}\ }\textbf {\bibinfo
  {volume} {405}},\ \bibinfo {pages} {143} (\bibinfo {year}
  {2000})}\BibitemShut {NoStop}%
\bibitem [{\citenamefont {van Waerbeke}\ \emph {et~al.}(2000)\citenamefont {van
  Waerbeke} \emph {et~al.}}]{van_Waerbeke:2000rm}%
  \BibitemOpen
  \bibfield  {author} {\bibinfo {author} {\bibfnamefont {L.}~\bibnamefont {van
  Waerbeke}} \emph {et~al.},\ }\href@noop {} {\bibfield  {journal} {\bibinfo
  {journal} {Astron. Astrophys.}\ }\textbf {\bibinfo {volume} {358}},\ \bibinfo
  {pages} {30} (\bibinfo {year} {2000})}\BibitemShut {NoStop}%
\bibitem [{\citenamefont {{Hoekstra}}\ \emph {et~al.}(2002)\citenamefont
  {{Hoekstra}}, \citenamefont {{Yee}}, \citenamefont {{Gladders}},
  \citenamefont {{Barrientos}}, \citenamefont {{Hall}},\ and\ \citenamefont
  {{Infante}}}]{Hoekstra02}%
  \BibitemOpen
  \bibfield  {author} {\bibinfo {author} {\bibfnamefont {H.}~\bibnamefont
  {{Hoekstra}}}, \bibinfo {author} {\bibfnamefont {H.~K.~C.}\ \bibnamefont
  {{Yee}}}, \bibinfo {author} {\bibfnamefont {M.~D.}\ \bibnamefont
  {{Gladders}}}, \bibinfo {author} {\bibfnamefont {L.~F.}\ \bibnamefont
  {{Barrientos}}}, \bibinfo {author} {\bibfnamefont {P.~B.}\ \bibnamefont
  {{Hall}}}, \ and\ \bibinfo {author} {\bibfnamefont {L.}~\bibnamefont
  {{Infante}}},\ }\href {\doibase 10.1086/340298} {\bibfield  {journal}
  {\bibinfo  {journal} {Astrophys. J.}\ }\textbf {\bibinfo {volume} {572}},\
  \bibinfo {pages} {55} (\bibinfo {year} {2002})}\BibitemShut {NoStop}%
\bibitem [{\citenamefont {{Van Waerbeke}}\ \emph {et~al.}(2005)\citenamefont
  {{Van Waerbeke}}, \citenamefont {{Mellier}},\ and\ \citenamefont
  {{Hoekstra}}}]{vanWaerbeke05}%
  \BibitemOpen
  \bibfield  {author} {\bibinfo {author} {\bibfnamefont {L.}~\bibnamefont {{Van
  Waerbeke}}}, \bibinfo {author} {\bibfnamefont {Y.}~\bibnamefont {{Mellier}}},
  \ and\ \bibinfo {author} {\bibfnamefont {H.}~\bibnamefont {{Hoekstra}}},\
  }\href {\doibase 10.1051/0004-6361:20041513} {\bibfield  {journal} {\bibinfo
  {journal} {Astron. Astrophys.}\ }\textbf {\bibinfo {volume} {429}},\ \bibinfo
  {pages} {75} (\bibinfo {year} {2005})}\BibitemShut {NoStop}%
\bibitem [{\citenamefont {{Jarvis}}\ \emph {et~al.}(2006)\citenamefont
  {{Jarvis}}, \citenamefont {{Jain}}, \citenamefont {{Bernstein}},\ and\
  \citenamefont {{Dolney}}}]{Jarvis06}%
  \BibitemOpen
  \bibfield  {author} {\bibinfo {author} {\bibfnamefont {M.}~\bibnamefont
  {{Jarvis}}}, \bibinfo {author} {\bibfnamefont {B.}~\bibnamefont {{Jain}}},
  \bibinfo {author} {\bibfnamefont {G.}~\bibnamefont {{Bernstein}}}, \ and\
  \bibinfo {author} {\bibfnamefont {D.}~\bibnamefont {{Dolney}}},\ }\href
  {\doibase 10.1086/503418} {\bibfield  {journal} {\bibinfo  {journal}
  {Astrophys. J.}\ }\textbf {\bibinfo {volume} {644}},\ \bibinfo {pages} {71}
  (\bibinfo {year} {2006})}\BibitemShut {NoStop}%
\bibitem [{\citenamefont {{Semboloni}}\ \emph {et~al.}(2006)\citenamefont
  {{Semboloni}}, \citenamefont {{Mellier}}, \citenamefont {{van Waerbeke}},
  \citenamefont {{Hoekstra}}, \citenamefont {{Tereno}}, \citenamefont
  {{Benabed}}, \citenamefont {{Gwyn}}, \citenamefont {{Fu}}, \citenamefont
  {{Hudson}}, \citenamefont {{Maoli}},\ and\ \citenamefont
  {{Parker}}}]{Semboloni06}%
  \BibitemOpen
  \bibfield  {author} {\bibinfo {author} {\bibfnamefont {E.}~\bibnamefont
  {{Semboloni}}}, \bibinfo {author} {\bibfnamefont {Y.}~\bibnamefont
  {{Mellier}}}, \bibinfo {author} {\bibfnamefont {L.}~\bibnamefont {{van
  Waerbeke}}}, \bibinfo {author} {\bibfnamefont {H.}~\bibnamefont
  {{Hoekstra}}}, \bibinfo {author} {\bibfnamefont {I.}~\bibnamefont
  {{Tereno}}}, \bibinfo {author} {\bibfnamefont {K.}~\bibnamefont {{Benabed}}},
  \bibinfo {author} {\bibfnamefont {S.~D.~J.}\ \bibnamefont {{Gwyn}}}, \bibinfo
  {author} {\bibfnamefont {L.}~\bibnamefont {{Fu}}}, \bibinfo {author}
  {\bibfnamefont {M.~J.}\ \bibnamefont {{Hudson}}}, \bibinfo {author}
  {\bibfnamefont {R.}~\bibnamefont {{Maoli}}}, \ and\ \bibinfo {author}
  {\bibfnamefont {L.~C.}\ \bibnamefont {{Parker}}},\ }\href {\doibase
  10.1051/0004-6361:20054479} {\bibfield  {journal} {\bibinfo  {journal}
  {Astron. Astrophys.}\ }\textbf {\bibinfo {volume} {452}},\ \bibinfo {pages}
  {51} (\bibinfo {year} {2006})}\BibitemShut {NoStop}%
\bibitem [{\citenamefont {{Benjamin}}\ \emph {et~al.}(2007)\citenamefont
  {{Benjamin}}, \citenamefont {{Heymans}}, \citenamefont {{Semboloni}},
  \citenamefont {{van Waerbeke}}, \citenamefont {{Hoekstra}}, \citenamefont
  {{Erben}}, \citenamefont {{Gladders}}, \citenamefont {{Hetterscheidt}},
  \citenamefont {{Mellier}},\ and\ \citenamefont
  {{Yee}}}]{2007MNRAS.381..702B}%
  \BibitemOpen
  \bibfield  {author} {\bibinfo {author} {\bibfnamefont {J.}~\bibnamefont
  {{Benjamin}}}, \bibinfo {author} {\bibfnamefont {C.}~\bibnamefont
  {{Heymans}}}, \bibinfo {author} {\bibfnamefont {E.}~\bibnamefont
  {{Semboloni}}}, \bibinfo {author} {\bibfnamefont {L.}~\bibnamefont {{van
  Waerbeke}}}, \bibinfo {author} {\bibfnamefont {H.}~\bibnamefont
  {{Hoekstra}}}, \bibinfo {author} {\bibfnamefont {T.}~\bibnamefont {{Erben}}},
  \bibinfo {author} {\bibfnamefont {M.~D.}\ \bibnamefont {{Gladders}}},
  \bibinfo {author} {\bibfnamefont {M.}~\bibnamefont {{Hetterscheidt}}},
  \bibinfo {author} {\bibfnamefont {Y.}~\bibnamefont {{Mellier}}}, \ and\
  \bibinfo {author} {\bibfnamefont {H.~K.~C.}\ \bibnamefont {{Yee}}},\ }\href
  {\doibase 10.1111/j.1365-2966.2007.12202.x} {\bibfield  {journal} {\bibinfo
  {journal} {Mon. Not. R. Astron. Soc.}\ }\textbf {\bibinfo {volume} {381}},\
  \bibinfo {pages} {702} (\bibinfo {year} {2007})}\BibitemShut {NoStop}%
\bibitem [{\citenamefont {{Massey}}\ \emph {et~al.}(2007)\citenamefont
  {{Massey}} \emph {et~al.}}]{Massey07}%
  \BibitemOpen
  \bibfield  {author} {\bibinfo {author} {\bibfnamefont {R.}~\bibnamefont
  {{Massey}}} \emph {et~al.},\ }\href {\doibase 10.1086/516599} {\bibfield
  {journal} {\bibinfo  {journal} {Astrophys. J.}\ }\textbf {\bibinfo {volume}
  {172}},\ \bibinfo {pages} {239} (\bibinfo {year} {2007})}\BibitemShut
  {NoStop}%
\bibitem [{\citenamefont {{Hetterscheidt}}\ \emph {et~al.}(2007)\citenamefont
  {{Hetterscheidt}}, \citenamefont {{Simon}}, \citenamefont {{Schirmer}},
  \citenamefont {{Hildebrandt}}, \citenamefont {{Schrabback}}, \citenamefont
  {{Erben}},\ and\ \citenamefont {{Schneider}}}]{Hetterscheidt07}%
  \BibitemOpen
  \bibfield  {author} {\bibinfo {author} {\bibfnamefont {M.}~\bibnamefont
  {{Hetterscheidt}}}, \bibinfo {author} {\bibfnamefont {P.}~\bibnamefont
  {{Simon}}}, \bibinfo {author} {\bibfnamefont {M.}~\bibnamefont {{Schirmer}}},
  \bibinfo {author} {\bibfnamefont {H.}~\bibnamefont {{Hildebrandt}}}, \bibinfo
  {author} {\bibfnamefont {T.}~\bibnamefont {{Schrabback}}}, \bibinfo {author}
  {\bibfnamefont {T.}~\bibnamefont {{Erben}}}, \ and\ \bibinfo {author}
  {\bibfnamefont {P.}~\bibnamefont {{Schneider}}},\ }\href {\doibase
  10.1051/0004-6361:20065885} {\bibfield  {journal} {\bibinfo  {journal}
  {Astron. Astrophys.}\ }\textbf {\bibinfo {volume} {468}},\ \bibinfo {pages}
  {859} (\bibinfo {year} {2007})}\BibitemShut {NoStop}%
\bibitem [{\citenamefont {{Schrabback}}\ \emph {et~al.}(2010)\citenamefont
  {{Schrabback}} \emph {et~al.}}]{Schrabback10}%
  \BibitemOpen
  \bibfield  {author} {\bibinfo {author} {\bibfnamefont {T.}~\bibnamefont
  {{Schrabback}}} \emph {et~al.},\ }\href {\doibase
  10.1051/0004-6361/200913577} {\bibfield  {journal} {\bibinfo  {journal}
  {Astron. Astrophys.}\ }\textbf {\bibinfo {volume} {516}},\ \bibinfo {eid}
  {A63} (\bibinfo {year} {2010})}\BibitemShut {NoStop}%
\bibitem [{\citenamefont {{Lin}}\ \emph {et~al.}(2012)\citenamefont {{Lin}},
  \citenamefont {{Dodelson}}, \citenamefont {{Seo}}, \citenamefont
  {{Soares-Santos}}, \citenamefont {{Annis}}, \citenamefont {{Hao}},
  \citenamefont {{Johnston}}, \citenamefont {{Kubo}}, \citenamefont {{Reis}},\
  and\ \citenamefont {{Simet}}}]{Lin12}%
  \BibitemOpen
  \bibfield  {author} {\bibinfo {author} {\bibfnamefont {H.}~\bibnamefont
  {{Lin}}}, \bibinfo {author} {\bibfnamefont {S.}~\bibnamefont {{Dodelson}}},
  \bibinfo {author} {\bibfnamefont {H.-J.}\ \bibnamefont {{Seo}}}, \bibinfo
  {author} {\bibfnamefont {M.}~\bibnamefont {{Soares-Santos}}}, \bibinfo
  {author} {\bibfnamefont {J.}~\bibnamefont {{Annis}}}, \bibinfo {author}
  {\bibfnamefont {J.}~\bibnamefont {{Hao}}}, \bibinfo {author} {\bibfnamefont
  {D.}~\bibnamefont {{Johnston}}}, \bibinfo {author} {\bibfnamefont {J.~M.}\
  \bibnamefont {{Kubo}}}, \bibinfo {author} {\bibfnamefont {R.~R.~R.}\
  \bibnamefont {{Reis}}}, \ and\ \bibinfo {author} {\bibfnamefont
  {M.}~\bibnamefont {{Simet}}},\ }\href {\doibase 10.1088/0004-637X/761/1/15}
  {\bibfield  {journal} {\bibinfo  {journal} {Astrophys. J.}\ }\textbf
  {\bibinfo {volume} {761}},\ \bibinfo {eid} {15} (\bibinfo {year}
  {2012})}\BibitemShut {NoStop}%
\bibitem [{\citenamefont {{Huff}}\ \emph {et~al.}(2014)\citenamefont {{Huff}},
  \citenamefont {{Eifler}}, \citenamefont {{Hirata}}, \citenamefont
  {{Mandelbaum}}, \citenamefont {{Schlegel}},\ and\ \citenamefont
  {{Seljak}}}]{Huff14}%
  \BibitemOpen
  \bibfield  {author} {\bibinfo {author} {\bibfnamefont {E.~M.}\ \bibnamefont
  {{Huff}}}, \bibinfo {author} {\bibfnamefont {T.}~\bibnamefont {{Eifler}}},
  \bibinfo {author} {\bibfnamefont {C.~M.}\ \bibnamefont {{Hirata}}}, \bibinfo
  {author} {\bibfnamefont {R.}~\bibnamefont {{Mandelbaum}}}, \bibinfo {author}
  {\bibfnamefont {D.}~\bibnamefont {{Schlegel}}}, \ and\ \bibinfo {author}
  {\bibfnamefont {U.}~\bibnamefont {{Seljak}}},\ }\href {\doibase
  10.1093/mnras/stu145} {\bibfield  {journal} {\bibinfo  {journal} {Mon. Not.
  R. Astron. Soc.}\ }\textbf {\bibinfo {volume} {440}},\ \bibinfo {pages}
  {1322} (\bibinfo {year} {2014})}\BibitemShut {NoStop}%
\bibitem [{\citenamefont {{Jee}}\ \emph {et~al.}(2013)\citenamefont {{Jee}},
  \citenamefont {{Tyson}}, \citenamefont {{Schneider}}, \citenamefont
  {{Wittman}}, \citenamefont {{Schmidt}},\ and\ \citenamefont
  {{Hilbert}}}]{jee2013}%
  \BibitemOpen
  \bibfield  {author} {\bibinfo {author} {\bibfnamefont {M.~J.}\ \bibnamefont
  {{Jee}}}, \bibinfo {author} {\bibfnamefont {J.~A.}\ \bibnamefont {{Tyson}}},
  \bibinfo {author} {\bibfnamefont {M.~D.}\ \bibnamefont {{Schneider}}},
  \bibinfo {author} {\bibfnamefont {D.}~\bibnamefont {{Wittman}}}, \bibinfo
  {author} {\bibfnamefont {S.}~\bibnamefont {{Schmidt}}}, \ and\ \bibinfo
  {author} {\bibfnamefont {S.}~\bibnamefont {{Hilbert}}},\ }\href {\doibase
  10.1088/0004-637X/765/1/74} {\bibfield  {journal} {\bibinfo  {journal}
  {Astrophys. J.}\ }\textbf {\bibinfo {volume} {765}},\ \bibinfo {eid} {74}
  (\bibinfo {year} {2013})}\BibitemShut {NoStop}%
\bibitem [{\citenamefont {{Jee}}\ \emph {et~al.}(2016)\citenamefont {{Jee}},
  \citenamefont {{Tyson}}, \citenamefont {{Hilbert}}, \citenamefont
  {{Schneider}}, \citenamefont {{Schmidt}},\ and\ \citenamefont
  {{Wittman}}}]{jee2015}%
  \BibitemOpen
  \bibfield  {author} {\bibinfo {author} {\bibfnamefont {M.~J.}\ \bibnamefont
  {{Jee}}}, \bibinfo {author} {\bibfnamefont {J.~A.}\ \bibnamefont {{Tyson}}},
  \bibinfo {author} {\bibfnamefont {S.}~\bibnamefont {{Hilbert}}}, \bibinfo
  {author} {\bibfnamefont {M.~D.}\ \bibnamefont {{Schneider}}}, \bibinfo
  {author} {\bibfnamefont {S.}~\bibnamefont {{Schmidt}}}, \ and\ \bibinfo
  {author} {\bibfnamefont {D.}~\bibnamefont {{Wittman}}},\ }\href {\doibase
  10.3847/0004-637X/824/2/77} {\bibfield  {journal} {\bibinfo  {journal}
  {Astrophys. J.}\ }\textbf {\bibinfo {volume} {824}},\ \bibinfo {eid} {77}
  (\bibinfo {year} {2016})}\BibitemShut {NoStop}%
\bibitem [{\citenamefont {{Heymans}}\ \emph {et~al.}(2012)\citenamefont
  {{Heymans}} \emph {et~al.}}]{heymans12}%
  \BibitemOpen
  \bibfield  {author} {\bibinfo {author} {\bibfnamefont {C.}~\bibnamefont
  {{Heymans}}} \emph {et~al.},\ }\href {\doibase
  10.1111/j.1365-2966.2012.21952.x} {\bibfield  {journal} {\bibinfo  {journal}
  {Mon. Not. R. Astron. Soc.}\ }\textbf {\bibinfo {volume} {427}},\ \bibinfo
  {pages} {146} (\bibinfo {year} {2012})}\BibitemShut {NoStop}%
\bibitem [{\citenamefont {{Kilbinger}}\ \emph {et~al.}(2013)\citenamefont
  {{Kilbinger}} \emph {et~al.}}]{kilbinger13}%
  \BibitemOpen
  \bibfield  {author} {\bibinfo {author} {\bibfnamefont {M.}~\bibnamefont
  {{Kilbinger}}} \emph {et~al.},\ }\href {\doibase 10.1093/mnras/stt041}
  {\bibfield  {journal} {\bibinfo  {journal} {Mon. Not. R. Astron. Soc.}\
  }\textbf {\bibinfo {volume} {430}},\ \bibinfo {pages} {2200} (\bibinfo {year}
  {2013})}\BibitemShut {NoStop}%
\bibitem [{\citenamefont {{Heymans}}\ \emph {et~al.}(2013)\citenamefont
  {{Heymans}} \emph {et~al.}}]{heymans13}%
  \BibitemOpen
  \bibfield  {author} {\bibinfo {author} {\bibfnamefont {C.}~\bibnamefont
  {{Heymans}}} \emph {et~al.},\ }\href {\doibase 10.1093/mnras/stt601}
  {\bibfield  {journal} {\bibinfo  {journal} {Mon. Not. R. Astron. Soc.}\
  }\textbf {\bibinfo {volume} {432}},\ \bibinfo {pages} {2433} (\bibinfo {year}
  {2013})}\BibitemShut {NoStop}%
\bibitem [{\citenamefont {Benjamin}\ \emph {et~al.}(2013)\citenamefont
  {Benjamin} \emph {et~al.}}]{joachimicfht}%
  \BibitemOpen
  \bibfield  {author} {\bibinfo {author} {\bibfnamefont {J.}~\bibnamefont
  {Benjamin}} \emph {et~al.},\ }\href {\doibase 10.1093/mnras/stt276}
  {\bibfield  {journal} {\bibinfo  {journal} {Mon. Not. R. Astron. Soc.}\
  }\textbf {\bibinfo {volume} {431}},\ \bibinfo {pages} {1547} (\bibinfo {year}
  {2013})}\BibitemShut {NoStop}%
\bibitem [{\citenamefont {{Kitching}}\ \emph {et~al.}(2014)\citenamefont
  {{Kitching}} \emph {et~al.}}]{kitching14}%
  \BibitemOpen
  \bibfield  {author} {\bibinfo {author} {\bibfnamefont {T.~D.}\ \bibnamefont
  {{Kitching}}} \emph {et~al.},\ }\href {\doibase 10.1093/mnras/stu934}
  {\bibfield  {journal} {\bibinfo  {journal} {Mon. Not. R. Astron. Soc.}\
  }\textbf {\bibinfo {volume} {442}},\ \bibinfo {pages} {1326} (\bibinfo {year}
  {2014})}\BibitemShut {NoStop}%
\bibitem [{\citenamefont {{Joudaki}}\ \emph
  {et~al.}(2017{\natexlab{a}})\citenamefont {{Joudaki}}, \citenamefont
  {{Blake}}, \citenamefont {{Heymans}}, \citenamefont {{Choi}}, \citenamefont
  {{Harnois-Deraps}}, \citenamefont {{Hildebrandt}}, \citenamefont
  {{Joachimi}}, \citenamefont {{Johnson}}, \citenamefont {{Mead}},
  \citenamefont {{Parkinson}}, \citenamefont {{Viola}},\ and\ \citenamefont
  {{van Waerbeke}}}]{joudaki17}%
  \BibitemOpen
  \bibfield  {author} {\bibinfo {author} {\bibfnamefont {S.}~\bibnamefont
  {{Joudaki}}}, \bibinfo {author} {\bibfnamefont {C.}~\bibnamefont {{Blake}}},
  \bibinfo {author} {\bibfnamefont {C.}~\bibnamefont {{Heymans}}}, \bibinfo
  {author} {\bibfnamefont {A.}~\bibnamefont {{Choi}}}, \bibinfo {author}
  {\bibfnamefont {J.}~\bibnamefont {{Harnois-Deraps}}}, \bibinfo {author}
  {\bibfnamefont {H.}~\bibnamefont {{Hildebrandt}}}, \bibinfo {author}
  {\bibfnamefont {B.}~\bibnamefont {{Joachimi}}}, \bibinfo {author}
  {\bibfnamefont {A.}~\bibnamefont {{Johnson}}}, \bibinfo {author}
  {\bibfnamefont {A.}~\bibnamefont {{Mead}}}, \bibinfo {author} {\bibfnamefont
  {D.}~\bibnamefont {{Parkinson}}}, \bibinfo {author} {\bibfnamefont
  {M.}~\bibnamefont {{Viola}}}, \ and\ \bibinfo {author} {\bibfnamefont
  {L.}~\bibnamefont {{van Waerbeke}}},\ }\href {\doibase 10.1093/mnras/stw2665}
  {\bibfield  {journal} {\bibinfo  {journal} {Mon. Not. R. Astron. Soc.}\
  }\textbf {\bibinfo {volume} {465}},\ \bibinfo {pages} {2033} (\bibinfo {year}
  {2017}{\natexlab{a}})}\BibitemShut {NoStop}%
\bibitem [{\citenamefont {{DES Collaboration}}\ \emph
  {et~al.}(2016)\citenamefont {{DES Collaboration}} \emph {et~al.}}]{des2016}%
  \BibitemOpen
  \bibfield  {author} {\bibinfo {author} {\bibnamefont {{DES Collaboration}}}
  \emph {et~al.},\ }\href {\doibase 10.1103/PhysRevD.94.022001} {\bibfield
  {journal} {\bibinfo  {journal} {Phys. Rev. D}\ }\textbf {\bibinfo {volume}
  {94}},\ \bibinfo {eid} {022001} (\bibinfo {year} {2016})}\BibitemShut
  {NoStop}%
\bibitem [{\citenamefont {{Kuijken}}\ \emph {et~al.}(2015)\citenamefont
  {{Kuijken}} \emph {et~al.}}]{kids2015}%
  \BibitemOpen
  \bibfield  {author} {\bibinfo {author} {\bibfnamefont {K.}~\bibnamefont
  {{Kuijken}}} \emph {et~al.},\ }\href {\doibase 10.1093/mnras/stv2140}
  {\bibfield  {journal} {\bibinfo  {journal} {Mon. Not. R. Astron. Soc.}\
  }\textbf {\bibinfo {volume} {454}},\ \bibinfo {pages} {3500} (\bibinfo {year}
  {2015})}\BibitemShut {NoStop}%
\bibitem [{\citenamefont {{Hildebrandt}}\ \emph {et~al.}(2017)\citenamefont
  {{Hildebrandt}} \emph {et~al.}}]{kids450}%
  \BibitemOpen
  \bibfield  {author} {\bibinfo {author} {\bibfnamefont {H.}~\bibnamefont
  {{Hildebrandt}}} \emph {et~al.},\ }\href {\doibase 10.1093/mnras/stw2805}
  {\bibfield  {journal} {\bibinfo  {journal} {Mon. Not. R. Astron. Soc.}\
  }\textbf {\bibinfo {volume} {465}},\ \bibinfo {pages} {1454} (\bibinfo {year}
  {2017})}\BibitemShut {NoStop}%
\bibitem [{\citenamefont {{Joudaki}}\ \emph
  {et~al.}(2017{\natexlab{b}})\citenamefont {{Joudaki}} \emph
  {et~al.}}]{kids450b}%
  \BibitemOpen
  \bibfield  {author} {\bibinfo {author} {\bibfnamefont {S.}~\bibnamefont
  {{Joudaki}}} \emph {et~al.},\ }\href {\doibase 10.1093/mnras/stx998}
  {\bibfield  {journal} {\bibinfo  {journal} {Mon. Not. R. Astron. Soc.}\
  }\textbf {\bibinfo {volume} {471}},\ \bibinfo {pages} {1259} (\bibinfo {year}
  {2017}{\natexlab{b}})}\BibitemShut {NoStop}%
\bibitem [{\citenamefont {{K{\"o}hlinger}}\ \emph {et~al.}(2017)\citenamefont
  {{K{\"o}hlinger}} \emph {et~al.}}]{kids450c}%
  \BibitemOpen
  \bibfield  {author} {\bibinfo {author} {\bibfnamefont {F.}~\bibnamefont
  {{K{\"o}hlinger}}} \emph {et~al.},\ }\href {\doibase 10.1093/mnras/stx1820}
  {\bibfield  {journal} {\bibinfo  {journal} {Mon. Not. R. Astron. Soc.}\
  }\textbf {\bibinfo {volume} {471}},\ \bibinfo {pages} {4412} (\bibinfo {year}
  {2017})}\BibitemShut {NoStop}%
\bibitem [{\citenamefont {{van Uitert}}\ \emph {et~al.}(2018)\citenamefont
  {{van Uitert}} \emph {et~al.}}]{kids450d}%
  \BibitemOpen
  \bibfield  {author} {\bibinfo {author} {\bibfnamefont {E.}~\bibnamefont {{van
  Uitert}}} \emph {et~al.},\ }\href {\doibase 10.1093/mnras/sty551} {\bibfield
  {journal} {\bibinfo  {journal} {Mon. Not. R. Astron. Soc.}\ }\textbf
  {\bibinfo {volume} {476}},\ \bibinfo {pages} {4662} (\bibinfo {year}
  {2018})}\BibitemShut {NoStop}%
\bibitem [{\citenamefont {{Joudaki}}\ \emph {et~al.}(2018)\citenamefont
  {{Joudaki}} \emph {et~al.}}]{2017arXiv170706627J}%
  \BibitemOpen
  \bibfield  {author} {\bibinfo {author} {\bibfnamefont {S.}~\bibnamefont
  {{Joudaki}}} \emph {et~al.},\ }\href {\doibase 10.1093/mnras/stx2820}
  {\bibfield  {journal} {\bibinfo  {journal} {Mon. Not. R. Astron. Soc.}\
  }\textbf {\bibinfo {volume} {474}},\ \bibinfo {pages} {4894} (\bibinfo {year}
  {2018})},\ \Eprint {http://arxiv.org/abs/1707.06627} {arXiv:1707.06627}
  \BibitemShut {NoStop}%
\bibitem [{\citenamefont {{Zuntz}}\ \emph {et~al.}(2017)\citenamefont {{Zuntz}}
  \emph {et~al.}}]{shearcat}%
  \BibitemOpen
  \bibfield  {author} {\bibinfo {author} {\bibfnamefont {J.}~\bibnamefont
  {{Zuntz}}} \emph {et~al.} (\bibinfo {collaboration} {DES Collaboration}),\
  }\href@noop {} {\bibfield  {journal} {\bibinfo  {journal} {ArXiv e-prints}\ }
  (\bibinfo {year} {2017})},\ \Eprint {http://arxiv.org/abs/1708.01533}
  {arXiv:1708.01533} \BibitemShut {NoStop}%
\bibitem [{\citenamefont {{Hoyle}}\ \emph {et~al.}(2018)\citenamefont {{Hoyle}}
  \emph {et~al.}}]{photoz}%
  \BibitemOpen
  \bibfield  {author} {\bibinfo {author} {\bibfnamefont {B.}~\bibnamefont
  {{Hoyle}}} \emph {et~al.} (\bibinfo {collaboration} {DES Collaboration}),\
  }\href {\doibase 10.1093/mnras/sty957} {\bibfield  {journal} {\bibinfo
  {journal} {\mnras}\ } (\bibinfo {year} {2018}),\
  10.1093/mnras/sty957}\BibitemShut {NoStop}%
\bibitem [{\citenamefont {{Cawthon}}\ \emph {et~al.}(2017)\citenamefont
  {{Cawthon}} \emph {et~al.}}]{redmagicpz}%
  \BibitemOpen
  \bibfield  {author} {\bibinfo {author} {\bibfnamefont {R.}~\bibnamefont
  {{Cawthon}}} \emph {et~al.} (\bibinfo {collaboration} {DES Collaboration}),\
  }\href@noop {} {\bibfield  {journal} {\bibinfo  {journal} {ArXiv e-prints}\ }
  (\bibinfo {year} {2017})},\ \Eprint {http://arxiv.org/abs/1712.07298}
  {arXiv:1712.07298} \BibitemShut {NoStop}%
\bibitem [{\citenamefont {{Gatti}}\ \emph {et~al.}(2018)\citenamefont {{Gatti}}
  \emph {et~al.}}]{xcorrtechnique}%
  \BibitemOpen
  \bibfield  {author} {\bibinfo {author} {\bibfnamefont {M.}~\bibnamefont
  {{Gatti}}} \emph {et~al.} (\bibinfo {collaboration} {DES Collaboration}),\
  }\href {\doibase 10.1093/mnras/sty466} {\bibfield  {journal} {\bibinfo
  {journal} {\mnras}\ } (\bibinfo {year} {2018}),\
  10.1093/mnras/sty466}\BibitemShut {NoStop}%
\bibitem [{\citenamefont {{Davis}}\ \emph {et~al.}(2017)\citenamefont {{Davis}}
  \emph {et~al.}}]{xcorr}%
  \BibitemOpen
  \bibfield  {author} {\bibinfo {author} {\bibfnamefont {C.}~\bibnamefont
  {{Davis}}} \emph {et~al.} (\bibinfo {collaboration} {DES Collaboration}),\
  }\href@noop {} {\bibfield  {journal} {\bibinfo  {journal} {ArXiv e-prints}\ }
  (\bibinfo {year} {2017})},\ \Eprint {http://arxiv.org/abs/1710.02517}
  {arXiv:1710.02517} \BibitemShut {NoStop}%
\bibitem [{\citenamefont {Troxel}\ and\ \citenamefont
  {Ishak}(2015)}]{Troxel20151}%
  \BibitemOpen
  \bibfield  {author} {\bibinfo {author} {\bibfnamefont {M.}~\bibnamefont
  {Troxel}}\ and\ \bibinfo {author} {\bibfnamefont {M.}~\bibnamefont {Ishak}},\
  }\href {\doibase http://dx.doi.org/10.1016/j.physrep.2014.11.001} {\bibfield
  {journal} {\bibinfo  {journal} {Physics Reports}\ }\textbf {\bibinfo {volume}
  {558}},\ \bibinfo {pages} {1 } (\bibinfo {year} {2015})}\BibitemShut
  {NoStop}%
\bibitem [{\citenamefont {{Joachimi}}\ \emph {et~al.}(2015)\citenamefont
  {{Joachimi}}, \citenamefont {{Cacciato}}, \citenamefont {{Kitching}},
  \citenamefont {{Leonard}}, \citenamefont {{Mandelbaum}}, \citenamefont
  {{Sch{\"a}fer}}, \citenamefont {{Sif{\'o}n}}, \citenamefont {{Hoekstra}},
  \citenamefont {{Kiessling}}, \citenamefont {{Kirk}},\ and\ \citenamefont
  {{Rassat}}}]{Joachimi2015}%
  \BibitemOpen
  \bibfield  {author} {\bibinfo {author} {\bibfnamefont {B.}~\bibnamefont
  {{Joachimi}}}, \bibinfo {author} {\bibfnamefont {M.}~\bibnamefont
  {{Cacciato}}}, \bibinfo {author} {\bibfnamefont {T.~D.}\ \bibnamefont
  {{Kitching}}}, \bibinfo {author} {\bibfnamefont {A.}~\bibnamefont
  {{Leonard}}}, \bibinfo {author} {\bibfnamefont {R.}~\bibnamefont
  {{Mandelbaum}}}, \bibinfo {author} {\bibfnamefont {B.~M.}\ \bibnamefont
  {{Sch{\"a}fer}}}, \bibinfo {author} {\bibfnamefont {C.}~\bibnamefont
  {{Sif{\'o}n}}}, \bibinfo {author} {\bibfnamefont {H.}~\bibnamefont
  {{Hoekstra}}}, \bibinfo {author} {\bibfnamefont {A.}~\bibnamefont
  {{Kiessling}}}, \bibinfo {author} {\bibfnamefont {D.}~\bibnamefont {{Kirk}}},
  \ and\ \bibinfo {author} {\bibfnamefont {A.}~\bibnamefont {{Rassat}}},\
  }\href {\doibase 10.1007/s11214-015-0177-4} {\bibfield  {journal} {\bibinfo
  {journal} {Sp. Sci. Rev.}\ }\textbf {\bibinfo {volume} {193}},\ \bibinfo
  {pages} {1} (\bibinfo {year} {2015})}\BibitemShut {NoStop}%
\bibitem [{\citenamefont {{van Daalen}}\ \emph {et~al.}(2011)\citenamefont
  {{van Daalen}}, \citenamefont {{Schaye}}, \citenamefont {{Booth}},\ and\
  \citenamefont {{Dalla Vecchia}}}]{vandalen11}%
  \BibitemOpen
  \bibfield  {author} {\bibinfo {author} {\bibfnamefont {M.~P.}\ \bibnamefont
  {{van Daalen}}}, \bibinfo {author} {\bibfnamefont {J.}~\bibnamefont
  {{Schaye}}}, \bibinfo {author} {\bibfnamefont {C.~M.}\ \bibnamefont
  {{Booth}}}, \ and\ \bibinfo {author} {\bibfnamefont {C.}~\bibnamefont {{Dalla
  Vecchia}}},\ }\href {\doibase 10.1111/j.1365-2966.2011.18981.x} {\bibfield
  {journal} {\bibinfo  {journal} {Mon. Not. R. Astron. Soc.}\ }\textbf
  {\bibinfo {volume} {415}},\ \bibinfo {pages} {3649} (\bibinfo {year}
  {2011})}\BibitemShut {NoStop}%
\bibitem [{\citenamefont {{Semboloni}}\ \emph {et~al.}(2011)\citenamefont
  {{Semboloni}}, \citenamefont {{Hoekstra}}, \citenamefont {{Schaye}},
  \citenamefont {{van Daalen}},\ and\ \citenamefont
  {{McCarthy}}}]{sembolini11}%
  \BibitemOpen
  \bibfield  {author} {\bibinfo {author} {\bibfnamefont {E.}~\bibnamefont
  {{Semboloni}}}, \bibinfo {author} {\bibfnamefont {H.}~\bibnamefont
  {{Hoekstra}}}, \bibinfo {author} {\bibfnamefont {J.}~\bibnamefont
  {{Schaye}}}, \bibinfo {author} {\bibfnamefont {M.~P.}\ \bibnamefont {{van
  Daalen}}}, \ and\ \bibinfo {author} {\bibfnamefont {I.~G.}\ \bibnamefont
  {{McCarthy}}},\ }\href {\doibase 10.1111/j.1365-2966.2011.19385.x} {\bibfield
   {journal} {\bibinfo  {journal} {Mon. Not. R. Astron. Soc.}\ }\textbf
  {\bibinfo {volume} {417}},\ \bibinfo {pages} {2020} (\bibinfo {year}
  {2011})}\BibitemShut {NoStop}%
\bibitem [{\citenamefont {{Harnois-D{\'e}raps}}\ \emph
  {et~al.}(2015)\citenamefont {{Harnois-D{\'e}raps}}, \citenamefont {{van
  Waerbeke}}, \citenamefont {{Viola}},\ and\ \citenamefont
  {{Heymans}}}]{harnois14}%
  \BibitemOpen
  \bibfield  {author} {\bibinfo {author} {\bibfnamefont {J.}~\bibnamefont
  {{Harnois-D{\'e}raps}}}, \bibinfo {author} {\bibfnamefont {L.}~\bibnamefont
  {{van Waerbeke}}}, \bibinfo {author} {\bibfnamefont {M.}~\bibnamefont
  {{Viola}}}, \ and\ \bibinfo {author} {\bibfnamefont {C.}~\bibnamefont
  {{Heymans}}},\ }\href {\doibase 10.1093/mnras/stv646} {\bibfield  {journal}
  {\bibinfo  {journal} {Mon. Not. R. Astron. Soc.}\ }\textbf {\bibinfo {volume}
  {450}},\ \bibinfo {pages} {1212} (\bibinfo {year} {2015})}\BibitemShut
  {NoStop}%
\bibitem [{\citenamefont {{Becker}}\ \emph {et~al.}(2016)\citenamefont
  {{Becker}} \emph {et~al.}}]{becker2016}%
  \BibitemOpen
  \bibfield  {author} {\bibinfo {author} {\bibfnamefont {M.~R.}\ \bibnamefont
  {{Becker}}} \emph {et~al.} (\bibinfo {collaboration} {DES Collaboration}),\
  }\href {\doibase 10.1103/PhysRevD.94.022002} {\bibfield  {journal} {\bibinfo
  {journal} {Phys. Rev. D}\ }\textbf {\bibinfo {volume} {94}},\ \bibinfo {eid}
  {022002} (\bibinfo {year} {2016})}\BibitemShut {NoStop}%
\bibitem [{\citenamefont {{Kacprzak}}\ \emph {et~al.}(2016)\citenamefont
  {{Kacprzak}} \emph {et~al.}}]{Kacprzak2016}%
  \BibitemOpen
  \bibfield  {author} {\bibinfo {author} {\bibfnamefont {T.}~\bibnamefont
  {{Kacprzak}}} \emph {et~al.} (\bibinfo {collaboration} {DES Collaboration}),\
  }\href {\doibase 10.1093/mnras/stw2070} {\bibfield  {journal} {\bibinfo
  {journal} {Mon. Not. R. Astron. Soc.}\ }\textbf {\bibinfo {volume} {463}},\
  \bibinfo {pages} {3653} (\bibinfo {year} {2016})}\BibitemShut {NoStop}%
\bibitem [{\citenamefont {{Kwan}}\ \emph {et~al.}(2017)\citenamefont {{Kwan}}
  \emph {et~al.}}]{kwan2016}%
  \BibitemOpen
  \bibfield  {author} {\bibinfo {author} {\bibfnamefont {J.}~\bibnamefont
  {{Kwan}}} \emph {et~al.} (\bibinfo {collaboration} {DES Collaboration}),\
  }\href {\doibase 10.1093/mnras/stw2464} {\bibfield  {journal} {\bibinfo
  {journal} {Mon. Not. R. Astron. Soc.}\ }\textbf {\bibinfo {volume} {464}},\
  \bibinfo {pages} {4045} (\bibinfo {year} {2017})}\BibitemShut {NoStop}%
\bibitem [{\citenamefont {{DES Collaboration}}\ \emph
  {et~al.}(2017)\citenamefont {{DES Collaboration}} \emph {et~al.}}]{keypaper}%
  \BibitemOpen
  \bibfield  {author} {\bibinfo {author} {\bibnamefont {{DES Collaboration}}}
  \emph {et~al.} (\bibinfo {collaboration} {DES Collaboration}),\ }\href@noop
  {} {\bibfield  {journal} {\bibinfo  {journal} {ArXiv e-prints}\ } (\bibinfo
  {year} {2017})},\ \Eprint {http://arxiv.org/abs/1708.01530}
  {arXiv:1708.01530} \BibitemShut {NoStop}%
\bibitem [{\citenamefont {{Drlica-Wagner}}\ \emph {et~al.}(2018)\citenamefont
  {{Drlica-Wagner}} \emph {et~al.}}]{y1gold}%
  \BibitemOpen
  \bibfield  {author} {\bibinfo {author} {\bibfnamefont {A.}~\bibnamefont
  {{Drlica-Wagner}}} \emph {et~al.} (\bibinfo {collaboration} {DES
  Collaboration}),\ }\href {\doibase 10.3847/1538-4365/aab4f5} {\bibfield
  {journal} {\bibinfo  {journal} {\apjs}\ }\textbf {\bibinfo {volume} {235}},\
  \bibinfo {eid} {33} (\bibinfo {year} {2018})}\BibitemShut {NoStop}%
\bibitem [{\citenamefont {{Elvin-Poole}}\ \emph {et~al.}(2017)\citenamefont
  {{Elvin-Poole}} \emph {et~al.}}]{wthetapaper}%
  \BibitemOpen
  \bibfield  {author} {\bibinfo {author} {\bibfnamefont {J.}~\bibnamefont
  {{Elvin-Poole}}} \emph {et~al.} (\bibinfo {collaboration} {DES
  Collaboration}),\ }\href@noop {} {\bibfield  {journal} {\bibinfo  {journal}
  {ArXiv e-prints}\ } (\bibinfo {year} {2017})},\ \Eprint
  {http://arxiv.org/abs/1708.01536} {arXiv:1708.01536} \BibitemShut {NoStop}%
\bibitem [{\citenamefont {{Samuroff}}\ \emph {et~al.}(2018)\citenamefont
  {{Samuroff}} \emph {et~al.}}]{des_sim_2017}%
  \BibitemOpen
  \bibfield  {author} {\bibinfo {author} {\bibfnamefont {S.}~\bibnamefont
  {{Samuroff}}} \emph {et~al.} (\bibinfo {collaboration} {DES Collaboration}),\
  }\href {\doibase 10.1093/mnras/stx3282} {\bibfield  {journal} {\bibinfo
  {journal} {\mnras}\ }\textbf {\bibinfo {volume} {475}},\ \bibinfo {pages}
  {4524} (\bibinfo {year} {2018})}\BibitemShut {NoStop}%
\bibitem [{\citenamefont {{Chang}}\ \emph {et~al.}(2018)\citenamefont {{Chang}}
  \emph {et~al.}}]{des_mm_2017}%
  \BibitemOpen
  \bibfield  {author} {\bibinfo {author} {\bibfnamefont {C.}~\bibnamefont
  {{Chang}}} \emph {et~al.} (\bibinfo {collaboration} {DES Collaboration}),\
  }\href {\doibase 10.1093/mnras/stx3363} {\bibfield  {journal} {\bibinfo
  {journal} {\mnras}\ }\textbf {\bibinfo {volume} {475}},\ \bibinfo {pages}
  {3165} (\bibinfo {year} {2018})}\BibitemShut {NoStop}%
\bibitem [{\citenamefont {{Prat}}\ \emph {et~al.}(2017)\citenamefont {{Prat}}
  \emph {et~al.}}]{gglpaper}%
  \BibitemOpen
  \bibfield  {author} {\bibinfo {author} {\bibfnamefont {J.}~\bibnamefont
  {{Prat}}} \emph {et~al.} (\bibinfo {collaboration} {DES Collaboration}),\
  }\href@noop {} {\bibfield  {journal} {\bibinfo  {journal} {ArXiv e-prints}\ }
  (\bibinfo {year} {2017})},\ \Eprint {http://arxiv.org/abs/1708.01537}
  {arXiv:1708.01537} \BibitemShut {NoStop}%
\bibitem [{\citenamefont {{Krause}}\ \emph {et~al.}(2017)\citenamefont
  {{Krause}} \emph {et~al.}}]{methodpaper}%
  \BibitemOpen
  \bibfield  {author} {\bibinfo {author} {\bibfnamefont {E.}~\bibnamefont
  {{Krause}}} \emph {et~al.} (\bibinfo {collaboration} {DES Collaboration}),\
  }\href@noop {} {\bibfield  {journal} {\bibinfo  {journal} {ArXiv e-prints}\ }
  (\bibinfo {year} {2017})},\ \Eprint {http://arxiv.org/abs/1706.09359}
  {arXiv:1706.09359} \BibitemShut {NoStop}%
\bibitem [{\citenamefont {{MacCrann}}\ \emph {et~al.}(2018)\citenamefont
  {{MacCrann}} \emph {et~al.}}]{simspaper}%
  \BibitemOpen
  \bibfield  {author} {\bibinfo {author} {\bibfnamefont {N.}~\bibnamefont
  {{MacCrann}}} \emph {et~al.} (\bibinfo {collaboration} {DES Collaboration}),\
  }\href@noop {} {\bibfield  {journal} {\bibinfo  {journal} {ArXiv e-prints}\ }
  (\bibinfo {year} {2018})},\ \Eprint {http://arxiv.org/abs/1803.09795}
  {arXiv:1803.09795} \BibitemShut {NoStop}%
\bibitem [{\citenamefont {{Flaugher}}\ \emph {et~al.}(2015)\citenamefont
  {{Flaugher}} \emph {et~al.}}]{decam}%
  \BibitemOpen
  \bibfield  {author} {\bibinfo {author} {\bibfnamefont {B.}~\bibnamefont
  {{Flaugher}}} \emph {et~al.} (\bibinfo {collaboration} {DES Collaboration}),\
  }\href {\doibase 10.1088/0004-6256/150/5/150} {\bibfield  {journal} {\bibinfo
   {journal} {Astron. J.}\ }\textbf {\bibinfo {volume} {150}},\ \bibinfo {eid}
  {150} (\bibinfo {year} {2015})}\BibitemShut {NoStop}%
\bibitem [{\citenamefont {{Jarvis}}\ \emph {et~al.}(2016)\citenamefont
  {{Jarvis}} \emph {et~al.}}]{jarvis2016}%
  \BibitemOpen
  \bibfield  {author} {\bibinfo {author} {\bibfnamefont {M.}~\bibnamefont
  {{Jarvis}}} \emph {et~al.} (\bibinfo {collaboration} {DES Collaboration}),\
  }\href {\doibase 10.1093/mnras/stw990} {\bibfield  {journal} {\bibinfo
  {journal} {Mon. Not. R. Astron. Soc.}\ }\textbf {\bibinfo {volume} {460}},\
  \bibinfo {pages} {2245} (\bibinfo {year} {2016})}\BibitemShut {NoStop}%
\bibitem [{\citenamefont {{Huff}}\ and\ \citenamefont
  {{Mandelbaum}}(2017)}]{HuffMandelbaum2017}%
  \BibitemOpen
  \bibfield  {author} {\bibinfo {author} {\bibfnamefont {E.}~\bibnamefont
  {{Huff}}}\ and\ \bibinfo {author} {\bibfnamefont {R.}~\bibnamefont
  {{Mandelbaum}}},\ }\href@noop {} {\bibfield  {journal} {\bibinfo  {journal}
  {ArXiv e-prints}\ } (\bibinfo {year} {2017})},\ \Eprint
  {http://arxiv.org/abs/1702.02600} {arXiv:1702.02600} \BibitemShut {NoStop}%
\bibitem [{\citenamefont {{Sheldon}}\ and\ \citenamefont
  {{Huff}}(2017)}]{SheldonHuff2017}%
  \BibitemOpen
  \bibfield  {author} {\bibinfo {author} {\bibfnamefont {E.~S.}\ \bibnamefont
  {{Sheldon}}}\ and\ \bibinfo {author} {\bibfnamefont {E.~M.}\ \bibnamefont
  {{Huff}}},\ }\href {\doibase 10.3847/1538-4357/aa704b} {\bibfield  {journal}
  {\bibinfo  {journal} {Astrophys. J.}\ }\textbf {\bibinfo {volume} {841}},\
  \bibinfo {eid} {24} (\bibinfo {year} {2017})}\BibitemShut {NoStop}%
\bibitem [{\citenamefont {{G{\'o}rski}}\ \emph {et~al.}(2005)\citenamefont
  {{G{\'o}rski}}, \citenamefont {{Hivon}}, \citenamefont {{Banday}},
  \citenamefont {{Wandelt}}, \citenamefont {{Hansen}}, \citenamefont
  {{Reinecke}},\ and\ \citenamefont {{Bartelmann}}}]{gorski2005}%
  \BibitemOpen
  \bibfield  {author} {\bibinfo {author} {\bibfnamefont {K.~M.}\ \bibnamefont
  {{G{\'o}rski}}}, \bibinfo {author} {\bibfnamefont {E.}~\bibnamefont
  {{Hivon}}}, \bibinfo {author} {\bibfnamefont {A.~J.}\ \bibnamefont
  {{Banday}}}, \bibinfo {author} {\bibfnamefont {B.~D.}\ \bibnamefont
  {{Wandelt}}}, \bibinfo {author} {\bibfnamefont {F.~K.}\ \bibnamefont
  {{Hansen}}}, \bibinfo {author} {\bibfnamefont {M.}~\bibnamefont
  {{Reinecke}}}, \ and\ \bibinfo {author} {\bibfnamefont {M.}~\bibnamefont
  {{Bartelmann}}},\ }\href {\doibase 10.1086/427976} {\bibfield  {journal}
  {\bibinfo  {journal} {Astrophys. J.}\ }\textbf {\bibinfo {volume} {622}},\
  \bibinfo {pages} {759} (\bibinfo {year} {2005})}\BibitemShut {NoStop}%
\bibitem [{\citenamefont {{Zuntz}}\ \emph {et~al.}(2013)\citenamefont
  {{Zuntz}}, \citenamefont {{Kacprzak}}, \citenamefont {{Voigt}}, \citenamefont
  {{Hirsch}}, \citenamefont {{Rowe}},\ and\ \citenamefont
  {{Bridle}}}]{zuntz2013}%
  \BibitemOpen
  \bibfield  {author} {\bibinfo {author} {\bibfnamefont {J.}~\bibnamefont
  {{Zuntz}}}, \bibinfo {author} {\bibfnamefont {T.}~\bibnamefont {{Kacprzak}}},
  \bibinfo {author} {\bibfnamefont {L.}~\bibnamefont {{Voigt}}}, \bibinfo
  {author} {\bibfnamefont {M.}~\bibnamefont {{Hirsch}}}, \bibinfo {author}
  {\bibfnamefont {B.}~\bibnamefont {{Rowe}}}, \ and\ \bibinfo {author}
  {\bibfnamefont {S.}~\bibnamefont {{Bridle}}},\ }\href {\doibase
  10.1093/mnras/stt1125} {\bibfield  {journal} {\bibinfo  {journal} {Mon. Not.
  R. Astron. Soc.}\ }\textbf {\bibinfo {volume} {434}},\ \bibinfo {pages}
  {1604} (\bibinfo {year} {2013})}\BibitemShut {NoStop}%
\bibitem [{\citenamefont {{Ben{\'{\i}}tez}}(2000)}]{Benitez2000}%
  \BibitemOpen
  \bibfield  {author} {\bibinfo {author} {\bibfnamefont {N.}~\bibnamefont
  {{Ben{\'{\i}}tez}}},\ }\href {\doibase 10.1086/308947} {\bibfield  {journal}
  {\bibinfo  {journal} {Astrophys. J.}\ }\textbf {\bibinfo {volume} {536}},\
  \bibinfo {pages} {571} (\bibinfo {year} {2000})}\BibitemShut {NoStop}%
\bibitem [{\citenamefont {{Chang}}\ \emph {et~al.}(2013)\citenamefont
  {{Chang}}, \citenamefont {{Jarvis}}, \citenamefont {{Jain}}, \citenamefont
  {{Kahn}}, \citenamefont {{Kirkby}}, \citenamefont {{Connolly}}, \citenamefont
  {{Krughoff}}, \citenamefont {{Peng}},\ and\ \citenamefont
  {{Peterson}}}]{C13}%
  \BibitemOpen
  \bibfield  {author} {\bibinfo {author} {\bibfnamefont {C.}~\bibnamefont
  {{Chang}}}, \bibinfo {author} {\bibfnamefont {M.}~\bibnamefont {{Jarvis}}},
  \bibinfo {author} {\bibfnamefont {B.}~\bibnamefont {{Jain}}}, \bibinfo
  {author} {\bibfnamefont {S.~M.}\ \bibnamefont {{Kahn}}}, \bibinfo {author}
  {\bibfnamefont {D.}~\bibnamefont {{Kirkby}}}, \bibinfo {author}
  {\bibfnamefont {A.}~\bibnamefont {{Connolly}}}, \bibinfo {author}
  {\bibfnamefont {S.}~\bibnamefont {{Krughoff}}}, \bibinfo {author}
  {\bibfnamefont {E.-H.}\ \bibnamefont {{Peng}}}, \ and\ \bibinfo {author}
  {\bibfnamefont {J.~R.}\ \bibnamefont {{Peterson}}},\ }\href {\doibase
  10.1093/mnras/stt1156} {\bibfield  {journal} {\bibinfo  {journal} {Mon. Not.
  R. Astron. Soc.}\ }\textbf {\bibinfo {volume} {434}},\ \bibinfo {pages}
  {2121} (\bibinfo {year} {2013})}\BibitemShut {NoStop}%
\bibitem [{\citenamefont {{Laigle}}\ \emph {et~al.}(2016)\citenamefont
  {{Laigle}}, \citenamefont {{McCracken}}, \citenamefont {{Ilbert}},
  \citenamefont {{Hsieh}}, \citenamefont {{Davidzon}}, \citenamefont
  {{Capak}},\ and\ \citenamefont {{others}}}]{Laigle}%
  \BibitemOpen
  \bibfield  {author} {\bibinfo {author} {\bibfnamefont {C.}~\bibnamefont
  {{Laigle}}}, \bibinfo {author} {\bibfnamefont {H.~J.}\ \bibnamefont
  {{McCracken}}}, \bibinfo {author} {\bibfnamefont {O.}~\bibnamefont
  {{Ilbert}}}, \bibinfo {author} {\bibfnamefont {B.~C.}\ \bibnamefont
  {{Hsieh}}}, \bibinfo {author} {\bibfnamefont {I.}~\bibnamefont {{Davidzon}}},
  \bibinfo {author} {\bibfnamefont {P.}~\bibnamefont {{Capak}}}, \ and\
  \bibinfo {author} {\bibnamefont {{others}}},\ }\href {\doibase
  10.3847/0067-0049/224/2/24} {\bibfield  {journal} {\bibinfo  {journal}
  {Astrophys. J.s}\ }\textbf {\bibinfo {volume} {224}},\ \bibinfo {eid} {24}
  (\bibinfo {year} {2016})}\BibitemShut {NoStop}%
\bibitem [{\citenamefont {{DeRose}}\ \emph {et~al.}(2017)\citenamefont
  {{DeRose}}, \citenamefont {{Wechsler}}, \citenamefont {{Rykoff}} \emph
  {et~al.}}]{DeRose2017}%
  \BibitemOpen
  \bibfield  {author} {\bibinfo {author} {\bibfnamefont {J.}~\bibnamefont
  {{DeRose}}}, \bibinfo {author} {\bibfnamefont {R.}~\bibnamefont
  {{Wechsler}}}, \bibinfo {author} {\bibfnamefont {E.}~\bibnamefont
  {{Rykoff}}},  \emph {et~al.},\ }\href@noop {} {\bibfield  {journal} {\bibinfo
   {journal} {in prep.}\ } (\bibinfo {year} {2017})}\BibitemShut {NoStop}%
\bibitem [{\citenamefont {{Wechsler}}\ \emph {et~al.}(2017)\citenamefont
  {{Wechsler}}, \citenamefont {{DeRose}}, \citenamefont {{Busha}} \emph
  {et~al.}}]{Wechsler2017}%
  \BibitemOpen
  \bibfield  {author} {\bibinfo {author} {\bibfnamefont {R.}~\bibnamefont
  {{Wechsler}}}, \bibinfo {author} {\bibfnamefont {J.}~\bibnamefont
  {{DeRose}}}, \bibinfo {author} {\bibfnamefont {M.}~\bibnamefont {{Busha}}},
  \emph {et~al.},\ }\href@noop {} {\bibfield  {journal} {\bibinfo  {journal}
  {in prep.}\ } (\bibinfo {year} {2017})}\BibitemShut {NoStop}%
\bibitem [{\citenamefont {{Springel}}(2005)}]{springel2005}%
  \BibitemOpen
  \bibfield  {author} {\bibinfo {author} {\bibfnamefont {V.}~\bibnamefont
  {{Springel}}},\ }\href {\doibase 10.1111/j.1365-2966.2005.09655.x} {\bibfield
   {journal} {\bibinfo  {journal} {Mon. Not. R. Astron. Soc.}\ }\textbf
  {\bibinfo {volume} {364}},\ \bibinfo {pages} {1105} (\bibinfo {year}
  {2005})}\BibitemShut {NoStop}%
\bibitem [{\citenamefont {{Crocce}}\ \emph {et~al.}(2006)\citenamefont
  {{Crocce}}, \citenamefont {{Pueblas}},\ and\ \citenamefont
  {{Scoccimarro}}}]{crocce2006}%
  \BibitemOpen
  \bibfield  {author} {\bibinfo {author} {\bibfnamefont {M.}~\bibnamefont
  {{Crocce}}}, \bibinfo {author} {\bibfnamefont {S.}~\bibnamefont {{Pueblas}}},
  \ and\ \bibinfo {author} {\bibfnamefont {R.}~\bibnamefont {{Scoccimarro}}},\
  }\href {\doibase 10.1111/j.1365-2966.2006.11040.x} {\bibfield  {journal}
  {\bibinfo  {journal} {Mon. Not. R. Astron. Soc.}\ }\textbf {\bibinfo {volume}
  {373}},\ \bibinfo {pages} {369} (\bibinfo {year} {2006})}\BibitemShut
  {NoStop}%
\bibitem [{\citenamefont {{Cooper}}\ \emph {et~al.}(2011)\citenamefont
  {{Cooper}}, \citenamefont {{Aird}}, \citenamefont {{Coil}}, \citenamefont
  {{Davis}}, \citenamefont {{Faber}}, \citenamefont {{Juneau}}, \citenamefont
  {{Lotz}}, \citenamefont {{Nandra}}, \citenamefont {{Newman}}, \citenamefont
  {{Willmer}},\ and\ \citenamefont {{Yan}}}]{Cooper2011}%
  \BibitemOpen
  \bibfield  {author} {\bibinfo {author} {\bibfnamefont {M.~C.}\ \bibnamefont
  {{Cooper}}}, \bibinfo {author} {\bibfnamefont {J.~A.}\ \bibnamefont
  {{Aird}}}, \bibinfo {author} {\bibfnamefont {A.~L.}\ \bibnamefont {{Coil}}},
  \bibinfo {author} {\bibfnamefont {M.}~\bibnamefont {{Davis}}}, \bibinfo
  {author} {\bibfnamefont {S.~M.}\ \bibnamefont {{Faber}}}, \bibinfo {author}
  {\bibfnamefont {S.}~\bibnamefont {{Juneau}}}, \bibinfo {author}
  {\bibfnamefont {J.~M.}\ \bibnamefont {{Lotz}}}, \bibinfo {author}
  {\bibfnamefont {K.}~\bibnamefont {{Nandra}}}, \bibinfo {author}
  {\bibfnamefont {J.~A.}\ \bibnamefont {{Newman}}}, \bibinfo {author}
  {\bibfnamefont {C.~N.~A.}\ \bibnamefont {{Willmer}}}, \ and\ \bibinfo
  {author} {\bibfnamefont {R.}~\bibnamefont {{Yan}}},\ }\href {\doibase
  10.1088/0067-0049/193/1/14} {\bibfield  {journal} {\bibinfo  {journal}
  {Astrophys. J.s}\ }\textbf {\bibinfo {volume} {193}},\ \bibinfo {eid} {14}
  (\bibinfo {year} {2011})}\BibitemShut {NoStop}%
\bibitem [{\citenamefont {{Becker}}(2013)}]{Becker2013b}%
  \BibitemOpen
  \bibfield  {author} {\bibinfo {author} {\bibfnamefont {M.~R.}\ \bibnamefont
  {{Becker}}},\ }\href {\doibase 10.1093/mnras/stt1352} {\bibfield  {journal}
  {\bibinfo  {journal} {Mon. Not. R. Astron. Soc.}\ }\textbf {\bibinfo {volume}
  {435}},\ \bibinfo {pages} {115} (\bibinfo {year} {2013})}\BibitemShut
  {NoStop}%
\bibitem [{\citenamefont {{Fosalba}}\ \emph {et~al.}(2008)\citenamefont
  {{Fosalba}}, \citenamefont {{Gazta{\~n}aga}}, \citenamefont {{Castander}},\
  and\ \citenamefont {{Manera}}}]{2008MNRAS.391..435F}%
  \BibitemOpen
  \bibfield  {author} {\bibinfo {author} {\bibfnamefont {P.}~\bibnamefont
  {{Fosalba}}}, \bibinfo {author} {\bibfnamefont {E.}~\bibnamefont
  {{Gazta{\~n}aga}}}, \bibinfo {author} {\bibfnamefont {F.~J.}\ \bibnamefont
  {{Castander}}}, \ and\ \bibinfo {author} {\bibfnamefont {M.}~\bibnamefont
  {{Manera}}},\ }\href {\doibase 10.1111/j.1365-2966.2008.13910.x} {\bibfield
  {journal} {\bibinfo  {journal} {Mon. Not. R. Astron. Soc.}\ }\textbf
  {\bibinfo {volume} {391}},\ \bibinfo {pages} {435} (\bibinfo {year}
  {2008})}\BibitemShut {NoStop}%
\bibitem [{\citenamefont {{Fosalba}}\ \emph
  {et~al.}(2015{\natexlab{a}})\citenamefont {{Fosalba}}, \citenamefont
  {{Gazta{\~n}aga}}, \citenamefont {{Castander}},\ and\ \citenamefont
  {{Crocce}}}]{MICEIII}%
  \BibitemOpen
  \bibfield  {author} {\bibinfo {author} {\bibfnamefont {P.}~\bibnamefont
  {{Fosalba}}}, \bibinfo {author} {\bibfnamefont {E.}~\bibnamefont
  {{Gazta{\~n}aga}}}, \bibinfo {author} {\bibfnamefont {F.~J.}\ \bibnamefont
  {{Castander}}}, \ and\ \bibinfo {author} {\bibfnamefont {M.}~\bibnamefont
  {{Crocce}}},\ }\href {\doibase 10.1093/mnras/stu2464} {\bibfield  {journal}
  {\bibinfo  {journal} {Mon. Not. R. Astron. Soc.}\ }\textbf {\bibinfo {volume}
  {447}},\ \bibinfo {pages} {1319} (\bibinfo {year}
  {2015}{\natexlab{a}})}\BibitemShut {NoStop}%
\bibitem [{\citenamefont {{Blanton}}\ \emph {et~al.}(2003)\citenamefont
  {{Blanton}} \emph {et~al.}}]{2003ApJ...592..819B}%
  \BibitemOpen
  \bibfield  {author} {\bibinfo {author} {\bibfnamefont {M.~R.}\ \bibnamefont
  {{Blanton}}} \emph {et~al.},\ }\href {\doibase 10.1086/375776} {\bibfield
  {journal} {\bibinfo  {journal} {Astrophys. J.}\ }\textbf {\bibinfo {volume}
  {592}},\ \bibinfo {pages} {819} (\bibinfo {year} {2003})}\BibitemShut
  {NoStop}%
\bibitem [{\citenamefont {{Zehavi}}\ \emph {et~al.}(2011)\citenamefont
  {{Zehavi}} \emph {et~al.}}]{2011ApJ...736...59Z}%
  \BibitemOpen
  \bibfield  {author} {\bibinfo {author} {\bibfnamefont {I.}~\bibnamefont
  {{Zehavi}}} \emph {et~al.},\ }\href {\doibase 10.1088/0004-637X/736/1/59}
  {\bibfield  {journal} {\bibinfo  {journal} {Astrophys. J.}\ }\textbf
  {\bibinfo {volume} {736}},\ \bibinfo {eid} {59} (\bibinfo {year}
  {2011})}\BibitemShut {NoStop}%
\bibitem [{\citenamefont {{Ilbert}}\ \emph {et~al.}(2009)\citenamefont
  {{Ilbert}} \emph {et~al.}}]{ilbert2009}%
  \BibitemOpen
  \bibfield  {author} {\bibinfo {author} {\bibfnamefont {O.}~\bibnamefont
  {{Ilbert}}} \emph {et~al.},\ }\href {\doibase 10.1088/0004-637X/690/2/1236}
  {\bibfield  {journal} {\bibinfo  {journal} {Astrophys. J.}\ }\textbf
  {\bibinfo {volume} {690}},\ \bibinfo {pages} {1236} (\bibinfo {year}
  {2009})}\BibitemShut {NoStop}%
\bibitem [{\citenamefont {{Fosalba}}\ \emph
  {et~al.}(2015{\natexlab{b}})\citenamefont {{Fosalba}}, \citenamefont
  {{Crocce}}, \citenamefont {{Gazta{\~n}aga}},\ and\ \citenamefont
  {{Castander}}}]{MICEI}%
  \BibitemOpen
  \bibfield  {author} {\bibinfo {author} {\bibfnamefont {P.}~\bibnamefont
  {{Fosalba}}}, \bibinfo {author} {\bibfnamefont {M.}~\bibnamefont {{Crocce}}},
  \bibinfo {author} {\bibfnamefont {E.}~\bibnamefont {{Gazta{\~n}aga}}}, \ and\
  \bibinfo {author} {\bibfnamefont {F.~J.}\ \bibnamefont {{Castander}}},\
  }\href {\doibase 10.1093/mnras/stv138} {\bibfield  {journal} {\bibinfo
  {journal} {Mon. Not. R. Astron. Soc.}\ }\textbf {\bibinfo {volume} {448}},\
  \bibinfo {pages} {2987} (\bibinfo {year} {2015}{\natexlab{b}})}\BibitemShut
  {NoStop}%
\bibitem [{\citenamefont {{Crocce}}\ \emph {et~al.}(2015)\citenamefont
  {{Crocce}}, \citenamefont {{Castander}}, \citenamefont {{Gazta{\~n}aga}},
  \citenamefont {{Fosalba}},\ and\ \citenamefont {{Carretero}}}]{MICEII}%
  \BibitemOpen
  \bibfield  {author} {\bibinfo {author} {\bibfnamefont {M.}~\bibnamefont
  {{Crocce}}}, \bibinfo {author} {\bibfnamefont {F.~J.}\ \bibnamefont
  {{Castander}}}, \bibinfo {author} {\bibfnamefont {E.}~\bibnamefont
  {{Gazta{\~n}aga}}}, \bibinfo {author} {\bibfnamefont {P.}~\bibnamefont
  {{Fosalba}}}, \ and\ \bibinfo {author} {\bibfnamefont {J.}~\bibnamefont
  {{Carretero}}},\ }\href {\doibase 10.1093/mnras/stv1708} {\bibfield
  {journal} {\bibinfo  {journal} {Mon. Not. R. Astron. Soc.}\ }\textbf
  {\bibinfo {volume} {453}},\ \bibinfo {pages} {1513} (\bibinfo {year}
  {2015})}\BibitemShut {NoStop}%
\bibitem [{\citenamefont {{Carretero}}\ \emph {et~al.}(2015)\citenamefont
  {{Carretero}}, \citenamefont {{Castander}}, \citenamefont {{Gazta{\~n}aga}},
  \citenamefont {{Crocce}},\ and\ \citenamefont
  {{Fosalba}}}]{2015MNRAS.447..646C}%
  \BibitemOpen
  \bibfield  {author} {\bibinfo {author} {\bibfnamefont {J.}~\bibnamefont
  {{Carretero}}}, \bibinfo {author} {\bibfnamefont {F.~J.}\ \bibnamefont
  {{Castander}}}, \bibinfo {author} {\bibfnamefont {E.}~\bibnamefont
  {{Gazta{\~n}aga}}}, \bibinfo {author} {\bibfnamefont {M.}~\bibnamefont
  {{Crocce}}}, \ and\ \bibinfo {author} {\bibfnamefont {P.}~\bibnamefont
  {{Fosalba}}},\ }\href {\doibase 10.1093/mnras/stu2402} {\bibfield  {journal}
  {\bibinfo  {journal} {Mon. Not. R. Astron. Soc.}\ }\textbf {\bibinfo {volume}
  {447}},\ \bibinfo {pages} {646} (\bibinfo {year} {2015})}\BibitemShut
  {NoStop}%
\bibitem [{\citenamefont {{Coles}}\ and\ \citenamefont
  {{Jones}}(1991)}]{ColesJones1991}%
  \BibitemOpen
  \bibfield  {author} {\bibinfo {author} {\bibfnamefont {P.}~\bibnamefont
  {{Coles}}}\ and\ \bibinfo {author} {\bibfnamefont {B.}~\bibnamefont
  {{Jones}}},\ }\href {\doibase 10.1093/mnras/248.1.1} {\bibfield  {journal}
  {\bibinfo  {journal} {"Mon. Not. R. Astron. Soc."}\ }\textbf {\bibinfo
  {volume} {248}},\ \bibinfo {pages} {1} (\bibinfo {year} {1991})}\BibitemShut
  {NoStop}%
\bibitem [{\citenamefont {Kayo}\ \emph {et~al.}(2001)\citenamefont {Kayo},
  \citenamefont {Taruya},\ and\ \citenamefont {Suto}}]{Kayoetal2001}%
  \BibitemOpen
  \bibfield  {author} {\bibinfo {author} {\bibfnamefont {I.}~\bibnamefont
  {Kayo}}, \bibinfo {author} {\bibfnamefont {A.}~\bibnamefont {Taruya}}, \ and\
  \bibinfo {author} {\bibfnamefont {Y.}~\bibnamefont {Suto}},\ }\href {\doibase
  10.1086/323227} {\bibfield  {journal} {\bibinfo  {journal} {Astrophys. J.}\
  }\textbf {\bibinfo {volume} {561}},\ \bibinfo {pages} {22} (\bibinfo {year}
  {2001})}\BibitemShut {NoStop}%
\bibitem [{\citenamefont {Lahav}\ and\ \citenamefont
  {Suto}(2004)}]{LahavSuto2004}%
  \BibitemOpen
  \bibfield  {author} {\bibinfo {author} {\bibfnamefont {O.}~\bibnamefont
  {Lahav}}\ and\ \bibinfo {author} {\bibfnamefont {Y.}~\bibnamefont {Suto}},\
  }\href {\doibase 10.12942/lrr-2004-8} {\bibfield  {journal} {\bibinfo
  {journal} {Living Rev. Rel.}\ }\textbf {\bibinfo {volume} {7}},\ \bibinfo
  {pages} {8} (\bibinfo {year} {2004})}\BibitemShut {NoStop}%
\bibitem [{\citenamefont {{Hilbert}}\ \emph {et~al.}(2011)\citenamefont
  {{Hilbert}}, \citenamefont {{Hartlap}},\ and\ \citenamefont
  {{Schneider}}}]{Hilbertetal2011}%
  \BibitemOpen
  \bibfield  {author} {\bibinfo {author} {\bibfnamefont {S.}~\bibnamefont
  {{Hilbert}}}, \bibinfo {author} {\bibfnamefont {J.}~\bibnamefont
  {{Hartlap}}}, \ and\ \bibinfo {author} {\bibfnamefont {P.}~\bibnamefont
  {{Schneider}}},\ }\href {\doibase 10.1051/0004-6361/201117294} {\bibfield
  {journal} {\bibinfo  {journal} {Astron. Astrophys.}\ }\textbf {\bibinfo
  {volume} {536}},\ \bibinfo {eid} {A85} (\bibinfo {year} {2011})}\BibitemShut
  {NoStop}%
\bibitem [{\citenamefont {Xavier}\ \emph {et~al.}(2016)\citenamefont {Xavier},
  \citenamefont {Abdalla},\ and\ \citenamefont {Joachimi}}]{Xavieretal2016}%
  \BibitemOpen
  \bibfield  {author} {\bibinfo {author} {\bibfnamefont {H.~S.}\ \bibnamefont
  {Xavier}}, \bibinfo {author} {\bibfnamefont {F.~B.}\ \bibnamefont {Abdalla}},
  \ and\ \bibinfo {author} {\bibfnamefont {B.}~\bibnamefont {Joachimi}},\
  }\href {\doibase 10.1093/mnras/stw874, 10.1093/mnras/459/4/3693} {\bibfield
  {journal} {\bibinfo  {journal} {Mon. Not. Roy. Astron. Soc.}\ }\textbf
  {\bibinfo {volume} {459}},\ \bibinfo {pages} {3693} (\bibinfo {year}
  {2016})}\BibitemShut {NoStop}%
\bibitem [{\citenamefont {{Schneider}}\ \emph {et~al.}(2002)\citenamefont
  {{Schneider}}, \citenamefont {{van Waerbeke}},\ and\ \citenamefont
  {{Mellier}}}]{schneider02}%
  \BibitemOpen
  \bibfield  {author} {\bibinfo {author} {\bibfnamefont {P.}~\bibnamefont
  {{Schneider}}}, \bibinfo {author} {\bibfnamefont {L.}~\bibnamefont {{van
  Waerbeke}}}, \ and\ \bibinfo {author} {\bibfnamefont {Y.}~\bibnamefont
  {{Mellier}}},\ }\href {\doibase 10.1051/0004-6361:20020626} {\bibfield
  {journal} {\bibinfo  {journal} {\aap}\ }\textbf {\bibinfo {volume} {389}},\
  \bibinfo {pages} {729} (\bibinfo {year} {2002})}\BibitemShut {NoStop}%
\bibitem [{\citenamefont {{Jarvis}}\ \emph {et~al.}(2004)\citenamefont
  {{Jarvis}}, \citenamefont {{Bernstein}},\ and\ \citenamefont
  {{Jain}}}]{treecorr}%
  \BibitemOpen
  \bibfield  {author} {\bibinfo {author} {\bibfnamefont {M.}~\bibnamefont
  {{Jarvis}}}, \bibinfo {author} {\bibfnamefont {G.}~\bibnamefont
  {{Bernstein}}}, \ and\ \bibinfo {author} {\bibfnamefont {B.}~\bibnamefont
  {{Jain}}},\ }\href {\doibase 10.1111/j.1365-2966.2004.07926.x} {\bibfield
  {journal} {\bibinfo  {journal} {Mon. Not. R. Astron. Soc.}\ }\textbf
  {\bibinfo {volume} {352}},\ \bibinfo {pages} {338} (\bibinfo {year}
  {2004})}\BibitemShut {NoStop}%
\bibitem [{\citenamefont {{Krause}}\ and\ \citenamefont
  {{Eifler}}(2017)}]{ke16}%
  \BibitemOpen
  \bibfield  {author} {\bibinfo {author} {\bibfnamefont {E.}~\bibnamefont
  {{Krause}}}\ and\ \bibinfo {author} {\bibfnamefont {T.}~\bibnamefont
  {{Eifler}}},\ }\href {\doibase 10.1093/mnras/stx1261} {\bibfield  {journal}
  {\bibinfo  {journal} {Mon. Not. R. Astron. Soc.}\ }\textbf {\bibinfo {volume}
  {470}},\ \bibinfo {pages} {2100} (\bibinfo {year} {2017})}\BibitemShut
  {NoStop}%
\bibitem [{\citenamefont {{Eifler}}\ \emph {et~al.}(2009)\citenamefont
  {{Eifler}}, \citenamefont {{Schneider}},\ and\ \citenamefont
  {{Hartlap}}}]{2009A&A...502..721E}%
  \BibitemOpen
  \bibfield  {author} {\bibinfo {author} {\bibfnamefont {T.}~\bibnamefont
  {{Eifler}}}, \bibinfo {author} {\bibfnamefont {P.}~\bibnamefont
  {{Schneider}}}, \ and\ \bibinfo {author} {\bibfnamefont {J.}~\bibnamefont
  {{Hartlap}}},\ }\href {\doibase 10.1051/0004-6361/200811276} {\bibfield
  {journal} {\bibinfo  {journal} {Astron. Astrophys.}\ }\textbf {\bibinfo
  {volume} {502}},\ \bibinfo {pages} {721} (\bibinfo {year}
  {2009})}\BibitemShut {NoStop}%
\bibitem [{\citenamefont {{Klein}}\ and\ \citenamefont
  {{Roodman}}(2005)}]{2005ARNPS..55..141K}%
  \BibitemOpen
  \bibfield  {author} {\bibinfo {author} {\bibfnamefont {J.~R.}\ \bibnamefont
  {{Klein}}}\ and\ \bibinfo {author} {\bibfnamefont {A.}~\bibnamefont
  {{Roodman}}},\ }\href {\doibase 10.1146/annurev.nucl.55.090704.151521}
  {\bibfield  {journal} {\bibinfo  {journal} {Ann. Rev. Nucl. Part. Sci.}\
  }\textbf {\bibinfo {volume} {55}},\ \bibinfo {pages} {141} (\bibinfo {year}
  {2005})}\BibitemShut {NoStop}%
\bibitem [{\citenamefont {{Kaiser}}(1992)}]{limberkaiser1}%
  \BibitemOpen
  \bibfield  {author} {\bibinfo {author} {\bibfnamefont {N.}~\bibnamefont
  {{Kaiser}}},\ }\href {\doibase 10.1086/171151} {\bibfield  {journal}
  {\bibinfo  {journal} {Astrophys. J.}\ }\textbf {\bibinfo {volume} {388}},\
  \bibinfo {pages} {272} (\bibinfo {year} {1992})}\BibitemShut {NoStop}%
\bibitem [{\citenamefont {{Kaiser}}(1998)}]{limberkaiser2}%
  \BibitemOpen
  \bibfield  {author} {\bibinfo {author} {\bibfnamefont {N.}~\bibnamefont
  {{Kaiser}}},\ }\href {\doibase 10.1086/305515} {\bibfield  {journal}
  {\bibinfo  {journal} {Astrophys. J.}\ }\textbf {\bibinfo {volume} {498}},\
  \bibinfo {pages} {26} (\bibinfo {year} {1998})}\BibitemShut {NoStop}%
\bibitem [{\citenamefont {{Limber}}(1953)}]{limber}%
  \BibitemOpen
  \bibfield  {author} {\bibinfo {author} {\bibfnamefont {D.~N.}\ \bibnamefont
  {{Limber}}},\ }\href {\doibase 10.1086/145672} {\bibfield  {journal}
  {\bibinfo  {journal} {Astrophys. J.}\ }\textbf {\bibinfo {volume} {117}},\
  \bibinfo {pages} {134} (\bibinfo {year} {1953})}\BibitemShut {NoStop}%
\bibitem [{\citenamefont {LoVerde}\ and\ \citenamefont
  {Afshordi}(2008)}]{PhysRevD.78.123506}%
  \BibitemOpen
  \bibfield  {author} {\bibinfo {author} {\bibfnamefont {M.}~\bibnamefont
  {LoVerde}}\ and\ \bibinfo {author} {\bibfnamefont {N.}~\bibnamefont
  {Afshordi}},\ }\href {\doibase 10.1103/PhysRevD.78.123506} {\bibfield
  {journal} {\bibinfo  {journal} {Phys. Rev. D}\ }\textbf {\bibinfo {volume}
  {78}},\ \bibinfo {pages} {123506} (\bibinfo {year} {2008})}\BibitemShut
  {NoStop}%
\bibitem [{\citenamefont {{Eifler}}\ \emph {et~al.}(2014)\citenamefont
  {{Eifler}}, \citenamefont {{Krause}}, \citenamefont {{Schneider}},\ and\
  \citenamefont {{Honscheid}}}]{cosmolike}%
  \BibitemOpen
  \bibfield  {author} {\bibinfo {author} {\bibfnamefont {T.}~\bibnamefont
  {{Eifler}}}, \bibinfo {author} {\bibfnamefont {E.}~\bibnamefont {{Krause}}},
  \bibinfo {author} {\bibfnamefont {P.}~\bibnamefont {{Schneider}}}, \ and\
  \bibinfo {author} {\bibfnamefont {K.}~\bibnamefont {{Honscheid}}},\ }\href
  {\doibase 10.1093/mnras/stu251} {\bibfield  {journal} {\bibinfo  {journal}
  {Mon. Not. R. Astron. Soc.}\ }\textbf {\bibinfo {volume} {440}},\ \bibinfo
  {pages} {1379} (\bibinfo {year} {2014})}\BibitemShut {NoStop}%
\bibitem [{\citenamefont {Zuntz}\ \emph {et~al.}(2015)\citenamefont {Zuntz},
  \citenamefont {Paterno}, \citenamefont {Jennings}, \citenamefont {Rudd},
  \citenamefont {Manzotti}, \citenamefont {Dodelson}, \citenamefont {Bridle},
  \citenamefont {Sehrish},\ and\ \citenamefont {Kowalkowski}}]{cosmosis}%
  \BibitemOpen
  \bibfield  {author} {\bibinfo {author} {\bibfnamefont {J.}~\bibnamefont
  {Zuntz}}, \bibinfo {author} {\bibfnamefont {M.}~\bibnamefont {Paterno}},
  \bibinfo {author} {\bibfnamefont {E.}~\bibnamefont {Jennings}}, \bibinfo
  {author} {\bibfnamefont {D.}~\bibnamefont {Rudd}}, \bibinfo {author}
  {\bibfnamefont {A.}~\bibnamefont {Manzotti}}, \bibinfo {author}
  {\bibfnamefont {S.}~\bibnamefont {Dodelson}}, \bibinfo {author}
  {\bibfnamefont {S.}~\bibnamefont {Bridle}}, \bibinfo {author} {\bibfnamefont
  {S.}~\bibnamefont {Sehrish}}, \ and\ \bibinfo {author} {\bibfnamefont
  {J.}~\bibnamefont {Kowalkowski}},\ }\href {\doibase
  http://dx.doi.org/10.1016/j.ascom.2015.05.005} {\bibfield  {journal}
  {\bibinfo  {journal} {Astronomy and Computing}\ }\textbf {\bibinfo {volume}
  {12}},\ \bibinfo {pages} {45 } (\bibinfo {year} {2015})}\BibitemShut
  {NoStop}%
\bibitem [{\citenamefont {{Blas}}\ \emph {et~al.}(2011)\citenamefont {{Blas}},
  \citenamefont {{Lesgourgues}},\ and\ \citenamefont {{Tram}}}]{class}%
  \BibitemOpen
  \bibfield  {author} {\bibinfo {author} {\bibfnamefont {D.}~\bibnamefont
  {{Blas}}}, \bibinfo {author} {\bibfnamefont {J.}~\bibnamefont
  {{Lesgourgues}}}, \ and\ \bibinfo {author} {\bibfnamefont {T.}~\bibnamefont
  {{Tram}}},\ }\href {\doibase 10.1088/1475-7516/2011/07/034} {\bibfield
  {journal} {\bibinfo  {journal} {\jcap}\ }\textbf {\bibinfo {volume} {7}},\
  \bibinfo {eid} {034} (\bibinfo {year} {2011})}\BibitemShut {NoStop}%
\bibitem [{\citenamefont {{Lewis}}\ \emph {et~al.}(2000)\citenamefont
  {{Lewis}}, \citenamefont {{Challinor}},\ and\ \citenamefont
  {{Lasenby}}}]{Lewis2000}%
  \BibitemOpen
  \bibfield  {author} {\bibinfo {author} {\bibfnamefont {A.}~\bibnamefont
  {{Lewis}}}, \bibinfo {author} {\bibfnamefont {A.}~\bibnamefont
  {{Challinor}}}, \ and\ \bibinfo {author} {\bibfnamefont {A.}~\bibnamefont
  {{Lasenby}}},\ }\href {\doibase 10.1086/309179} {\bibfield  {journal}
  {\bibinfo  {journal} {Astrophys. J.}\ }\textbf {\bibinfo {volume} {538}},\
  \bibinfo {pages} {473} (\bibinfo {year} {2000})}\BibitemShut {NoStop}%
\bibitem [{\citenamefont {{Howlett}}\ \emph {et~al.}(2012)\citenamefont
  {{Howlett}}, \citenamefont {{Lewis}}, \citenamefont {{Hall}},\ and\
  \citenamefont {{Challinor}}}]{howlett2012}%
  \BibitemOpen
  \bibfield  {author} {\bibinfo {author} {\bibfnamefont {C.}~\bibnamefont
  {{Howlett}}}, \bibinfo {author} {\bibfnamefont {A.}~\bibnamefont {{Lewis}}},
  \bibinfo {author} {\bibfnamefont {A.}~\bibnamefont {{Hall}}}, \ and\ \bibinfo
  {author} {\bibfnamefont {A.}~\bibnamefont {{Challinor}}},\ }\href {\doibase
  10.1088/1475-7516/2012/04/027} {\bibfield  {journal} {\bibinfo  {journal}
  {\jcap}\ }\textbf {\bibinfo {volume} {4}},\ \bibinfo {eid} {027} (\bibinfo
  {year} {2012})}\BibitemShut {NoStop}%
\bibitem [{\citenamefont {{Feroz}}\ \emph {et~al.}(2009)\citenamefont
  {{Feroz}}, \citenamefont {{Hobson}},\ and\ \citenamefont {{Bridges}}}]{mn1}%
  \BibitemOpen
  \bibfield  {author} {\bibinfo {author} {\bibfnamefont {F.}~\bibnamefont
  {{Feroz}}}, \bibinfo {author} {\bibfnamefont {M.~P.}\ \bibnamefont
  {{Hobson}}}, \ and\ \bibinfo {author} {\bibfnamefont {M.}~\bibnamefont
  {{Bridges}}},\ }\href {\doibase 10.1111/j.1365-2966.2009.14548.x} {\bibfield
  {journal} {\bibinfo  {journal} {Mon. Not. R. Astron. Soc.}\ }\textbf
  {\bibinfo {volume} {398}},\ \bibinfo {pages} {1601} (\bibinfo {year}
  {2009})}\BibitemShut {NoStop}%
\bibitem [{\citenamefont {{Feroz}}\ and\ \citenamefont {{Hobson}}(2008)}]{mn2}%
  \BibitemOpen
  \bibfield  {author} {\bibinfo {author} {\bibfnamefont {F.}~\bibnamefont
  {{Feroz}}}\ and\ \bibinfo {author} {\bibfnamefont {M.~P.}\ \bibnamefont
  {{Hobson}}},\ }\href {\doibase 10.1111/j.1365-2966.2007.12353.x} {\bibfield
  {journal} {\bibinfo  {journal} {Mon. Not. R. Astron. Soc.}\ }\textbf
  {\bibinfo {volume} {384}},\ \bibinfo {pages} {449} (\bibinfo {year}
  {2008})}\BibitemShut {NoStop}%
\bibitem [{\citenamefont {{Feroz}}\ \emph {et~al.}(2013)\citenamefont
  {{Feroz}}, \citenamefont {{Hobson}}, \citenamefont {{Cameron}},\ and\
  \citenamefont {{Pettitt}}}]{mn3}%
  \BibitemOpen
  \bibfield  {author} {\bibinfo {author} {\bibfnamefont {F.}~\bibnamefont
  {{Feroz}}}, \bibinfo {author} {\bibfnamefont {M.~P.}\ \bibnamefont
  {{Hobson}}}, \bibinfo {author} {\bibfnamefont {E.}~\bibnamefont {{Cameron}}},
  \ and\ \bibinfo {author} {\bibfnamefont {A.~N.}\ \bibnamefont {{Pettitt}}},\
  }\href@noop {} {\bibfield  {journal} {\bibinfo  {journal} {ArXiv e-prints}\ }
  (\bibinfo {year} {2013})},\ \Eprint {http://arxiv.org/abs/1306.2144}
  {arXiv:1306.2144} \BibitemShut {NoStop}%
\bibitem [{\citenamefont {{Foreman-Mackey}}\ \emph {et~al.}(2013)\citenamefont
  {{Foreman-Mackey}}, \citenamefont {{Hogg}}, \citenamefont {{Lang}},\ and\
  \citenamefont {{Goodman}}}]{emcee}%
  \BibitemOpen
  \bibfield  {author} {\bibinfo {author} {\bibfnamefont {D.}~\bibnamefont
  {{Foreman-Mackey}}}, \bibinfo {author} {\bibfnamefont {D.~W.}\ \bibnamefont
  {{Hogg}}}, \bibinfo {author} {\bibfnamefont {D.}~\bibnamefont {{Lang}}}, \
  and\ \bibinfo {author} {\bibfnamefont {J.}~\bibnamefont {{Goodman}}},\ }\href
  {\doibase 10.1086/670067} {\bibfield  {journal} {\bibinfo  {journal} {Publ.
  Astron. Soc. Pac.}\ }\textbf {\bibinfo {volume} {125}},\ \bibinfo {pages}
  {306} (\bibinfo {year} {2013})}\BibitemShut {NoStop}%
\bibitem [{\citenamefont {Patrignani}\ \emph {et~al.}(2016)\citenamefont
  {Patrignani} \emph {et~al.}}]{pdg}%
  \BibitemOpen
  \bibfield  {author} {\bibinfo {author} {\bibfnamefont {C.}~\bibnamefont
  {Patrignani}} \emph {et~al.} (\bibinfo {collaboration} {Particle Data
  Group}),\ }\href {\doibase 10.1088/1674-1137/40/10/100001} {\bibfield
  {journal} {\bibinfo  {journal} {Chin. Phys.}\ }\textbf {\bibinfo {volume}
  {C40}},\ \bibinfo {pages} {100001} (\bibinfo {year} {2016})}\BibitemShut
  {NoStop}%
\bibitem [{\citenamefont {{Planck Collaboration}}\ \emph
  {et~al.}(2016)\citenamefont {{Planck Collaboration}}, \citenamefont {{Ade}}
  \emph {et~al.}}]{planck2015cosmo}%
  \BibitemOpen
  \bibfield  {author} {\bibinfo {author} {\bibnamefont {{Planck
  Collaboration}}}, \bibinfo {author} {\bibfnamefont {P.~A.~R.}\ \bibnamefont
  {{Ade}}},  \emph {et~al.},\ }\href {\doibase 10.1051/0004-6361/201525830}
  {\bibfield  {journal} {\bibinfo  {journal} {\aap}\ }\textbf {\bibinfo
  {volume} {594}},\ \bibinfo {eid} {A13} (\bibinfo {year} {2016})},\ \Eprint
  {http://arxiv.org/abs/1502.01589} {arXiv:1502.01589} \BibitemShut {NoStop}%
\bibitem [{\citenamefont {{Huterer}}\ and\ \citenamefont
  {{Turner}}(2001)}]{HutTur01}%
  \BibitemOpen
  \bibfield  {author} {\bibinfo {author} {\bibfnamefont {D.}~\bibnamefont
  {{Huterer}}}\ and\ \bibinfo {author} {\bibfnamefont {M.~S.}\ \bibnamefont
  {{Turner}}},\ }\href {\doibase 10.1103/PhysRevD.64.123527} {\bibfield
  {journal} {\bibinfo  {journal} {Phys. Rev. D}\ }\textbf {\bibinfo {volume}
  {64}},\ \bibinfo {pages} {123527} (\bibinfo {year} {2001})}\BibitemShut
  {NoStop}%
\bibitem [{\citenamefont {{Wang}}(2008)}]{wang08}%
  \BibitemOpen
  \bibfield  {author} {\bibinfo {author} {\bibfnamefont {Y.}~\bibnamefont
  {{Wang}}},\ }\href {\doibase 10.1103/PhysRevD.77.123525} {\bibfield
  {journal} {\bibinfo  {journal} {Phys. Rev. D}\ }\textbf {\bibinfo {volume}
  {77}},\ \bibinfo {eid} {123525} (\bibinfo {year} {2008})}\BibitemShut
  {NoStop}%
\bibitem [{\citenamefont {{Smith}}\ \emph {et~al.}(2003)\citenamefont
  {{Smith}}, \citenamefont {{Peacock}}, \citenamefont {{Jenkins}},
  \citenamefont {{White}}, \citenamefont {{Frenk}}, \citenamefont {{Pearce}},
  \citenamefont {{Thomas}}, \citenamefont {{Efstathiou}},\ and\ \citenamefont
  {{Couchman}}}]{SPJ+03}%
  \BibitemOpen
  \bibfield  {author} {\bibinfo {author} {\bibfnamefont {R.~E.}\ \bibnamefont
  {{Smith}}}, \bibinfo {author} {\bibfnamefont {J.~A.}\ \bibnamefont
  {{Peacock}}}, \bibinfo {author} {\bibfnamefont {A.}~\bibnamefont
  {{Jenkins}}}, \bibinfo {author} {\bibfnamefont {S.~D.~M.}\ \bibnamefont
  {{White}}}, \bibinfo {author} {\bibfnamefont {C.~S.}\ \bibnamefont
  {{Frenk}}}, \bibinfo {author} {\bibfnamefont {F.~R.}\ \bibnamefont
  {{Pearce}}}, \bibinfo {author} {\bibfnamefont {P.~A.}\ \bibnamefont
  {{Thomas}}}, \bibinfo {author} {\bibfnamefont {G.}~\bibnamefont
  {{Efstathiou}}}, \ and\ \bibinfo {author} {\bibfnamefont {H.~M.~P.}\
  \bibnamefont {{Couchman}}},\ }\href {\doibase
  10.1046/j.1365-8711.2003.06503.x} {\bibfield  {journal} {\bibinfo  {journal}
  {Mon. Not. R. Astron. Soc.}\ }\textbf {\bibinfo {volume} {341}},\ \bibinfo
  {pages} {1311} (\bibinfo {year} {2003})}\BibitemShut {NoStop}%
\bibitem [{\citenamefont {{Takahashi}}\ \emph {et~al.}(2012)\citenamefont
  {{Takahashi}}, \citenamefont {{Sato}}, \citenamefont {{Nishimichi}},
  \citenamefont {{Taruya}},\ and\ \citenamefont {{Oguri}}}]{takahashi2012}%
  \BibitemOpen
  \bibfield  {author} {\bibinfo {author} {\bibfnamefont {R.}~\bibnamefont
  {{Takahashi}}}, \bibinfo {author} {\bibfnamefont {M.}~\bibnamefont {{Sato}}},
  \bibinfo {author} {\bibfnamefont {T.}~\bibnamefont {{Nishimichi}}}, \bibinfo
  {author} {\bibfnamefont {A.}~\bibnamefont {{Taruya}}}, \ and\ \bibinfo
  {author} {\bibfnamefont {M.}~\bibnamefont {{Oguri}}},\ }\href {\doibase
  10.1088/0004-637X/761/2/152} {\bibfield  {journal} {\bibinfo  {journal}
  {Astrophys. J.}\ }\textbf {\bibinfo {volume} {761}},\ \bibinfo {eid} {152}
  (\bibinfo {year} {2012})}\BibitemShut {NoStop}%
\bibitem [{\citenamefont {{Bird}}\ \emph {et~al.}(2012)\citenamefont {{Bird}},
  \citenamefont {{Viel}},\ and\ \citenamefont {{Haehnelt}}}]{nu}%
  \BibitemOpen
  \bibfield  {author} {\bibinfo {author} {\bibfnamefont {S.}~\bibnamefont
  {{Bird}}}, \bibinfo {author} {\bibfnamefont {M.}~\bibnamefont {{Viel}}}, \
  and\ \bibinfo {author} {\bibfnamefont {M.~G.}\ \bibnamefont {{Haehnelt}}},\
  }\href {\doibase 10.1111/j.1365-2966.2011.20222.x} {\bibfield  {journal}
  {\bibinfo  {journal} {Mon. Not. R. Astron. Soc.}\ }\textbf {\bibinfo {volume}
  {420}},\ \bibinfo {pages} {2551} (\bibinfo {year} {2012})}\BibitemShut
  {NoStop}%
\bibitem [{\citenamefont {{Heitmann}}\ \emph {et~al.}(2010)\citenamefont
  {{Heitmann}}, \citenamefont {{White}}, \citenamefont {{Wagner}},
  \citenamefont {{Habib}},\ and\ \citenamefont
  {{Higdon}}}]{2010ApJ...715..104H}%
  \BibitemOpen
  \bibfield  {author} {\bibinfo {author} {\bibfnamefont {K.}~\bibnamefont
  {{Heitmann}}}, \bibinfo {author} {\bibfnamefont {M.}~\bibnamefont {{White}}},
  \bibinfo {author} {\bibfnamefont {C.}~\bibnamefont {{Wagner}}}, \bibinfo
  {author} {\bibfnamefont {S.}~\bibnamefont {{Habib}}}, \ and\ \bibinfo
  {author} {\bibfnamefont {D.}~\bibnamefont {{Higdon}}},\ }\href {\doibase
  10.1088/0004-637X/715/1/104} {\bibfield  {journal} {\bibinfo  {journal}
  {\apj}\ }\textbf {\bibinfo {volume} {715}},\ \bibinfo {pages} {104} (\bibinfo
  {year} {2010})}\BibitemShut {NoStop}%
\bibitem [{\citenamefont {{Schneider}}\ \emph {et~al.}(2016)\citenamefont
  {{Schneider}}, \citenamefont {{Teyssier}}, \citenamefont {{Potter}},
  \citenamefont {{Stadel}}, \citenamefont {{Onions}}, \citenamefont {{Reed}},
  \citenamefont {{Smith}}, \citenamefont {{Springel}}, \citenamefont
  {{Pearce}},\ and\ \citenamefont {{Scoccimarro}}}]{2016JCAP...04..047S}%
  \BibitemOpen
  \bibfield  {author} {\bibinfo {author} {\bibfnamefont {A.}~\bibnamefont
  {{Schneider}}}, \bibinfo {author} {\bibfnamefont {R.}~\bibnamefont
  {{Teyssier}}}, \bibinfo {author} {\bibfnamefont {D.}~\bibnamefont
  {{Potter}}}, \bibinfo {author} {\bibfnamefont {J.}~\bibnamefont {{Stadel}}},
  \bibinfo {author} {\bibfnamefont {J.}~\bibnamefont {{Onions}}}, \bibinfo
  {author} {\bibfnamefont {D.~S.}\ \bibnamefont {{Reed}}}, \bibinfo {author}
  {\bibfnamefont {R.~E.}\ \bibnamefont {{Smith}}}, \bibinfo {author}
  {\bibfnamefont {V.}~\bibnamefont {{Springel}}}, \bibinfo {author}
  {\bibfnamefont {F.~R.}\ \bibnamefont {{Pearce}}}, \ and\ \bibinfo {author}
  {\bibfnamefont {R.}~\bibnamefont {{Scoccimarro}}},\ }\href {\doibase
  10.1088/1475-7516/2016/04/047} {\bibfield  {journal} {\bibinfo  {journal}
  {\jcap}\ }\textbf {\bibinfo {volume} {4}},\ \bibinfo {eid} {047} (\bibinfo
  {year} {2016})}\BibitemShut {NoStop}%
\bibitem [{\citenamefont {{Lawrence}}\ \emph {et~al.}(2017)\citenamefont
  {{Lawrence}}, \citenamefont {{Heitmann}}, \citenamefont {{Kwan}},
  \citenamefont {{Upadhye}}, \citenamefont {{Bingham}}, \citenamefont
  {{Habib}}, \citenamefont {{Higdon}}, \citenamefont {{Pope}}, \citenamefont
  {{Finkel}},\ and\ \citenamefont {{Frontiere}}}]{2017ApJ...847...50L}%
  \BibitemOpen
  \bibfield  {author} {\bibinfo {author} {\bibfnamefont {E.}~\bibnamefont
  {{Lawrence}}}, \bibinfo {author} {\bibfnamefont {K.}~\bibnamefont
  {{Heitmann}}}, \bibinfo {author} {\bibfnamefont {J.}~\bibnamefont {{Kwan}}},
  \bibinfo {author} {\bibfnamefont {A.}~\bibnamefont {{Upadhye}}}, \bibinfo
  {author} {\bibfnamefont {D.}~\bibnamefont {{Bingham}}}, \bibinfo {author}
  {\bibfnamefont {S.}~\bibnamefont {{Habib}}}, \bibinfo {author} {\bibfnamefont
  {D.}~\bibnamefont {{Higdon}}}, \bibinfo {author} {\bibfnamefont
  {A.}~\bibnamefont {{Pope}}}, \bibinfo {author} {\bibfnamefont
  {H.}~\bibnamefont {{Finkel}}}, \ and\ \bibinfo {author} {\bibfnamefont
  {N.}~\bibnamefont {{Frontiere}}},\ }\href {\doibase 10.3847/1538-4357/aa86a9}
  {\bibfield  {journal} {\bibinfo  {journal} {\apj}\ }\textbf {\bibinfo
  {volume} {847}},\ \bibinfo {eid} {50} (\bibinfo {year} {2017})}\BibitemShut
  {NoStop}%
\bibitem [{\citenamefont {{Schaye}}\ \emph {et~al.}(2010)\citenamefont
  {{Schaye}}, \citenamefont {{Dalla Vecchia}}, \citenamefont {{Booth}},
  \citenamefont {{Wiersma}}, \citenamefont {{Theuns}}, \citenamefont {{Haas}},
  \citenamefont {{Bertone}}, \citenamefont {{Duffy}}, \citenamefont
  {{McCarthy}},\ and\ \citenamefont {{van de Voort}}}]{schaye10}%
  \BibitemOpen
  \bibfield  {author} {\bibinfo {author} {\bibfnamefont {J.}~\bibnamefont
  {{Schaye}}}, \bibinfo {author} {\bibfnamefont {C.}~\bibnamefont {{Dalla
  Vecchia}}}, \bibinfo {author} {\bibfnamefont {C.~M.}\ \bibnamefont
  {{Booth}}}, \bibinfo {author} {\bibfnamefont {R.~P.~C.}\ \bibnamefont
  {{Wiersma}}}, \bibinfo {author} {\bibfnamefont {T.}~\bibnamefont {{Theuns}}},
  \bibinfo {author} {\bibfnamefont {M.~R.}\ \bibnamefont {{Haas}}}, \bibinfo
  {author} {\bibfnamefont {S.}~\bibnamefont {{Bertone}}}, \bibinfo {author}
  {\bibfnamefont {A.~R.}\ \bibnamefont {{Duffy}}}, \bibinfo {author}
  {\bibfnamefont {I.~G.}\ \bibnamefont {{McCarthy}}}, \ and\ \bibinfo {author}
  {\bibfnamefont {F.}~\bibnamefont {{van de Voort}}},\ }\href {\doibase
  10.1111/j.1365-2966.2009.16029.x} {\bibfield  {journal} {\bibinfo  {journal}
  {Mon. Not. R. Astron. Soc.}\ }\textbf {\bibinfo {volume} {402}},\ \bibinfo
  {pages} {1536} (\bibinfo {year} {2010})}\BibitemShut {NoStop}%
\bibitem [{\citenamefont {{McCarthy}}\ \emph {et~al.}(2010)\citenamefont
  {{McCarthy}}, \citenamefont {{Schaye}}, \citenamefont {{Ponman}},
  \citenamefont {{Bower}}, \citenamefont {{Booth}}, \citenamefont {{Dalla
  Vecchia}}, \citenamefont {{Crain}}, \citenamefont {{Springel}}, \citenamefont
  {{Theuns}},\ and\ \citenamefont {{Wiersma}}}]{mccarthy11}%
  \BibitemOpen
  \bibfield  {author} {\bibinfo {author} {\bibfnamefont {I.~G.}\ \bibnamefont
  {{McCarthy}}}, \bibinfo {author} {\bibfnamefont {J.}~\bibnamefont
  {{Schaye}}}, \bibinfo {author} {\bibfnamefont {T.~J.}\ \bibnamefont
  {{Ponman}}}, \bibinfo {author} {\bibfnamefont {R.~G.}\ \bibnamefont
  {{Bower}}}, \bibinfo {author} {\bibfnamefont {C.~M.}\ \bibnamefont
  {{Booth}}}, \bibinfo {author} {\bibfnamefont {C.}~\bibnamefont {{Dalla
  Vecchia}}}, \bibinfo {author} {\bibfnamefont {R.~A.}\ \bibnamefont
  {{Crain}}}, \bibinfo {author} {\bibfnamefont {V.}~\bibnamefont {{Springel}}},
  \bibinfo {author} {\bibfnamefont {T.}~\bibnamefont {{Theuns}}}, \ and\
  \bibinfo {author} {\bibfnamefont {R.~P.~C.}\ \bibnamefont {{Wiersma}}},\
  }\href {\doibase 10.1111/j.1365-2966.2010.16750.x} {\bibfield  {journal}
  {\bibinfo  {journal} {Mon. Not. R. Astron. Soc.}\ }\textbf {\bibinfo {volume}
  {406}},\ \bibinfo {pages} {822} (\bibinfo {year} {2010})}\BibitemShut
  {NoStop}%
\bibitem [{\citenamefont {{Mead}}\ \emph {et~al.}(2015)\citenamefont {{Mead}},
  \citenamefont {{Peacock}}, \citenamefont {{Heymans}}, \citenamefont
  {{Joudaki}},\ and\ \citenamefont {{Heavens}}}]{mead}%
  \BibitemOpen
  \bibfield  {author} {\bibinfo {author} {\bibfnamefont {A.~J.}\ \bibnamefont
  {{Mead}}}, \bibinfo {author} {\bibfnamefont {J.~A.}\ \bibnamefont
  {{Peacock}}}, \bibinfo {author} {\bibfnamefont {C.}~\bibnamefont
  {{Heymans}}}, \bibinfo {author} {\bibfnamefont {S.}~\bibnamefont
  {{Joudaki}}}, \ and\ \bibinfo {author} {\bibfnamefont {A.~F.}\ \bibnamefont
  {{Heavens}}},\ }\href {\doibase 10.1093/mnras/stv2036} {\bibfield  {journal}
  {\bibinfo  {journal} {Mon. Not. R. Astron. Soc.}\ }\textbf {\bibinfo {volume}
  {454}},\ \bibinfo {pages} {1958} (\bibinfo {year} {2015})}\BibitemShut
  {NoStop}%
\bibitem [{\citenamefont {Heitmann}\ \emph {et~al.}(2010)\citenamefont
  {Heitmann}, \citenamefont {White}, \citenamefont {Wagner}, \citenamefont
  {Habib},\ and\ \citenamefont {Higdon}}]{emu}%
  \BibitemOpen
  \bibfield  {author} {\bibinfo {author} {\bibfnamefont {K.}~\bibnamefont
  {Heitmann}}, \bibinfo {author} {\bibfnamefont {M.}~\bibnamefont {White}},
  \bibinfo {author} {\bibfnamefont {C.}~\bibnamefont {Wagner}}, \bibinfo
  {author} {\bibfnamefont {S.}~\bibnamefont {Habib}}, \ and\ \bibinfo {author}
  {\bibfnamefont {D.}~\bibnamefont {Higdon}},\ }\href
  {http://stacks.iop.org/0004-637X/715/i=1/a=104} {\bibfield  {journal}
  {\bibinfo  {journal} {The Astrophysical Journal}\ }\textbf {\bibinfo {volume}
  {715}},\ \bibinfo {pages} {104} (\bibinfo {year} {2010})}\BibitemShut
  {NoStop}%
\bibitem [{\citenamefont {Heitmann}\ \emph {et~al.}(2014)\citenamefont
  {Heitmann}, \citenamefont {Lawrence}, \citenamefont {Kwan}, \citenamefont
  {Habib},\ and\ \citenamefont {Higdon}}]{emu2}%
  \BibitemOpen
  \bibfield  {author} {\bibinfo {author} {\bibfnamefont {K.}~\bibnamefont
  {Heitmann}}, \bibinfo {author} {\bibfnamefont {E.}~\bibnamefont {Lawrence}},
  \bibinfo {author} {\bibfnamefont {J.}~\bibnamefont {Kwan}}, \bibinfo {author}
  {\bibfnamefont {S.}~\bibnamefont {Habib}}, \ and\ \bibinfo {author}
  {\bibfnamefont {D.}~\bibnamefont {Higdon}},\ }\href
  {http://stacks.iop.org/0004-637X/780/i=1/a=111} {\bibfield  {journal}
  {\bibinfo  {journal} {The Astrophysical Journal}\ }\textbf {\bibinfo {volume}
  {780}},\ \bibinfo {pages} {111} (\bibinfo {year} {2014})}\BibitemShut
  {NoStop}%
\bibitem [{\citenamefont {{Heavens}}\ \emph {et~al.}(2000)\citenamefont
  {{Heavens}}, \citenamefont {{Refregier}},\ and\ \citenamefont
  {{Heymans}}}]{HRH2000}%
  \BibitemOpen
  \bibfield  {author} {\bibinfo {author} {\bibfnamefont {A.}~\bibnamefont
  {{Heavens}}}, \bibinfo {author} {\bibfnamefont {A.}~\bibnamefont
  {{Refregier}}}, \ and\ \bibinfo {author} {\bibfnamefont {C.}~\bibnamefont
  {{Heymans}}},\ }\href {\doibase 10.1046/j.1365-8711.2000.03907.x} {\bibfield
  {journal} {\bibinfo  {journal} {Mon. Not. R. Astron. Soc.}\ }\textbf
  {\bibinfo {volume} {319}},\ \bibinfo {pages} {649} (\bibinfo {year}
  {2000})}\BibitemShut {NoStop}%
\bibitem [{\citenamefont {{Croft}}\ and\ \citenamefont
  {{Metzler}}(2000)}]{croft00}%
  \BibitemOpen
  \bibfield  {author} {\bibinfo {author} {\bibfnamefont {R.~A.~C.}\
  \bibnamefont {{Croft}}}\ and\ \bibinfo {author} {\bibfnamefont {C.~A.}\
  \bibnamefont {{Metzler}}},\ }\href {\doibase 10.1086/317856} {\bibfield
  {journal} {\bibinfo  {journal} {Astrophys. J.}\ }\textbf {\bibinfo {volume}
  {545}},\ \bibinfo {pages} {561} (\bibinfo {year} {2000})}\BibitemShut
  {NoStop}%
\bibitem [{\citenamefont {{Catelan}}\ \emph {et~al.}(2001)\citenamefont
  {{Catelan}}, \citenamefont {{Kamionkowski}},\ and\ \citenamefont
  {{Blandford}}}]{CKB01}%
  \BibitemOpen
  \bibfield  {author} {\bibinfo {author} {\bibfnamefont {P.}~\bibnamefont
  {{Catelan}}}, \bibinfo {author} {\bibfnamefont {M.}~\bibnamefont
  {{Kamionkowski}}}, \ and\ \bibinfo {author} {\bibfnamefont {R.~D.}\
  \bibnamefont {{Blandford}}},\ }\href {\doibase
  10.1046/j.1365-8711.2001.04105.x} {\bibfield  {journal} {\bibinfo  {journal}
  {Mon. Not. R. Astron. Soc.}\ }\textbf {\bibinfo {volume} {320}},\ \bibinfo
  {pages} {L7} (\bibinfo {year} {2001})}\BibitemShut {NoStop}%
\bibitem [{\citenamefont {{Crittenden}}\ \emph {et~al.}(2001)\citenamefont
  {{Crittenden}}, \citenamefont {{Natarajan}}, \citenamefont {{Pen}},\ and\
  \citenamefont {{Theuns}}}]{CNP+01}%
  \BibitemOpen
  \bibfield  {author} {\bibinfo {author} {\bibfnamefont {R.~G.}\ \bibnamefont
  {{Crittenden}}}, \bibinfo {author} {\bibfnamefont {P.}~\bibnamefont
  {{Natarajan}}}, \bibinfo {author} {\bibfnamefont {U.-L.}\ \bibnamefont
  {{Pen}}}, \ and\ \bibinfo {author} {\bibfnamefont {T.}~\bibnamefont
  {{Theuns}}},\ }\href {\doibase 10.1086/322370} {\bibfield  {journal}
  {\bibinfo  {journal} {Astrophys. J.}\ }\textbf {\bibinfo {volume} {559}},\
  \bibinfo {pages} {552} (\bibinfo {year} {2001})}\BibitemShut {NoStop}%
\bibitem [{\citenamefont {{Hirata}}\ and\ \citenamefont
  {{Seljak}}(2004)}]{HS04}%
  \BibitemOpen
  \bibfield  {author} {\bibinfo {author} {\bibfnamefont {C.~M.}\ \bibnamefont
  {{Hirata}}}\ and\ \bibinfo {author} {\bibfnamefont {U.}~\bibnamefont
  {{Seljak}}},\ }\href {\doibase 10.1103/PhysRevD.70.063526} {\bibfield
  {journal} {\bibinfo  {journal} {Phys. Rev. D}\ }\textbf {\bibinfo {volume}
  {70}},\ \bibinfo {pages} {063526} (\bibinfo {year} {2004})}\BibitemShut
  {NoStop}%
\bibitem [{\citenamefont {{Joachimi}}\ \emph {et~al.}(2011)\citenamefont
  {{Joachimi}}, \citenamefont {{Mandelbaum}}, \citenamefont {{Abdalla}},\ and\
  \citenamefont {{Bridle}}}]{JMA+11}%
  \BibitemOpen
  \bibfield  {author} {\bibinfo {author} {\bibfnamefont {B.}~\bibnamefont
  {{Joachimi}}}, \bibinfo {author} {\bibfnamefont {R.}~\bibnamefont
  {{Mandelbaum}}}, \bibinfo {author} {\bibfnamefont {F.~B.}\ \bibnamefont
  {{Abdalla}}}, \ and\ \bibinfo {author} {\bibfnamefont {S.~L.}\ \bibnamefont
  {{Bridle}}},\ }\href {\doibase 10.1051/0004-6361/201015621} {\bibfield
  {journal} {\bibinfo  {journal} {Astron. Astrophys.}\ }\textbf {\bibinfo
  {volume} {527}},\ \bibinfo {eid} {A26} (\bibinfo {year} {2011})}\BibitemShut
  {NoStop}%
\bibitem [{\citenamefont {{Kirk}}\ \emph {et~al.}(2012)\citenamefont {{Kirk}},
  \citenamefont {{Rassat}}, \citenamefont {{Host}},\ and\ \citenamefont
  {{Bridle}}}]{KRH+12}%
  \BibitemOpen
  \bibfield  {author} {\bibinfo {author} {\bibfnamefont {D.}~\bibnamefont
  {{Kirk}}}, \bibinfo {author} {\bibfnamefont {A.}~\bibnamefont {{Rassat}}},
  \bibinfo {author} {\bibfnamefont {O.}~\bibnamefont {{Host}}}, \ and\ \bibinfo
  {author} {\bibfnamefont {S.}~\bibnamefont {{Bridle}}},\ }\href {\doibase
  10.1111/j.1365-2966.2012.21099.x} {\bibfield  {journal} {\bibinfo  {journal}
  {Mon. Not. R. Astron. Soc.}\ }\textbf {\bibinfo {volume} {424}},\ \bibinfo
  {pages} {1647} (\bibinfo {year} {2012})}\BibitemShut {NoStop}%
\bibitem [{\citenamefont {{Krause}}\ \emph {et~al.}(2016)\citenamefont
  {{Krause}}, \citenamefont {{Eifler}},\ and\ \citenamefont
  {{Blazek}}}]{KEB16}%
  \BibitemOpen
  \bibfield  {author} {\bibinfo {author} {\bibfnamefont {E.}~\bibnamefont
  {{Krause}}}, \bibinfo {author} {\bibfnamefont {T.}~\bibnamefont {{Eifler}}},
  \ and\ \bibinfo {author} {\bibfnamefont {J.}~\bibnamefont {{Blazek}}},\
  }\href {\doibase 10.1093/mnras/stv2615} {\bibfield  {journal} {\bibinfo
  {journal} {Mon. Not. R. Astron. Soc.}\ }\textbf {\bibinfo {volume} {456}},\
  \bibinfo {pages} {207} (\bibinfo {year} {2016})}\BibitemShut {NoStop}%
\bibitem [{\citenamefont {{Blazek}}\ \emph {et~al.}(2011)\citenamefont
  {{Blazek}}, \citenamefont {{McQuinn}},\ and\ \citenamefont
  {{Seljak}}}]{BMS11}%
  \BibitemOpen
  \bibfield  {author} {\bibinfo {author} {\bibfnamefont {J.}~\bibnamefont
  {{Blazek}}}, \bibinfo {author} {\bibfnamefont {M.}~\bibnamefont {{McQuinn}}},
  \ and\ \bibinfo {author} {\bibfnamefont {U.}~\bibnamefont {{Seljak}}},\
  }\href {\doibase 10.1088/1475-7516/2011/05/010} {\bibfield  {journal}
  {\bibinfo  {journal} {\jcap}\ }\textbf {\bibinfo {volume} {5}},\ \bibinfo
  {eid} {010} (\bibinfo {year} {2011})}\BibitemShut {NoStop}%
\bibitem [{\citenamefont {{Blazek}}\ \emph {et~al.}(2015)\citenamefont
  {{Blazek}}, \citenamefont {{Vlah}},\ and\ \citenamefont {{Seljak}}}]{BVS15}%
  \BibitemOpen
  \bibfield  {author} {\bibinfo {author} {\bibfnamefont {J.}~\bibnamefont
  {{Blazek}}}, \bibinfo {author} {\bibfnamefont {Z.}~\bibnamefont {{Vlah}}}, \
  and\ \bibinfo {author} {\bibfnamefont {U.}~\bibnamefont {{Seljak}}},\ }\href
  {\doibase 10.1088/1475-7516/2015/08/015} {\bibfield  {journal} {\bibinfo
  {journal} {\jcap}\ }\textbf {\bibinfo {volume} {8}},\ \bibinfo {eid} {015}
  (\bibinfo {year} {2015})}\BibitemShut {NoStop}%
\bibitem [{\citenamefont {{Lee}}\ and\ \citenamefont {{Pen}}(2000)}]{LP00}%
  \BibitemOpen
  \bibfield  {author} {\bibinfo {author} {\bibfnamefont {J.}~\bibnamefont
  {{Lee}}}\ and\ \bibinfo {author} {\bibfnamefont {U.-L.}\ \bibnamefont
  {{Pen}}},\ }\href {\doibase 10.1086/312556} {\bibfield  {journal} {\bibinfo
  {journal} {Astrophys. J. Lett.}\ }\textbf {\bibinfo {volume} {532}},\
  \bibinfo {pages} {L5} (\bibinfo {year} {2000})}\BibitemShut {NoStop}%
\bibitem [{\citenamefont {{Mandelbaum}}\ \emph {et~al.}(2011)\citenamefont
  {{Mandelbaum}} \emph {et~al.}}]{MBB+11}%
  \BibitemOpen
  \bibfield  {author} {\bibinfo {author} {\bibfnamefont {R.}~\bibnamefont
  {{Mandelbaum}}} \emph {et~al.},\ }\href {\doibase
  10.1111/j.1365-2966.2010.17485.x} {\bibfield  {journal} {\bibinfo  {journal}
  {Mon. Not. R. Astron. Soc.}\ }\textbf {\bibinfo {volume} {410}},\ \bibinfo
  {pages} {844} (\bibinfo {year} {2011})}\BibitemShut {NoStop}%
\bibitem [{\citenamefont {{Blazek}}\ \emph {et~al.}(2017)\citenamefont
  {{Blazek}}, \citenamefont {{MacCrann}}, \citenamefont {{Troxel}},\ and\
  \citenamefont {{Fang}}}]{bmt17}%
  \BibitemOpen
  \bibfield  {author} {\bibinfo {author} {\bibfnamefont {J.}~\bibnamefont
  {{Blazek}}}, \bibinfo {author} {\bibfnamefont {M.}~\bibnamefont
  {{MacCrann}}}, \bibinfo {author} {\bibfnamefont {M.~A.}\ \bibnamefont
  {{Troxel}}}, \ and\ \bibinfo {author} {\bibfnamefont {X.}~\bibnamefont
  {{Fang}}},\ }\href@noop {} {\bibfield  {journal} {\bibinfo  {journal} {ArXiv
  e-prints}\ } (\bibinfo {year} {2017})},\ \Eprint
  {http://arxiv.org/abs/1708.09247} {arXiv:1708.09247} \BibitemShut {NoStop}%
\bibitem [{\citenamefont {{Bridle}}\ and\ \citenamefont {{King}}(2007)}]{bk07}%
  \BibitemOpen
  \bibfield  {author} {\bibinfo {author} {\bibfnamefont {S.}~\bibnamefont
  {{Bridle}}}\ and\ \bibinfo {author} {\bibfnamefont {L.}~\bibnamefont
  {{King}}},\ }\href {\doibase 10.1088/1367-2630/9/12/444} {\bibfield
  {journal} {\bibinfo  {journal} {New Journal of Physics}\ }\textbf {\bibinfo
  {volume} {9}},\ \bibinfo {pages} {444} (\bibinfo {year} {2007})}\BibitemShut
  {NoStop}%
\bibitem [{\citenamefont {{Heymans}}\ \emph {et~al.}(2006)\citenamefont
  {{Heymans}} \emph {et~al.}}]{Heymans06}%
  \BibitemOpen
  \bibfield  {author} {\bibinfo {author} {\bibfnamefont {C.}~\bibnamefont
  {{Heymans}}} \emph {et~al.},\ }\href {\doibase
  10.1111/j.1365-2966.2006.10198.x} {\bibfield  {journal} {\bibinfo  {journal}
  {Mon. Not. R. Astron. Soc.}\ }\textbf {\bibinfo {volume} {368}},\ \bibinfo
  {pages} {1323} (\bibinfo {year} {2006})}\BibitemShut {NoStop}%
\bibitem [{\citenamefont {{Huterer}}\ \emph {et~al.}(2006)\citenamefont
  {{Huterer}}, \citenamefont {{Takada}}, \citenamefont {{Bernstein}},\ and\
  \citenamefont {{Jain}}}]{HTBJ06}%
  \BibitemOpen
  \bibfield  {author} {\bibinfo {author} {\bibfnamefont {D.}~\bibnamefont
  {{Huterer}}}, \bibinfo {author} {\bibfnamefont {M.}~\bibnamefont {{Takada}}},
  \bibinfo {author} {\bibfnamefont {G.}~\bibnamefont {{Bernstein}}}, \ and\
  \bibinfo {author} {\bibfnamefont {B.}~\bibnamefont {{Jain}}},\ }\href
  {\doibase 10.1111/j.1365-2966.2005.09782.x} {\bibfield  {journal} {\bibinfo
  {journal} {Mon. Not. R. Astron. Soc.}\ }\textbf {\bibinfo {volume} {366}},\
  \bibinfo {pages} {101} (\bibinfo {year} {2006})}\BibitemShut {NoStop}%
\bibitem [{\citenamefont {Jeffreys}(1961)}]{jeffreys61}%
  \BibitemOpen
  \bibfield  {author} {\bibinfo {author} {\bibfnamefont {H.}~\bibnamefont
  {Jeffreys}},\ }\href@noop {} {\emph {\bibinfo {title} {Theory of
  Probability}}},\ \bibinfo {edition} {3rd}\ ed.\ (\bibinfo  {publisher}
  {Oxford},\ \bibinfo {address} {Oxford, England},\ \bibinfo {year}
  {1961})\BibitemShut {NoStop}%
\bibitem [{\citenamefont {Kass}\ and\ \citenamefont
  {Raftery}(1995)}]{doi:10.1080/01621459.1995.10476572}%
  \BibitemOpen
  \bibfield  {author} {\bibinfo {author} {\bibfnamefont {R.~E.}\ \bibnamefont
  {Kass}}\ and\ \bibinfo {author} {\bibfnamefont {A.~E.}\ \bibnamefont
  {Raftery}},\ }\href {\doibase 10.1080/01621459.1995.10476572} {\bibfield
  {journal} {\bibinfo  {journal} {Journal of the American Statistical
  Association}\ }\textbf {\bibinfo {volume} {90}},\ \bibinfo {pages} {773}
  (\bibinfo {year} {1995})}\BibitemShut {NoStop}%
\bibitem [{\citenamefont {{Troxel}}\ \emph {et~al.}(2018)\citenamefont
  {{Troxel}} \emph {et~al.}}]{troxel2018}%
  \BibitemOpen
  \bibfield  {author} {\bibinfo {author} {\bibfnamefont {M.~A.}\ \bibnamefont
  {{Troxel}}} \emph {et~al.},\ }\href@noop {} {\bibfield  {journal} {\bibinfo
  {journal} {in prep.}\ } (\bibinfo {year} {2018})}\BibitemShut {NoStop}%
\bibitem [{\citenamefont {Beutler}\ \emph {et~al.}(2011)\citenamefont
  {Beutler}, \citenamefont {Blake}, \citenamefont {Colless}, \citenamefont
  {Jones}, \citenamefont {Staveley-Smith}, \citenamefont {Campbell},
  \citenamefont {Parker}, \citenamefont {Saunders},\ and\ \citenamefont
  {Watson}}]{Beutler:2011hx}%
  \BibitemOpen
  \bibfield  {author} {\bibinfo {author} {\bibfnamefont {F.}~\bibnamefont
  {Beutler}}, \bibinfo {author} {\bibfnamefont {C.}~\bibnamefont {Blake}},
  \bibinfo {author} {\bibfnamefont {M.}~\bibnamefont {Colless}}, \bibinfo
  {author} {\bibfnamefont {D.~H.}\ \bibnamefont {Jones}}, \bibinfo {author}
  {\bibfnamefont {L.}~\bibnamefont {Staveley-Smith}}, \bibinfo {author}
  {\bibfnamefont {L.}~\bibnamefont {Campbell}}, \bibinfo {author}
  {\bibfnamefont {Q.}~\bibnamefont {Parker}}, \bibinfo {author} {\bibfnamefont
  {W.}~\bibnamefont {Saunders}}, \ and\ \bibinfo {author} {\bibfnamefont
  {F.}~\bibnamefont {Watson}},\ }\href {\doibase
  10.1111/j.1365-2966.2011.19250.x} {\bibfield  {journal} {\bibinfo  {journal}
  {Mon. Not. Roy. Astron. Soc.}\ }\textbf {\bibinfo {volume} {416}},\ \bibinfo
  {pages} {3017} (\bibinfo {year} {2011})}\BibitemShut {NoStop}%
\bibitem [{\citenamefont {Ross}\ \emph {et~al.}(2015)\citenamefont {Ross},
  \citenamefont {Samushia}, \citenamefont {Howlett}, \citenamefont {Percival},
  \citenamefont {Burden},\ and\ \citenamefont {Manera}}]{Ross:2014qpa}%
  \BibitemOpen
  \bibfield  {author} {\bibinfo {author} {\bibfnamefont {A.~J.}\ \bibnamefont
  {Ross}}, \bibinfo {author} {\bibfnamefont {L.}~\bibnamefont {Samushia}},
  \bibinfo {author} {\bibfnamefont {C.}~\bibnamefont {Howlett}}, \bibinfo
  {author} {\bibfnamefont {W.~J.}\ \bibnamefont {Percival}}, \bibinfo {author}
  {\bibfnamefont {A.}~\bibnamefont {Burden}}, \ and\ \bibinfo {author}
  {\bibfnamefont {M.}~\bibnamefont {Manera}},\ }\href {\doibase
  10.1093/mnras/stv154} {\bibfield  {journal} {\bibinfo  {journal} {Mon. Not.
  Roy. Astron. Soc.}\ }\textbf {\bibinfo {volume} {449}},\ \bibinfo {pages}
  {835} (\bibinfo {year} {2015})}\BibitemShut {NoStop}%
\bibitem [{\citenamefont {Alam}\ \emph {et~al.}(2016)\citenamefont {Alam} \emph
  {et~al.}}]{Alam:2016hwk}%
  \BibitemOpen
  \bibfield  {author} {\bibinfo {author} {\bibfnamefont {S.}~\bibnamefont
  {Alam}} \emph {et~al.},\ }\href@noop {} {\bibfield  {journal} {\bibinfo
  {journal} {submitted to Mon. Not. Roy. Astron. Soc.}\ } (\bibinfo {year}
  {2016})}\BibitemShut {NoStop}%
\bibitem [{\citenamefont {Betoule}\ \emph {et~al.}(2014)\citenamefont {Betoule}
  \emph {et~al.}}]{Betoule:2014frx}%
  \BibitemOpen
  \bibfield  {author} {\bibinfo {author} {\bibfnamefont {M.}~\bibnamefont
  {Betoule}} \emph {et~al.},\ }\href {\doibase 10.1051/0004-6361/201423413}
  {\bibfield  {journal} {\bibinfo  {journal} {Astron. Astrophys.}\ }\textbf
  {\bibinfo {volume} {568}},\ \bibinfo {pages} {A22} (\bibinfo {year}
  {2014})}\BibitemShut {NoStop}%
\bibitem [{\citenamefont {Ade}\ \emph {et~al.}(2016)\citenamefont {Ade} \emph
  {et~al.}}]{Ade:2015xua}%
  \BibitemOpen
  \bibfield  {author} {\bibinfo {author} {\bibfnamefont {P.~A.~R.}\
  \bibnamefont {Ade}} \emph {et~al.},\ }\href {\doibase
  10.1051/0004-6361/201525830} {\bibfield  {journal} {\bibinfo  {journal}
  {Astron. Astrophys.}\ }\textbf {\bibinfo {volume} {594}},\ \bibinfo {pages}
  {A13} (\bibinfo {year} {2016})}\BibitemShut {NoStop}%
\bibitem [{\citenamefont {{Kitching}}\ \emph {et~al.}(2017)\citenamefont
  {{Kitching}}, \citenamefont {{Alsing}}, \citenamefont {{Heavens}},
  \citenamefont {{Jimenez}}, \citenamefont {{McEwen}},\ and\ \citenamefont
  {{Verde}}}]{kitching17}%
  \BibitemOpen
  \bibfield  {author} {\bibinfo {author} {\bibfnamefont {T.~D.}\ \bibnamefont
  {{Kitching}}}, \bibinfo {author} {\bibfnamefont {J.}~\bibnamefont
  {{Alsing}}}, \bibinfo {author} {\bibfnamefont {A.~F.}\ \bibnamefont
  {{Heavens}}}, \bibinfo {author} {\bibfnamefont {R.}~\bibnamefont
  {{Jimenez}}}, \bibinfo {author} {\bibfnamefont {J.~D.}\ \bibnamefont
  {{McEwen}}}, \ and\ \bibinfo {author} {\bibfnamefont {L.}~\bibnamefont
  {{Verde}}},\ }\href {\doibase 10.1093/mnras/stx1039} {\bibfield  {journal}
  {\bibinfo  {journal} {Mon. Not. R. Astron. Soc.}\ }\textbf {\bibinfo {volume}
  {469}},\ \bibinfo {pages} {2737} (\bibinfo {year} {2017})}\BibitemShut
  {NoStop}%
\bibitem [{\citenamefont {{Bonnett}}\ \emph {et~al.}(2016)\citenamefont
  {{Bonnett}} \emph {et~al.}}]{bonnett2016}%
  \BibitemOpen
  \bibfield  {author} {\bibinfo {author} {\bibfnamefont {C.}~\bibnamefont
  {{Bonnett}}} \emph {et~al.} (\bibinfo {collaboration} {DES Collaboration}),\
  }\href {\doibase 10.1103/PhysRevD.94.042005} {\bibfield  {journal} {\bibinfo
  {journal} {Phys. Rev. D}\ }\textbf {\bibinfo {volume} {94}},\ \bibinfo {eid}
  {042005} (\bibinfo {year} {2016})}\BibitemShut {NoStop}%
\bibitem [{\citenamefont {{Chisari}}\ \emph {et~al.}(2016)\citenamefont
  {{Chisari}}, \citenamefont {{Laigle}}, \citenamefont {{Codis}}, \citenamefont
  {{Dubois}}, \citenamefont {{Devriendt}}, \citenamefont {{Miller}},
  \citenamefont {{Benabed}}, \citenamefont {{Slyz}}, \citenamefont
  {{Gavazzi}},\ and\ \citenamefont {{Pichon}}}]{clc16}%
  \BibitemOpen
  \bibfield  {author} {\bibinfo {author} {\bibfnamefont {N.}~\bibnamefont
  {{Chisari}}}, \bibinfo {author} {\bibfnamefont {C.}~\bibnamefont {{Laigle}}},
  \bibinfo {author} {\bibfnamefont {S.}~\bibnamefont {{Codis}}}, \bibinfo
  {author} {\bibfnamefont {Y.}~\bibnamefont {{Dubois}}}, \bibinfo {author}
  {\bibfnamefont {J.}~\bibnamefont {{Devriendt}}}, \bibinfo {author}
  {\bibfnamefont {L.}~\bibnamefont {{Miller}}}, \bibinfo {author}
  {\bibfnamefont {K.}~\bibnamefont {{Benabed}}}, \bibinfo {author}
  {\bibfnamefont {A.}~\bibnamefont {{Slyz}}}, \bibinfo {author} {\bibfnamefont
  {R.}~\bibnamefont {{Gavazzi}}}, \ and\ \bibinfo {author} {\bibfnamefont
  {C.}~\bibnamefont {{Pichon}}},\ }\href {\doibase 10.1093/mnras/stw1409}
  {\bibfield  {journal} {\bibinfo  {journal} {Mon. Not. R. Astron. Soc.}\
  }\textbf {\bibinfo {volume} {461}},\ \bibinfo {pages} {2702} (\bibinfo {year}
  {2016})}\BibitemShut {NoStop}%
\bibitem [{\citenamefont {{Tenneti}}\ \emph {et~al.}(2016)\citenamefont
  {{Tenneti}}, \citenamefont {{Mandelbaum}},\ and\ \citenamefont {{Di
  Matteo}}}]{tmd16}%
  \BibitemOpen
  \bibfield  {author} {\bibinfo {author} {\bibfnamefont {A.}~\bibnamefont
  {{Tenneti}}}, \bibinfo {author} {\bibfnamefont {R.}~\bibnamefont
  {{Mandelbaum}}}, \ and\ \bibinfo {author} {\bibfnamefont {T.}~\bibnamefont
  {{Di Matteo}}},\ }\href {\doibase 10.1093/mnras/stw1823} {\bibfield
  {journal} {\bibinfo  {journal} {Mon. Not. R. Astron. Soc.}\ }\textbf
  {\bibinfo {volume} {462}},\ \bibinfo {pages} {2668} (\bibinfo {year}
  {2016})}\BibitemShut {NoStop}%
\bibitem [{\citenamefont {{Hilbert}}\ \emph {et~al.}(2017)\citenamefont
  {{Hilbert}}, \citenamefont {{Xu}}, \citenamefont {{Schneider}}, \citenamefont
  {{Springel}}, \citenamefont {{Vogelsberger}},\ and\ \citenamefont
  {{Hernquist}}}]{hxs17}%
  \BibitemOpen
  \bibfield  {author} {\bibinfo {author} {\bibfnamefont {S.}~\bibnamefont
  {{Hilbert}}}, \bibinfo {author} {\bibfnamefont {D.}~\bibnamefont {{Xu}}},
  \bibinfo {author} {\bibfnamefont {P.}~\bibnamefont {{Schneider}}}, \bibinfo
  {author} {\bibfnamefont {V.}~\bibnamefont {{Springel}}}, \bibinfo {author}
  {\bibfnamefont {M.}~\bibnamefont {{Vogelsberger}}}, \ and\ \bibinfo {author}
  {\bibfnamefont {L.}~\bibnamefont {{Hernquist}}},\ }\href {\doibase
  10.1093/mnras/stx482} {\bibfield  {journal} {\bibinfo  {journal} {Mon. Not.
  R. Astron. Soc.}\ }\textbf {\bibinfo {volume} {468}},\ \bibinfo {pages} {790}
  (\bibinfo {year} {2017})}\BibitemShut {NoStop}%
\bibitem [{\citenamefont {{Fenech Conti}}\ \emph {et~al.}(2017)\citenamefont
  {{Fenech Conti}}, \citenamefont {{Herbonnet}}, \citenamefont {{Hoekstra}},
  \citenamefont {{Merten}}, \citenamefont {{Miller}},\ and\ \citenamefont
  {{Viola}}}]{2017MNRAS.467.1627F}%
  \BibitemOpen
  \bibfield  {author} {\bibinfo {author} {\bibfnamefont {I.}~\bibnamefont
  {{Fenech Conti}}}, \bibinfo {author} {\bibfnamefont {R.}~\bibnamefont
  {{Herbonnet}}}, \bibinfo {author} {\bibfnamefont {H.}~\bibnamefont
  {{Hoekstra}}}, \bibinfo {author} {\bibfnamefont {J.}~\bibnamefont
  {{Merten}}}, \bibinfo {author} {\bibfnamefont {L.}~\bibnamefont {{Miller}}},
  \ and\ \bibinfo {author} {\bibfnamefont {M.}~\bibnamefont {{Viola}}},\ }\href
  {\doibase 10.1093/mnras/stx200} {\bibfield  {journal} {\bibinfo  {journal}
  {Mon. Not. R. Astron. Soc.}\ }\textbf {\bibinfo {volume} {467}},\ \bibinfo
  {pages} {1627} (\bibinfo {year} {2017})}\BibitemShut {NoStop}%
\bibitem [{\citenamefont {{Erben}}\ \emph {et~al.}(2013)\citenamefont {{Erben}}
  \emph {et~al.}}]{2013MNRAS.433.2545E}%
  \BibitemOpen
  \bibfield  {author} {\bibinfo {author} {\bibfnamefont {T.}~\bibnamefont
  {{Erben}}} \emph {et~al.},\ }\href {\doibase 10.1093/mnras/stt928} {\bibfield
   {journal} {\bibinfo  {journal} {Mon. Not. R. Astron. Soc.}\ }\textbf
  {\bibinfo {volume} {433}},\ \bibinfo {pages} {2545} (\bibinfo {year}
  {2013})}\BibitemShut {NoStop}%
\bibitem [{\citenamefont {{Begeman}}\ \emph {et~al.}(2013)\citenamefont
  {{Begeman}}, \citenamefont {{Belikov}}, \citenamefont {{Boxhoorn}},\ and\
  \citenamefont {{Valentijn}}}]{2013ExA....35....1B}%
  \BibitemOpen
  \bibfield  {author} {\bibinfo {author} {\bibfnamefont {K.}~\bibnamefont
  {{Begeman}}}, \bibinfo {author} {\bibfnamefont {A.~N.}\ \bibnamefont
  {{Belikov}}}, \bibinfo {author} {\bibfnamefont {D.~R.}\ \bibnamefont
  {{Boxhoorn}}}, \ and\ \bibinfo {author} {\bibfnamefont {E.~A.}\ \bibnamefont
  {{Valentijn}}},\ }\href {\doibase 10.1007/s10686-012-9311-4} {\bibfield
  {journal} {\bibinfo  {journal} {Experimental Astronomy}\ }\textbf {\bibinfo
  {volume} {35}},\ \bibinfo {pages} {1} (\bibinfo {year} {2013})}\BibitemShut
  {NoStop}%
\bibitem [{\citenamefont {{de Jong}}\ \emph {et~al.}(2015)\citenamefont {{de
  Jong}} \emph {et~al.}}]{2015A&A...582A..62D}%
  \BibitemOpen
  \bibfield  {author} {\bibinfo {author} {\bibfnamefont {J.~T.~A.}\
  \bibnamefont {{de Jong}}} \emph {et~al.},\ }\href {\doibase
  10.1051/0004-6361/201526601} {\bibfield  {journal} {\bibinfo  {journal}
  {\aap}\ }\textbf {\bibinfo {volume} {582}},\ \bibinfo {eid} {A62} (\bibinfo
  {year} {2015})}\BibitemShut {NoStop}%
\bibitem [{\citenamefont {{Miller}}\ \emph {et~al.}(2013)\citenamefont
  {{Miller}} \emph {et~al.}}]{2013MNRAS.429.2858M}%
  \BibitemOpen
  \bibfield  {author} {\bibinfo {author} {\bibfnamefont {L.}~\bibnamefont
  {{Miller}}} \emph {et~al.},\ }\href {\doibase 10.1093/mnras/sts454}
  {\bibfield  {journal} {\bibinfo  {journal} {Mon. Not. R. Astron. Soc.}\
  }\textbf {\bibinfo {volume} {429}},\ \bibinfo {pages} {2858} (\bibinfo {year}
  {2013})}\BibitemShut {NoStop}%
\bibitem [{\citenamefont {Hunter}(2007)}]{Hunter:2007}%
  \BibitemOpen
  \bibfield  {author} {\bibinfo {author} {\bibfnamefont {J.~D.}\ \bibnamefont
  {Hunter}},\ }\href {\doibase 10.1109/MCSE.2007.55} {\bibfield  {journal}
  {\bibinfo  {journal} {Computing In Science \& Engineering}\ }\textbf
  {\bibinfo {volume} {9}},\ \bibinfo {pages} {90} (\bibinfo {year}
  {2007})}\BibitemShut {NoStop}%
\bibitem [{\citenamefont {OSC}(1987)}]{OhioSupercomputerCenter1987}%
  \BibitemOpen
  \bibfield  {author} {\bibinfo {author} {\bibnamefont {OSC}},\ }\href@noop {}
  {\enquote {\bibinfo {title} {Ohio supercomputer center},}\ }\bibinfo
  {howpublished} {\url{http://osc.edu/ark:/19495/f5s1ph73}} (\bibinfo {year}
  {1987})\BibitemShut {NoStop}%
\bibitem [{\citenamefont {{Paulin-Henriksson}}\ \emph
  {et~al.}(2008)\citenamefont {{Paulin-Henriksson}}, \citenamefont {{Amara}},
  \citenamefont {{Voigt}}, \citenamefont {{Refregier}},\ and\ \citenamefont
  {{Bridle}}}]{paulinhenriksson08}%
  \BibitemOpen
  \bibfield  {author} {\bibinfo {author} {\bibfnamefont {S.}~\bibnamefont
  {{Paulin-Henriksson}}}, \bibinfo {author} {\bibfnamefont {A.}~\bibnamefont
  {{Amara}}}, \bibinfo {author} {\bibfnamefont {L.}~\bibnamefont {{Voigt}}},
  \bibinfo {author} {\bibfnamefont {A.}~\bibnamefont {{Refregier}}}, \ and\
  \bibinfo {author} {\bibfnamefont {S.~L.}\ \bibnamefont {{Bridle}}},\ }\href
  {\doibase 10.1051/0004-6361:20079150} {\bibfield  {journal} {\bibinfo
  {journal} {Astron. Astrophys.}\ }\textbf {\bibinfo {volume} {484}},\ \bibinfo
  {pages} {67} (\bibinfo {year} {2008})}\BibitemShut {NoStop}%
\bibitem [{\citenamefont {{Rowe}}(2010)}]{rowe2010}%
  \BibitemOpen
  \bibfield  {author} {\bibinfo {author} {\bibfnamefont {B.}~\bibnamefont
  {{Rowe}}},\ }\href {\doibase 10.1111/j.1365-2966.2010.16277.x} {\bibfield
  {journal} {\bibinfo  {journal} {Mon. Not. R. Astron. Soc.}\ }\textbf
  {\bibinfo {volume} {404}},\ \bibinfo {pages} {350} (\bibinfo {year}
  {2010})}\BibitemShut {NoStop}%
\end{thebibliography}%

\appendix

\section{Residual PSF Model Bias and Mean Shear}\label{sec:psf}

\subsection{Residual PSF Model Errors}

A robust treatment of the PSF is crucial for unbiased cosmic shear measurements. Imperfect modeling or deconvolution of the PSF can produce coherent additive and multiplicative shear biases, both of which contaminate the cosmic shear signal \citep{paulinhenriksson08}. In \cite{shearcat}, we identified spatially correlated ellipticity errors in the PSF modeling. We model the impact of PSF model ellipticity errors on the inferred shear using the linear relation
\be
\delta e_i^{\mathrm{sys}} = \beta (e_i^{\mathrm{p}} - e_i^{\mathrm{*}}) = \beta q_i
\ee
where $i$ denotes the shear component, $\epsilon_i^{\mathrm{p}}$ is the PSF model ellipticity, $\epsilon_i^{\mathrm{*}}$ is the true PSF ellipticity and therefore $q_i$ is the $i$th component of the PSF model ellipticity residual. This relation is exact in the case of an unweighted quadrupole ellipticity estimator, if both the galaxy and PSF profiles are Gaussian. While neither of these conditions are satisfied in our case, we use this model as a first-order approximation. If, as well as PSF modeling errors, there are also errors in the deconvolution of the PSF model from the galaxy image, one might also expect a systematic bias that is proportional to the PSF model ellipticity (sometimes this term is referred to as PSF \emph{leakage}), such that the model for the shear bias becomes
\be
\delta e_i^{\mathrm{sys}} = \alpha_i e_i^{\mathrm{p}} + \beta^i q_i. \label{eq:de_sys}
\ee
Note that we have no reason to expect non-zero $\alpha_i$ from either shape measurement method; \metacal\ uses a circularized PSF, and \imshape\ uses calibration simulations to remove any correlation with the PSF ellipticity. On the other hand we expect all shear estimation algorithms to have a non-zero $\beta$; even a `perfect' shear estimator has to assume a PSF model, and errors in that PSF model will propagate to errors in the shear estimation (\citet{paulinhenriksson08} estimate $\beta_i$ for an un-weighted moments shear estimator). 
In \cite{shearcat}, we measure a significant correlation between the estimated shear and the PSF model ellipticity. This could be evidence for non-zero $\alpha$, but could also arise from correlations between the PSF model ellipticity and the PSF ellipticity residuals even for $\alpha=0$. We demonstrate below that the latter is the most likely explanation. 

While we have an estimate of $e^{\mathrm{p}_i}$ at each galaxy position, we can only estimate $q_i$ at the position of stars (where we can evaluate the PSF model \emph{and} directly measure the star's profile). Therefore, we use cross-correlations between the galaxy and star samples in order to simultaneously estimate $\alpha$ and $\beta$ (we assume from now on that $\alpha_1=\alpha_2, \beta_1=\beta_2$). To do this, we use the following cross-correlations
\begin{align}
\left< e^{\mathrm{obs}} e^{\mathrm{p}} \right> &= \alpha \left<e^{\mathrm{p}} e^{\mathrm{p}} \right> + \beta \left<q e^{\mathrm{p}} \right> \label{eq:epeq1}\\
\left< e^{\mathrm{obs}} q \right> &= \alpha \left<e^{\mathrm{p}} q \right> + \beta \left<q q \right>.\label{eq:epeq2}
\end{align}
Note that in the above, the angle-brackets denote correlations of spin-2 quantities; we use the $\xi_+(\theta)$ statistic for all of these. Eqs. (\ref{eq:epeq1}) and (\ref{eq:epeq2}) follow from Eq. (\ref{eq:de_sys}), and provide a means to find $\alpha$ and $\beta$, which are taken to be free parameters (for each redshift bin). The correlations $\left<e^{\mathrm{p}} e^{\mathrm{p}} \right>$, $\left<q e^{\mathrm{p}} \right>$,  $\left<e^{\mathrm{p}} q \right>$, and $\left<q q \right>$ are measured from the star catalog described in \cite{shearcat}. 

\begin{figure}
\begin{center}
\includegraphics[width=\columnwidth]{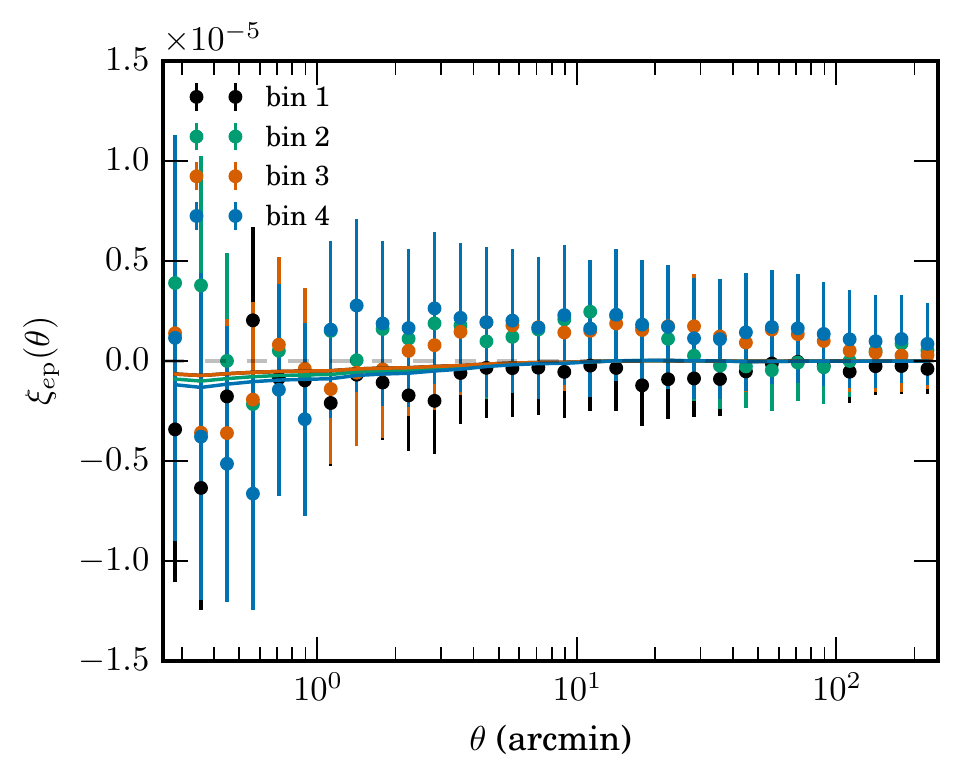}
\includegraphics[width=\columnwidth]{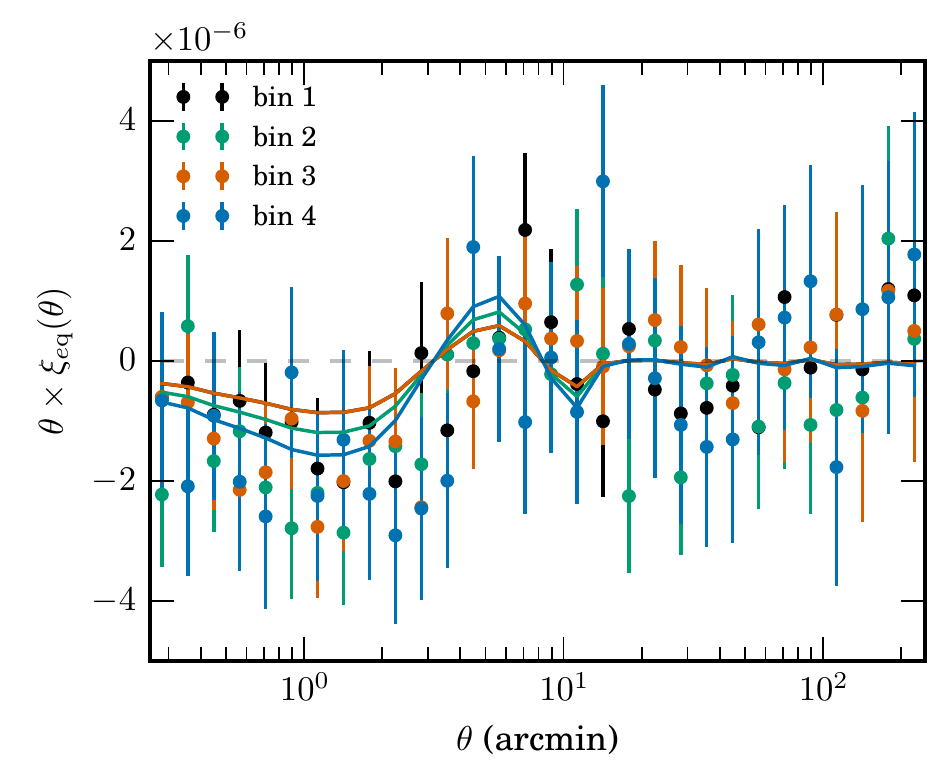}
\end{center}
\caption[]{The measured correlation functions $\xi_{ep}$ (Eq. (\ref{eq:epeq1})) and $\xi_{eq}$ (Eq. (\ref{eq:epeq2})). The resulting model from the best-fit $\alpha$ and $\beta$ from Eqs. (\ref{eq:epeq1}) \& (\ref{eq:epeq2}) for $\alpha$ and $\beta$ is shown as a solid line for each tomographic bin. The best-fit $\alpha$ is consistent with zero, while the values of $\beta_{\metacal}$ used are shown in Table \ref{tab:beta_constraints}.
\label{fig:mcal_ep_eq}}
\end{figure}

Figure \ref{fig:mcal_ep_eq} shows these two measured correlations for the \metacal\ catalog, for each redshift bin. Uncertainties are estimated using the lognormal mock shear catalogs described in Sec. \ref{sec:lognormal}. For both catalogs, we find the $\alpha$ values to be consistent with zero. Given this, and the fact we have no a priori reason to expect a non-zero $\alpha$, we assume $\alpha=0$ from now on.
As expected we find $\beta \sim -1$ for all redshift bins; constraints are shown in Table \ref{tab:beta_constraints}. Solid lines show the model predictions with the best-fit $\beta$. Note that these are actually constraints on the mean $\beta$ for each redshift bin - the value of $\beta$ for a specific galaxy will depend on the specific galaxy and PSF properties.

Given these estimates of $\beta$, we estimate the impact of PSF model ellipticity errors on our cosmological parameter inference as follows. The expected systematic contamination of $\xi_+^{ij}$, where $ij$ denotes the redshift bin pair, is 
\be
\xi_{+}^{ij,\mathrm{sys}} = \left< \beta^i \beta^j \right> \left(\left<q q \right> - \left<q_1\right>^2 - \left<q_2\right>^2\right)
\ee
where the second and third terms on the RHS arise because we are subtracting the mean ellipticity from each tomographic bin to correct for scale-independent additive biases (see Sec. \ref{sec:meanshear}).
We expect that on large scales (where additive biases are most significant), $\beta$ is uncorrelated between galaxies, and therefore make the assumption that 
$\left<\beta^i \beta^j \right> = \left< \beta^i \right> \left< \beta^j \right>$.
Using the measured $\left<q q \right>$ (also known as the first `$\rho$-statistic', $\rho_1(\theta)$ \citep{rowe2010}), and the best-fit $\beta$ values from Table \ref{tab:beta_constraints}, we produce a contaminated prediction of our data vector, which we then analyze using our parameter estimation framework to check for biases in cosmological parameters that this level of contamination would induce. We thus verify that the level of impact on cosmological parameters is entirely negligible, following the subtraction of the mean shear discussed below.

\begin{table}
\caption{Constraints on `$\beta$', the proportionality constant when assuming a linear relationship between inferred shear and PSF model ellipticity residual (see Eqn. \ref{eq:de_sys}). Errors quoted are 68\% confidence intervals.}
\label{tab:beta_constraints}
\begin{center}
\begin{ruledtabular}
\begin{tabular}{ lcccccc }
  Redshift bin 		& $\beta_{\metacal}$ & $\beta_{\imshape}$ \\
  \hline
  1     & $-0.72 \pm 0.26$ & $-1.65 \pm 0.54$ \\
  2 	& $-0.99 \pm 0.32$ & $-2.45 \pm 0.64$ \\
  3 	& $-0.72 \pm 0.32$ & $-1.4 \pm 0.60$ \\
  4     & $-1.31 \pm 0.43$ & $-0.92 \pm 0.78$ \\
\end{tabular}
\end{ruledtabular}
\end{center}
\end{table}

\subsection{Mean Residual Shear}\label{sec:meanshear}

While we conclude that the propagation of PSF model errors into the shape measurement does not produce a significant bias on cosmological constraints with cosmic shear, we are left with an unidentified mean shear that is too large in some tomographic bins to be consistent with shape noise and cosmic variance. The values of the mean shear are listed in Table \ref{table:meane}. Three of the eight $\langle e_i^j\rangle$ for \metacal\ have a magnitude greater than 2.5 times the predicted shear variance calculated using the lognormal mock catalogs described in Sec. \ref{sec:lognormal}. We cannot rule this out as being sourced by some unidentified additive bias or failure in our PSF bias model, so we test what impact this could have on our cosmological constraints. The contribution to $\xi_{+}$ of some additive shear bias $c_i^j = \langle e^j_i \rangle$ that is constant within a tomographic bin is
\begin{equation}
\xi^{ij}_{+,\mathrm{pred}} =  \xi^{ij}_{+,\mathrm{true}}+(\langle e^i_1 \rangle\langle e^j_1 \rangle + \langle e^i_2 \rangle\langle e^j_2 \rangle).
\end{equation}
Artificially introducing this bias due to mean shear to an unbiased simulated data vector results in non-negligible biases to cosmological parameters, and so we subtract the impact of the mean shear from the ellipticity before measuring $\xi_{\pm}$, as described in Sec. \ref{sec:2pt}. This mean shear is different for both shape catalogs, and its origin remains unclear. To verify that this is a sufficient correction for any large-scale problems, we also test that our inferred cosmological results are unchanged when we add a constant factor to the covariance that is equal to the impact of the mean shear on each $\xi^{ij}_{+}$. This factor is much larger in many blocks than the impact of cosmic variance, and thus removes most of the large-scale information from the signal for this test.

\begin{table}
\caption{Values for the mean shear of the \metacal\ and \imshape\ shape catalogs for each redshift bin.}
\label{table:meane}
\begin{center}
\begin{ruledtabular}
\begin{tabular}{ lcccc }
 & \multicolumn{2}{c}{\metacal\ } & \multicolumn{2}{c}{\imshape\ } \\
  Redshift bin 		& $\langle e^i_1\rangle \times10^{4}$ & $\langle e^i_2\rangle \times10^{4}$ & $\langle e^i_1\rangle \times10^{4}$ & $\langle e^i_2\rangle \times10^{4}$ \\
  \hline
  1     & $0.8$ & $4.2$ & $-2.1$ & $2.7$ \\
  2 	& $1.5$ & $4.6$  & $1.7$ & $1.6$ \\
  3 	& $3.8$ & $-1.8$  & $4.5$ & $0.2$ \\
  4     & $9.1$ & $-1.3$  & $5.9$ & $6.6$ \\
\end{tabular}
\end{ruledtabular}
\end{center}
\end{table}

\section{Testing for residual systematics in the cosmic shear signal}\label{sec:syssurvey}

Even after targeted tests for systematic bias in shape and \photoz\ measurements, there can persist biases that have not been identified that may impact the cosmic shear signal. In the DES SV cosmic shear analysis \cite{becker2016}, we tested for this by splitting the shear catalog into halves along a large number of either catalog or survey properties
that could correlate with shape or \photoz\ systematics. These tests assume that the quantity being used to split the shear catalog either should not be correlated with gravitational shear or is simply correlated with redshift, and thus any non-cosmological signal induced by the selection can be corrected to first order by reweighting the redshift distribution. This allows us to identify potential residual systematics that may either impact our cosmological constraints or indicate limitations in analyses that are sensitive to subsets of the shape or \photoz\ catalogs. 

\begin{figure*}
\begin{center}
\includegraphics[width=\textwidth]{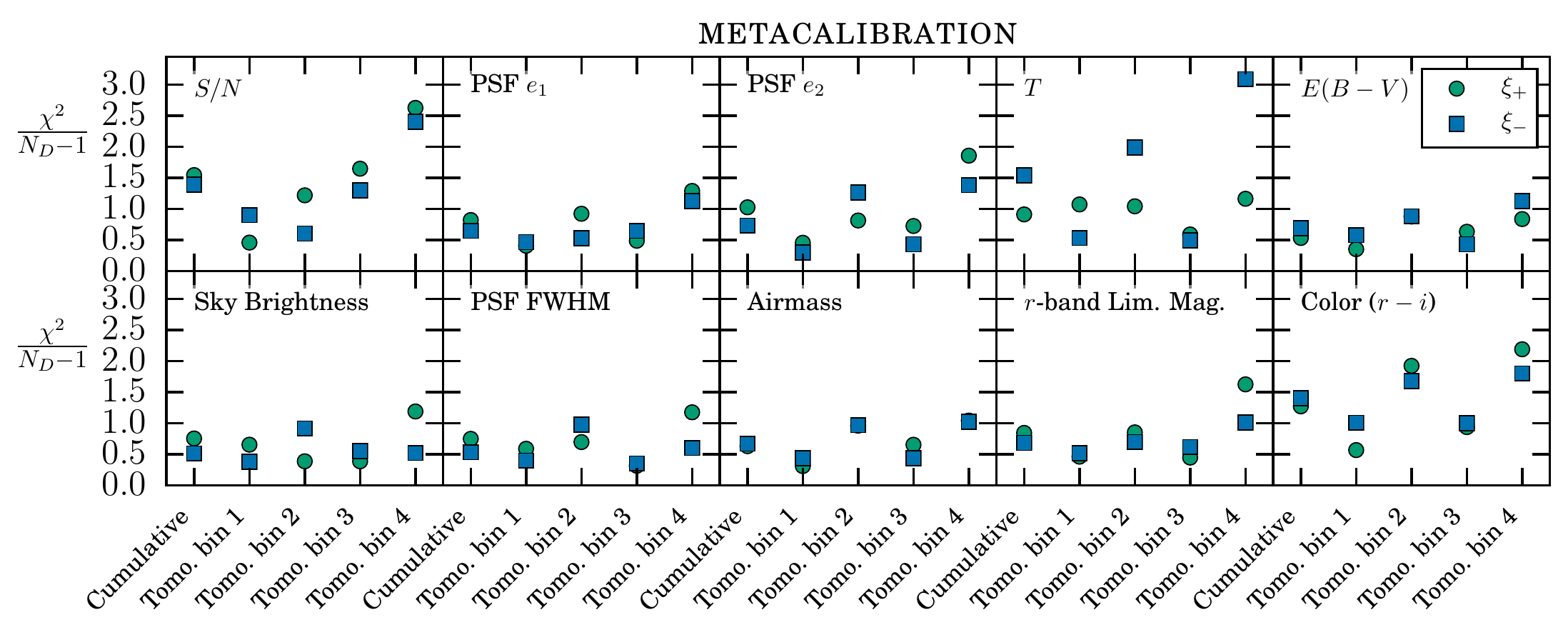}\\
\includegraphics[width=\textwidth]{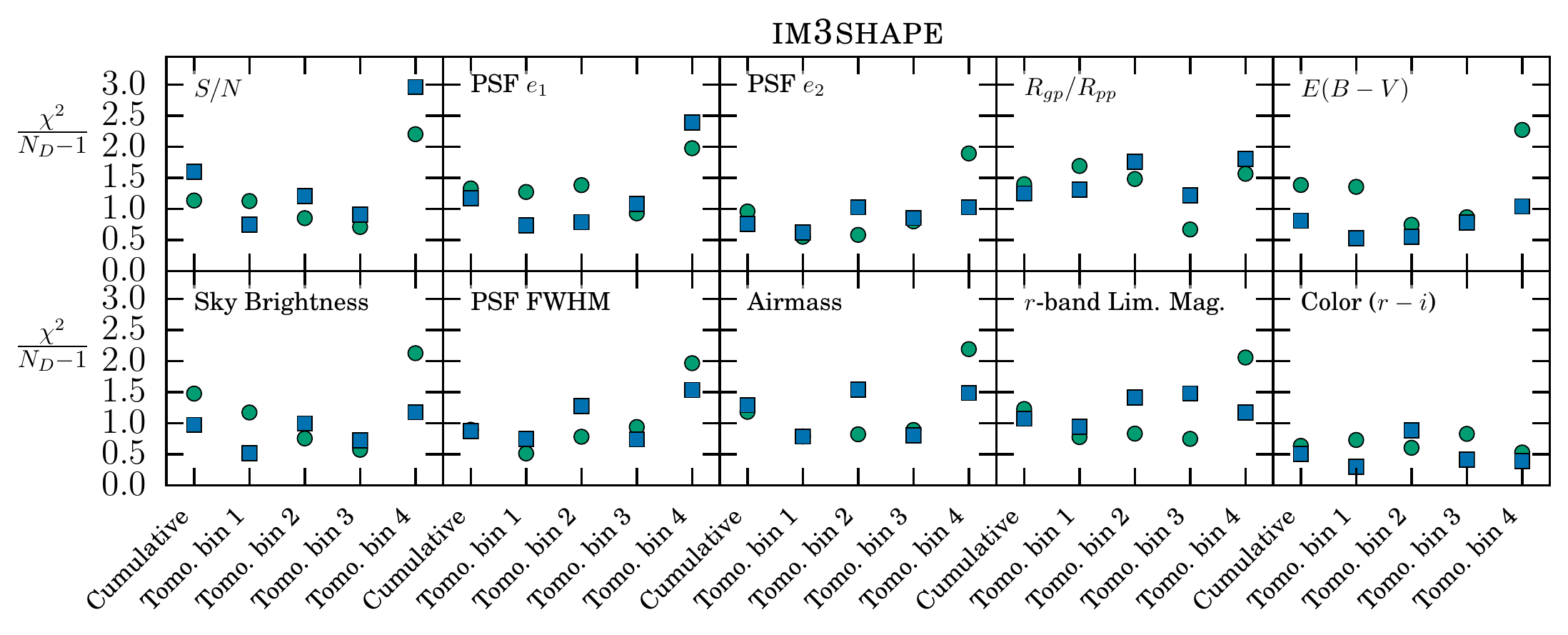}
\end{center}
\caption[]{The significance of differences in the cosmic shear signal, $\Delta \xi_{\pm}$, between subsets of the shape catalogs split by ten quantities that are most likely to be correlated with residual shear systematics. These are: signal-to-noise ($S/N$), PSF ellipticity (PSF $e_1$, $e_2$), galaxy size ($T$ - \metacal, $R_{gp}/R_{pp}$ - \imshape ), $r-i$ color, dust extinction ($E(B-V)$), sky brightness, PSF FWHM, airmass, and $r$-band limiting magnitude. We report the normalized $\chi^2$ of $\Delta \xi_{\pm}$ with the null hypothesis in each case for both $\xi_{\pm}$. The most interesting differences occur in the highest redshift bin, and when taking selection biases in the \photoz\ catalog into account, are not concerning as a contaminant to the cosmic shear signal. In all cases, the cumulative impact on the data vector combining all four auto-correlations is consistent with there being no significant residual systematic effects in the fiducial cosmic shear analysis. \label{fig:systests}}
\end{figure*}

\subsection{Methodology}\label{sec:sysmethods}

The method of implementing these tests is similar to that described in Section VI.C of \cite{becker2016}. For each shape catalog, we select ten quantities that are most likely to be correlated with residual shear systematics. These are: signal-to-noise ($S/N$), PSF ellipticity (PSF $e_1$, $e_2$), galaxy size ($T$ - \metacal, $R_{gp}/R_{pp}$ - \imshape ), $r-i$ color, dust extinction ($E(B-V)$), sky brightness, PSF FWHM, airmass, and $r$-band limiting magnitude. The first five are intrinsic properties of each galaxy image measured by our shape and PSF measurement pipelines, while the last five are taken to be the mean value of each property across exposures at a given position on the sky in \healpix~cells of $N_{\mathrm{side}}$ 4096. We exclude several properties tested in the DES SV analysis that are highly degenerate with the remaining quantities, and also do not include surface brightness, since we have identified no need to make an explicit surface brightness cut in the shape catalogs with improved DES Y1 data.

For each quantity and shape catalog, we split the objects used in the cosmic shear measurements into two halves, each with the same effective shape weight. We then correct any resulting differences in the redshift distribution of each half of the catalogs by constructing weights for each galaxy that match each redshift distribution to that of the full catalog. For an example of the impact of the redshift reweighting procedure, see Fig. 8 of \cite{becker2016}. We also recalculate $R_S$ for \metacal\ in each half, which should correct for any shear selection effects. Finally, we measure $\xi_{\pm}$ with the same binning as used in the fiducial measurement for each half using a weight that is the product of this redshift weight with any shape weight. Unlike in \cite{becker2016}, we use a single $z_{mc}$ drawn from the \photoz\ pdf of each galaxy, rather than construct the redshift weights using the full pdfs. For Gaussian pdfs, this procedure reproduces the correct peak of the original redshift distribution, but the reweighted distribution remains skewed. The effect of skewness in the reweighted redshift distribution should be subdominant to selection effects in the \photoz\ bias correction, however, and we ignore it.

We construct a covariance for these tests using 250 of the simulated shear maps described in Sec. \ref{sec:lognormal}. We build a mock catalog by sampling from the simulated shear maps an appropriate number density of objects and shape noise based on a mapping of the weights by position on the sky for each catalog subset to each simulated map. This captures the impact of added effective shape noise in the redshift reweighting procedure, but does not capture any correlation of the selection or weights with redshift in each tomographic bin. We then calculate $\Delta\xi_{\pm}$ for each quantity and its covariance. This requires approximately 200,000 measurements of $\xi_{\pm}$ in total.

A failure of these tests may not indicate a bias in any cosmological parameter constraints, but rather a failure due to selection effects not accounted for in catalog creation that does not impact the full sample. In the case of \metacal, we can explicitly correct for selection biases caused by splitting the catalog by any quantity measured during the metacalibration procedure. However, we have measured, using the resampled \cosmos~ catalog, significant selection biases in the \photoz\ distribution when splitting the catalog by, for example, $S/N$ and size. Finally, it is worth noting that a non-null detection in a test does not necessarily indicate a systematic that translates into a significant bias in cosmological parameter constraints, even though we may have the statistical power to detect it in this test, since we are taking the difference of two signals that share cosmic variance.

\subsection{Results}\label{sec:sysresults}

We report the resulting significance ($\chi^2/(N_D-1)$) for each quantity in Fig. \ref{fig:systests} for each tomographic auto-correlation (Bin. 1-4) and their combination (Cumulative), where $N_D$ is the length of the data vector. We show only the impact on auto-correlations, since any correlated systematic should be strongest there. The mean $\chi^2/(N_D-1)$ falls close to 1, as generally expected: 0.91 for \metacal~and 1.1 for \imshape. There is no significant indication of bias in either shape catalog for the `cumulative' points that combine the impact of all four tomographic bins. However, there does seem to be a higher significance for the $\Delta\xi_{\pm}$ in the highest tomographic bin in splits along several quantities for both catalogs. For \metacal, the highest significance detections of $\Delta\xi_{\pm}$ all occur with quantities that require significant redshift reweighting, and we have confirmed that a sufficient component of this $\chi^2$ is due to \photoz\ selection bias in splitting the samples, such that the apparent non-null result is not significant. We do not explicitly correct this in the figures, because there is a large uncertainty on the relative bias of subsamples.

\section{Full 1D Parameter Constraints}\label{sec:fullresults}

\begin{figure}
\includegraphics[width=\columnwidth]{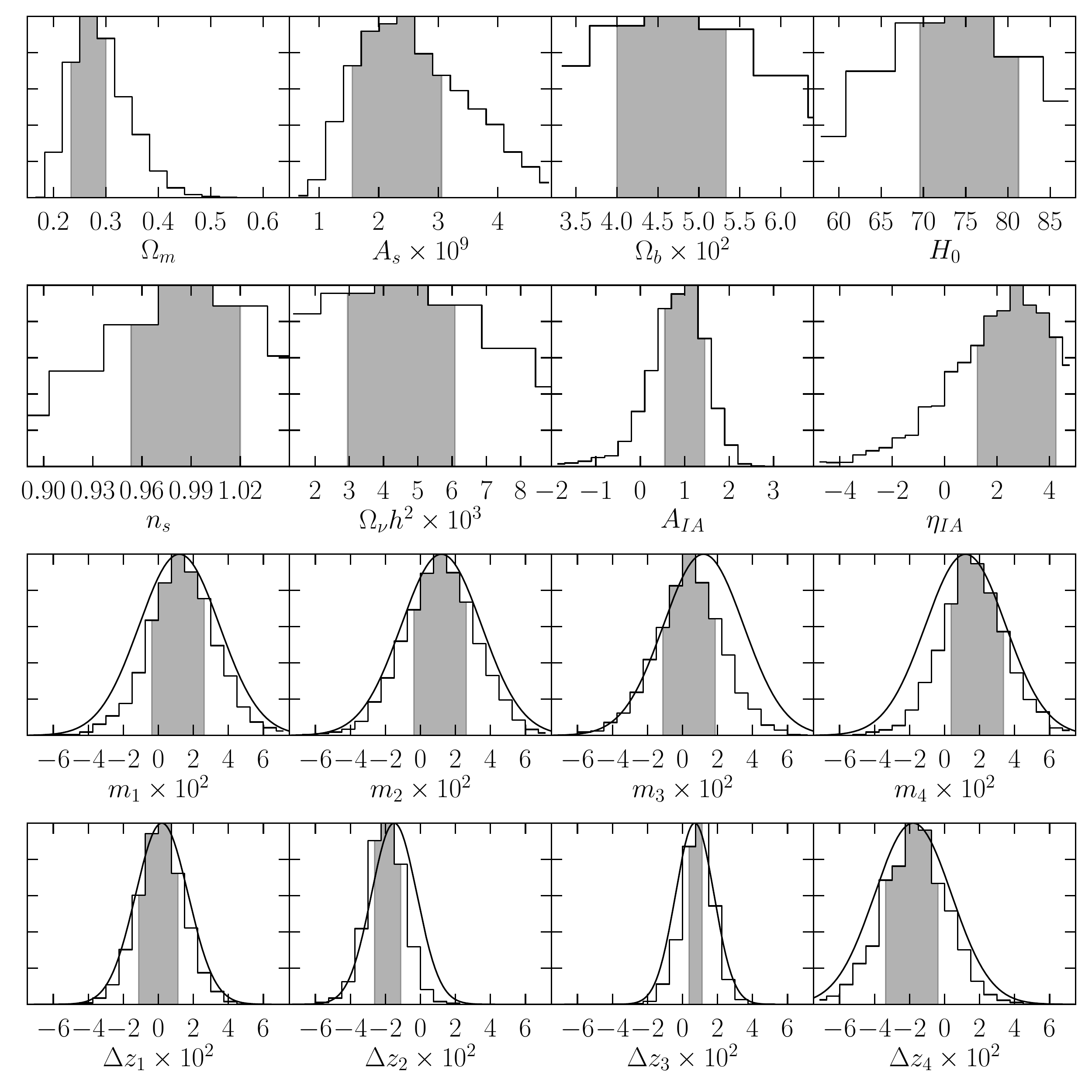}
\caption[]{The full 1D posteriors on all 16 parameters in our $\Lambda$CDM model space. Shaded regions show the approximate 68\% confidence intervals. The smooth curves show the Gaussian priors on systematic parameters.  \label{fig:1dtable}}
\end{figure}

We show 1D posteriors for the full $\Lambda$CDM parameter space in Fig. \ref{fig:1dtable}, including cosmological, astrophysical, and systematic parameters. We find no significant constraints beyond the prior imposed on the parameters $\Omega_b$, $H_0$, $n_s$, and $\Omega_{\nu}h^2$. The priors for all 16 parameters are listed in Table \ref{table:params} and a quantitative comparison of priors and posteriors of the non-cosmological parameters in Table \ref{table:syspost}. 

\label{lastpage}

\end{document}